\documentclass[a4paper,12pt]{article}
\usepackage{amssymb}
\usepackage{float}
\RequirePackage{amsthm}
\RequirePackage{enumerate}
\RequirePackage{color}
\RequirePackage{placeins}
\RequirePackage{pdfsync}
\RequirePackage{natbib}
\RequirePackage{amsmath}
\RequirePackage[hidelinks]{hyperref}
\RequirePackage{xcolor}
\RequirePackage{array}
\RequirePackage{booktabs}
\RequirePackage{fullpage}

\usepackage{enumitem}

\newlist{problems}{enumerate}{10}
\setlist[problems]{label = {(Q\arabic*)}}

\RequirePackage{multirow}
\RequirePackage{multicol}

\theoremstyle{remark}

\newcommand{\mathd}{\mathrm{d}}
\newcommand{\mathe}{\mathrm{e}}

\newcommand{\esp}{\ensuremath{\mathbb{E}}}

{\scriptsize }

\usepackage{graphicx}

\begin{document}
\title{Real-time prediction of severe influenza epidemics using Extreme Value Statistics}
\author{Maud Thomas\\
   Sorbonne Universit\'e, CNRS, LPSM, 4 place Jussieu, F-75005 Paris, France, \\maud.thomas@sorbonne-universite.fr\\
    and \\
    Holger Rootz\'en \\
    Mathematical Sciences, Chalmers University and University of Gothenburg,\\ Gothenburg, Sweden}

\date{}

\maketitle

\paragraph{Summary:} Each year, seasonal influenza epidemics cause hundreds of thousands of deaths worldwide and put high loads on health care systems. A main concern for resource planning is the risk of exceptionally severe epidemics. Taking advantage of  recent results on multivariate Generalized Pareto models in Extreme Value Statistics we develop methods for real-time prediction of the risk that an ongoing influenza epidemic will be exceptionally severe and for real-time detection of anomalous epidemics and use them for prediction and detection of anomalies for influenza epidemics in France.  Quality of predictions is assessed on observed and simulated data.  

\vfill

\section{Introduction}
\label{sec:intro}

Every year, seasonal influenza epidemics cause 250,000--500,000 deaths worldwide \citep{rambaut2008genomic} and put high strain on public health care systems in a short time frame, due essentially to increased doctor visits, and overcrowded emergency departments and intensive care units. Predicting the likelihood of an exceptionally severe epidemic in the future is of paramount importance for health resource planning \citep{breseehayden2013,khan2014health}. The three following questions are thus of central interest to public health  policy makers.
\begin{problems}
   \item \label{Q1} estimation of risks of occurrence of a very severe epidemic during the following years,
     \item \label{Q2} real-time prediction of the risk that an ongoing epidemic will be exceptionally severe, 
     and
  \item \label{Q3} real-time detection of unusual, thus potentially dangerous,  epidemics.
\end{problems}

In France, countrywide data on seasonal influenza morbidity have been available since 1985 (see Figure~\ref{fig:ILIdata} below). The Sentinelles network monitors cases of Influenza-like Illness (ILI) defined by the presence of fever in excess of 39$^{\circ}$C, respiratory symptoms, and muscle pains \citep{Sentinelles2019}. Though only a fraction of the ILI reported visits are in fact caused by influenza, their total number reflects the burden on the health care system.  
Figure~\ref{fig:ILIdata} suggests that influenza epidemics, or at least a proportion of them, may be conceptualised as extreme episodes that occur within the ILI time series. This led us to lean on   Extreme Value Statistics (EVS) to address the questions above. EVS have been developed to handle extreme events, such as extreme floods, heat waves or episodes with huge financial losses \citep[e.g.][]{katz2002statistics,embrechtsklupperberg1997}. EVS allow for prediction of risks of episodes outside of the observed range.  {The contributions of this paper are to develop  methods for real-time prediction of severe epidemics through estimation of conditional probabilities of exceedances of very high levels;  methods for detection of anomalous epidemics;  and to use the new methods for prediction  of influenza epidemics in France.}

Question~\ref{Q1} may be answered by standard and well established EVS methods { such as the univariate Peaks over Threshold method described in Section \ref{seq:predfutureyears} (for a more thorough presentation see \cite{coles2001introduction})}. For instance, they were applied by \citet{chen2015using}  to avian influenza, and by \citet{thomas2016applications} to  influenza mortality, and emergency department visits.

Previous approaches to Question~\ref{Q2} include high dimensional times series prediction  \citep{davis2016sparse} and the method of analogues \citep{Sentinelles2019}. The US Centre for Disease Control initiated a data challenge to predict the 2013--2014 US influenza epidemic. Their conclusion was ``Forecasting has become technically feasible, but further efforts are needed to improve forecast accuracy so that policy makers can reliably use these predictions'' \citep{biggerstaff2016results}. Building on recently developed EVS results and methods based on multivariate Generalized Pareto (GP) distributions \citep{rootzen2018multivariate,kiriliouk2019peaks}, we extend the toolbox of available techniques and develop methods to tackle Question~\ref{Q2}. 

 Question~\ref{Q3} is an anomaly detection problem. Recent publications have used EVS in this context  \citep{guillou2014extreme,thomas2017anomaly,goix2016apprentissage,chiapino2019multivariate}. In the present paper,  the multivariate GP models estimated in the course of solving Question~\ref{Q2} are used to detect anomalous epidemics. 

For predictions to be useful in practice, it is necessary to have an understanding of their reliability. Methods for evaluating the quality of prediction of extremes { have only very recently started to be approached in the literature.} { Here  we use a strategy that uses standardized Brier scores, Precision-Recall curves and Average Precision scores \citep{steyerberg2010assessing,brownlee2020imbalanced,saito2015precision}}. 

 { Section~\ref{sec:methods} describes the methods used in this study and in particular the multivariate GP distributions. Section~\ref{sec:ILI} presents the Sentinelles network data and the definition of epidemics. In Section~\ref{sec:results} the EVS methods are applied to the Sentinelles ILI data. Finally, Section~\ref{sec:conclusions} contains the conclusion.}

\section{Methods: EVS modelling}\label{sec:methods}

The main interest of EVS methods is to allow prediction outside of the range of the observations. In this section, we describe { and develop} the EVS tools required to answer the questions raised in this paper.  

Section~\ref{seq:predfutureyears} presents the univariate Peaks over Thresholds (PoT) method which is used for Question\ref{Q1}.  
Our approach to Question~\ref{Q2} is described in Section~\ref{subsec:realtimepredictiont}: it is based on the multivariate PoT method, { building} on recent results on multivariate EVS models \citep{rootzen2018multivariate,kiriliouk2019peaks}. Section~\ref{sec:anomaly} deals with Question~\ref{Q3} and combines a general framework for anomaly detection with Generalized Pareto  modelling. { Finally, a strategy for assessing real-time prediction of exceedances of very high levels is proposed in Section \ref{sub:assess:methods}. }


\subsection{Univariate PoT method}
\label{seq:predfutureyears}

The one-dimensional PoT method, introduced in the hydrological literature { and combined} with the { univariate GP distribution in \citet{smith1984threshold}}
 consists in selecting a suitably high threshold $u$ and defining excesses above $u$ as the differences between the observations and $u$. Under general conditions, the conditional distribution of the excesses, given that they are positive, is asymptotically as $u \to \infty$ a GP distribution with cumulative distribution function (cdf) 
\begin{equation}
\label{eq:GP}
  H(x)
  =
  \begin{cases}
    1 - ( 1 + \frac{\gamma}{\sigma}x  )^{-1/\gamma}_+
      & \text{if $\gamma \ne 0, \hspace{1mm} x> 0$} \\[1ex]
     1-\exp ( - \frac{x}{ \sigma})
      & \text{if $\gamma = 0, \hspace{1mm}x> 0.$}
  \end{cases}
\end{equation}
Here  $\sigma > 0$ is a scale parameter, and $\gamma \in \mathbb R$ is a shape parameter. For $a \in \mathbb R$, $(a)_+ =a$ if $a\geq0$ and $0$ if $a<0$. The parametrisation is chosen so that the cdf for $\gamma = 0$ is the limit as $\gamma \to 0$ of the cdf with non-zero $\gamma$. A useful account of this method is given in \citep{coles2001introduction}.

Assuming the observations are independent and identically distributed with common cdf $F$, the parameters of the cdf  \eqref{eq:GP} are  estimated from the observed excesses. For $y>u$, $F(y)$ is estimated,  by 
\[
\widehat F(y) = 1-\widehat{p}_u (1-\widehat{H}(y-u)),   
\]
where $\widehat{H}$ is the GP cdf~\eqref{eq:GP} with parameters replaced by their estimated values, and $\widehat{p}_u$ is the empirical frequency of observations which exceed the threshold $u$ \citep{coles2001introduction}.



 The level that the maximum of $n$ observations will exceed  with probability $1-\alpha$ 
is estimated by the $\alpha$-th quantile $\widehat y_\alpha$ of $\widehat F(y)^n$. For example, if $\gamma=0$, $H$ is an exponential distribution then
\begin{equation}\label{eq:quantile}
 \widehat y_\alpha = u + \widehat{\sigma} \{\log \widehat{p}_u - \log (1-\alpha^{1/n})\},
\end{equation}
for $\alpha$ such that $\widehat{p}_u \geq 1-\alpha^{1/n}$.


\subsection{Multivariate PoT method}
\label{subsec:realtimepredictiont}





The multivariate PoT method was introduced by { \citep{rootzen-tajvidi2006,michel2009parametric,brodin2009univariate}}. A suitably high threshold is chosen for each component of { a $d$-dimensional random vector $\boldsymbol Y = (Y_1,\ldots,Y_d)$. Let denote $\boldsymbol u = (u_1,\ldots,u_d)$ the $d$-dimensional vector of thresholds}. Excess vectors $\boldsymbol X = (X_1,\ldots,X_d)$ are defined as the component-wise differences between observations and thresholds and considered as positive if at least one of the components exceeds its threshold. Under general conditions, the joint distribution of the positive excess vectors is asymptotically a multivariate GP distribution as the thresholds tend to $\infty$. For the sake of completeness, we briefly recall the definitions and properties of multivariate GP distributions that will be needed in this paper \citep[for further details, see][]{rootzen2018multivariate,kiriliouk2019peaks}.  { For simplicity, we describe the method for $d=3$, but it can easily extended to any dimension $d$.}


Unlike in the univariate case, the family of multivariate GP distributions cannot be described as a parametric family. \citet{rootzen2018multivariate} have developed four representations of multivariate GP distributions for which closed formulas of densities are available. In this paper, we use the $U$-representation since it presents nicer properties across the dimension, which are useful for prediction. Moreover, at the price of standardization, the marginals of the distributions can be assumed to be standard exponential distributions \citep[see Section~3 in][]{kiriliouk2019peaks}. 
    Let $\boldsymbol{U}= (U_1, U_2, U_3)$ be a 3-dimensional random vector such that $\mathbb{E}[\mathe^{\max \boldsymbol{U}}]< \infty$, where $\max \boldsymbol{U} = \max\{U_1, U_2, U_3\}$, and let $ f_{\boldsymbol{U}} $ be the probability density function of $\boldsymbol{U} $. For $i=1,2,3$, let $f_i$ and $F_i$ denote the density and distribution functions of $U_i$, respectively. The vector $\boldsymbol U$ or its distribution is referred to as the generator.


According to Equation~(3.4) in \citep[][]{kiriliouk2019peaks}, the density  function $h_{\boldsymbol{U}}$ of the 3-dimensional GP distribution with standard exponential margins  generated by $\boldsymbol{U}$ is
\begin{equation}
\label{eq:urepresentation}
h_{\boldsymbol{U}}(\boldsymbol{x})= \frac{1_{\{\boldsymbol{x} \nleq \boldsymbol{0}\} }}{\esp[\mathe^{\max\boldsymbol{U}}]} \int_{0}^\infty f_{\boldsymbol{U}}(
\boldsymbol{x} +\log t) \mathd t,
\end{equation}
where $\boldsymbol{x}=(x_1,x_2,x_3)$ and $\boldsymbol{x} + \log t  = (x_1+ \log t, x_2 +\log t, x_3 + \log t)$. The indicator function $1_{\{\boldsymbol{x} \nleq \boldsymbol{0}\}}$ equals one if at least one of the components of $ \boldsymbol{x}$ are positive, and is zero otherwise. Different choices of distributions for $\mathbf U$ yield different GP models.

Our answer to Question \ref{Q2} is to estimate the conditional probability that $Y_3$ will exceed some level $\ell>u_3$ given $Y_1$ and $Y_2$, or equivalently, for $\mathbf{X } = (X_1, X_2, X_3) = (Y_1-u_1, Y_2-u_2, Y_3-u_3)$, the conditional probability that $X_3>\ell - u_3$ given $X_1$ and $X_2$. For this we assume that $\mathbf{X }$ is positive and follows a multivariate GP distribution with generator $\mathbf U$, 
\begin{equation}\label{eq:conditionalprobability}
P[X_3 \geq\ell | X_2=x_2, X_1=x_1] =
 \frac{ \int_{x_3 =\ell}^{\infty} 1_{\{\boldsymbol{x} \nleq \boldsymbol{0}\}} \int_{0}^\infty f_{\boldsymbol{U}}(\boldsymbol{x} +\log t) \mathd t \mathd x_3}
{\int_{x_3=-\infty}^{\infty} 1_{\{\boldsymbol{x} \nleq \boldsymbol{0}\}} \int_{0}^\infty f_{\boldsymbol{U}}(
\boldsymbol{x} +\log t) \mathd t \mathd x_3}.
\end{equation}

Formulas for general $\mathbf U$ distributions are given in the Supplementary Material (Proposition~1, Section~3).

\subsection{Anomaly detection}\label{sec:anomaly}

Question~\ref{Q3} belongs to the field of anomaly detection. In the present context, a natural approach  is to use the estimated GP model from Section~\ref{subsec:realtimepredictiont} to detect { new observations} that exhibit a significantly different pattern from the data used to fit the model. A statistical test for anomalous { observations} may be based on the GP negative log-likelihood, with a very large value suggesting that the new observation might be anomalous. The quantiles of the GP negative log-likelihood distribution were estimated by simulation in order to define the decision region of the test \citep[e.g. Section~2 of][]{root2015learning}. However, it must be stressed that ``anomalous'' has to be understood with respect to the fitted GP model. 

\subsection{Assessment strategy for real-time predictions}\label{sub:assess:methods}


There is a substantial literature on assessment of forecasting \citep[see][and references therein]{lerch2017forecaster}. { This literature provides metrics aimed at comparing predictions with observations. However, standard assessment metrics are not appropriate when the frequency of exceedances in the data is small and sometimes} { zero. This issue was tackled in \citep{renard2013data}, but they aimed at a rather different context, spatial environmental data, where prediction is done at not just one, but many stations. Here, we wish to compare estimated prediction probabilities and the observed outcome for just one data set, not many.}


{ For this purpose}, we use (i) standardized Brier scores; (ii) Precision-Recall Curves; and (iii) Average Precision scores, together with simulations from estimated models. { We also illustrate prediction quality visually with stratified plots of predicted probabilities of exceedance.} 

(i) The standardized Brier score is defined as
\[
1 - \frac{\frac{1}{N}\sum_{i=1}^N \left(\widehat{p}_i - o_i\right)^2}{p(1-p)},
\]
where $N$ is the number of predictions, $\widehat{p}_i$ is the prediction probability of exceedance, $o_i =1$ if an exceedance was observed, and $0$ otherwise,  and $p=\frac{1}{N}\sum_{i=1}^{N}o_i$, see e.g. \citep{steyerberg2010assessing}. The score  is bounded by $1$, with larger values indicating better prediction. Using the predictors $\widehat p_i = p$ gives the value $0$.

(ii) A question such as {``will  the maximum of the past observations be exceeded in the future?''} may be formulated as a binary classification problem. The data are divided into two classes: Positives (exceedances) and Negatives (no exceedances). The strategy consists first in computing $\widehat p_i$ and then in comparing this estimate to some cut-off probability value $p_c$. If $\widehat p_i \geq p_c$ then the observation is assigned to the Positives class, and to the Negatives class otherwise. The ``true positives'' (``false positives'') correspond to the observations that are correctly (incorrectly) assigned  to the Positives class, and the ``true negatives'' (``false negatives'') to the observations that are correctly (incorrectly) assigned to the Negatives class. 

Common methods to assess the performance of binary classifiers include true positive and true negative rates, and ROC (Receiver Operating Characteristics) curves. These methods, however, are uninformative when the classes are severely imbalanced, which is the case when predicting very high level exceedances which are rare by nature.  In this context, \citet{saito2015precision} have argued that Precision-Recall curves are more informative. These curves display the values of
$$
\mathrm{Precision} (p_c)= \frac{\mbox{true positives}}{\mbox{true positives + false positives}}
$$
against the values of
$$
\mathrm{Recall} (p_c) = \frac{\mbox{true positives}}{\mbox{true positives + false negatives}} \, .
$$
as the cut-off probability $p_c$ varies from $0$ to $1$. 
Precision quantifies the number of correct positive predictions out all positive predictions made; and Recall {    (often also called Sensitivity)} quantifies the number of correct positive predictions out of all positive predictions that could have been made. Both focus on the Positives class (the minority class) and are unconcerned with the Negatives (the majority class). The Precision-Recall curve of a skillful model bows towards the point with coordinates $(1,1)$. The curve of a no-skill classifier will be a horizontal line on the plot with a y-coordinate  proportional to the number of Positives in the dataset. For a balanced dataset this will be 0.5 \citep{brownlee2020imbalanced}. 

(iii) The Average Precision score is an approximation to the area under the Precision-Recall curve \citep{su2015relationship}. A perfect prediction  model would have an Average Precision score equal to 1, and the closer the score is to 1, the better the prediction performance of the model. 


\section{The Sentinelles network data}
\label{sec:ILI}

The French nationwide Sentinelles network consists of approximately 1,500 general practitioners in France who participate on a voluntary basis in the ILI surveillance, and report new cases of ILI observed in their practice. Based on these data, nationwide weekly ILI incidence rates---i.e. numbers of new cases in France per week per 100,000 individuals---are estimated { from a Poisson model which also provides confidence intervals \citep[for further details, see][]{Sentinelles2019,souty2014improving}.} { Estimating incidence is difficult, especially since there is no available gold standard for comparison. The quality of the data produced by the Sentinelles network has been assessed in comparison with ILI incidence estimated from independent sources \citep[see e.g.][]{kalimeri2019unsupervised,carling2013risks,pelat2017improving}.}

In the Sentinelles network, epidemics are identified using the Serfling method  \citep{serfling1963methods}. It consists in fitting a cyclic regression model to the weekly ILI rates and setting the start of the epidemic at the first of the first two consecutive weeks during which the ILI incidence rates exceed the upper bound of the 90\% prediction interval \citep[for further details see][]{Sentinelles2019}. 

\begin{figure}
	\begin{center}
		\includegraphics[width=0.85\textwidth]{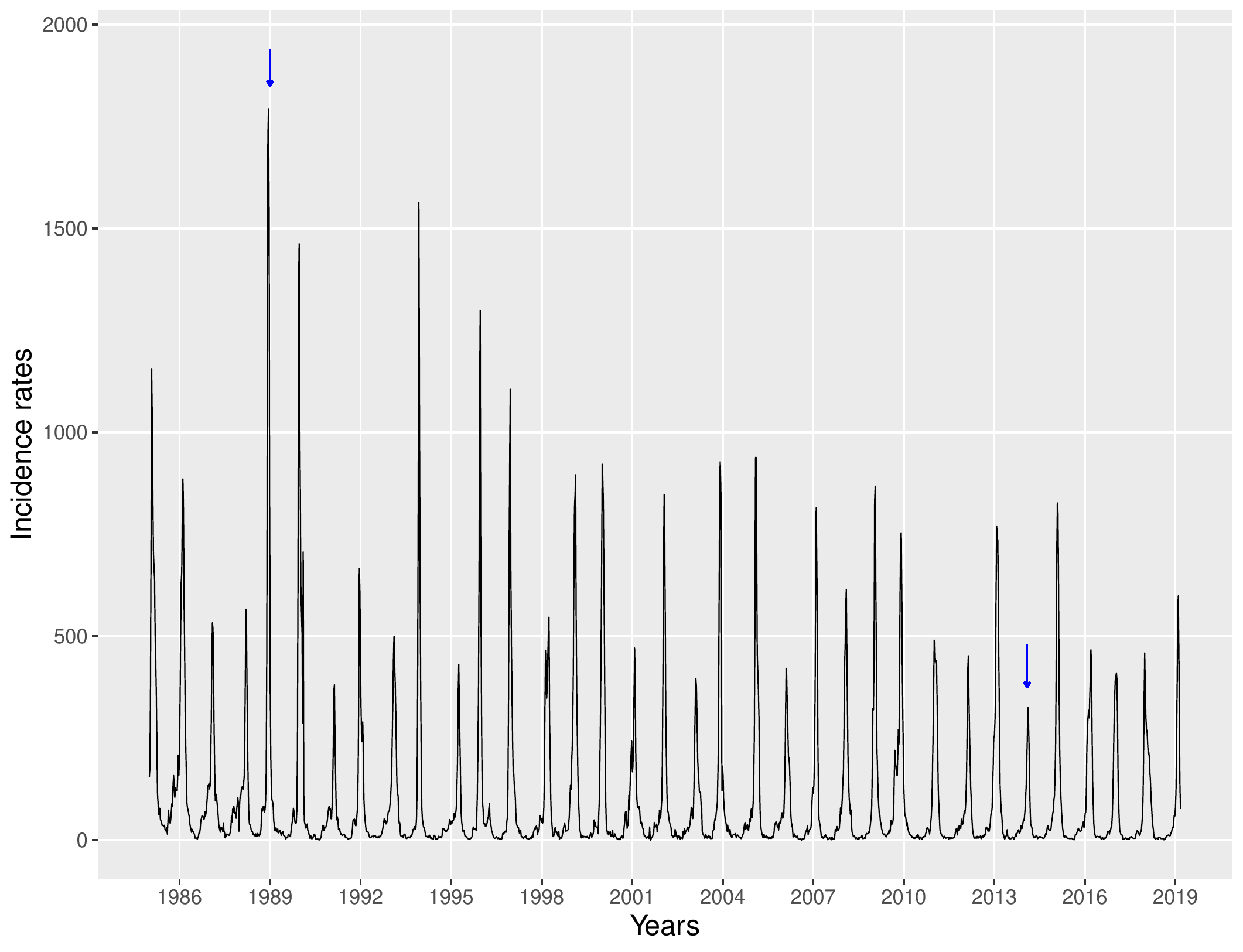}
	\end{center}
	\caption{Weekly ILI incidence rates in metropolitan France from January 1985 to February 2019. Arrows indicate epidemics with the highest and the lowest peaks.}
	\label{fig:ILIdata}
\end{figure}

Weekly ILI incidence rates from January 1985 to February 2019 were downloaded for analysis. Figure~\ref{fig:ILIdata} shows the time series of weekly ILI incidence rates, which includes 35 epidemics. The epidemic with the highest peak corresponds to the 1989-epidemic, with a value of 1,793. The lowest peak was observed in 2014, with a value of 325. The durations of the Serfling epidemics range from 5 to 16 weeks, in 1991{/2014} and 2010, respectively.  

\begin{figure}
	\begin{center}
		\begin{tabular}{cc}
			\includegraphics[width = 0.45\textwidth]{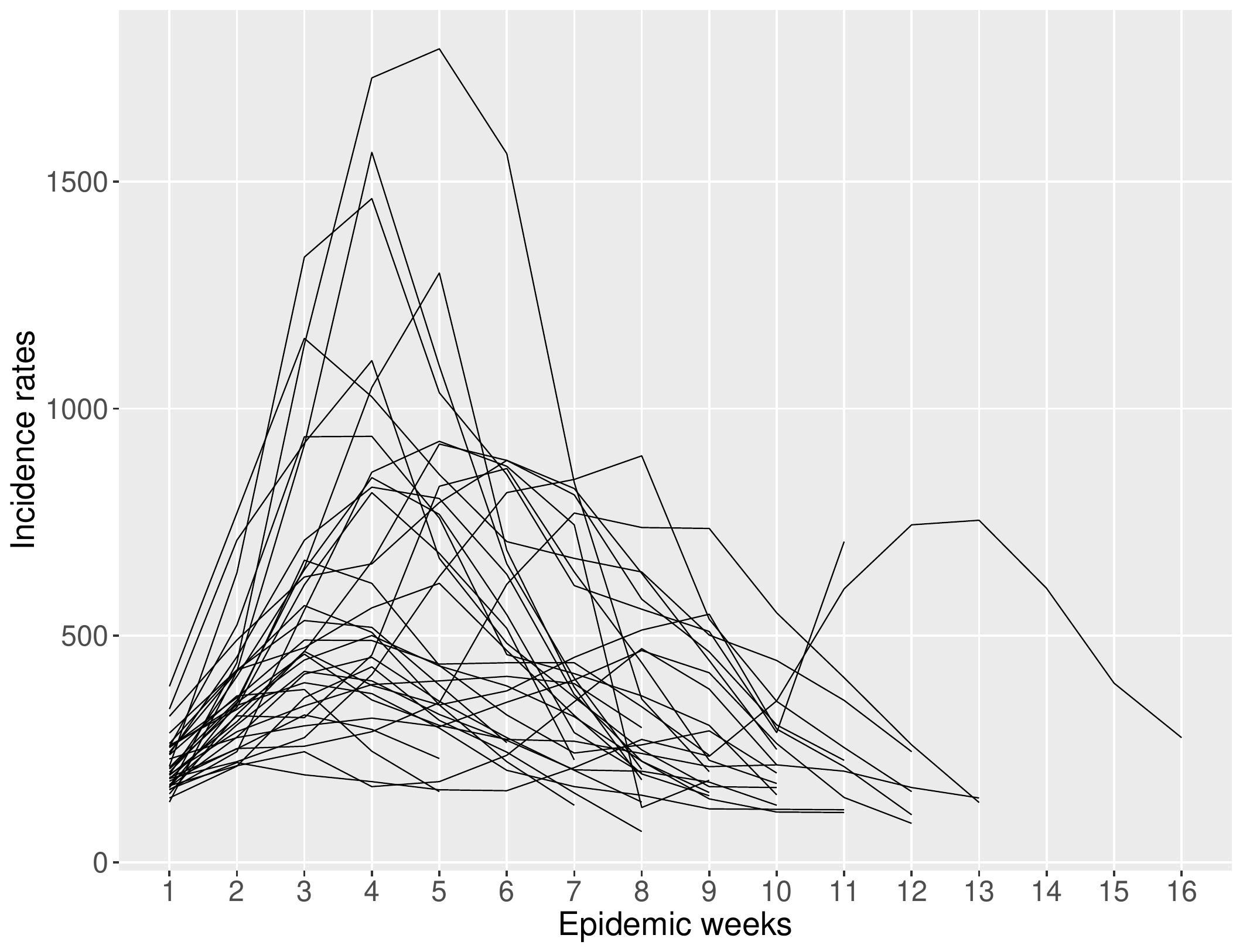}&
			\includegraphics[width = 0.45\textwidth]{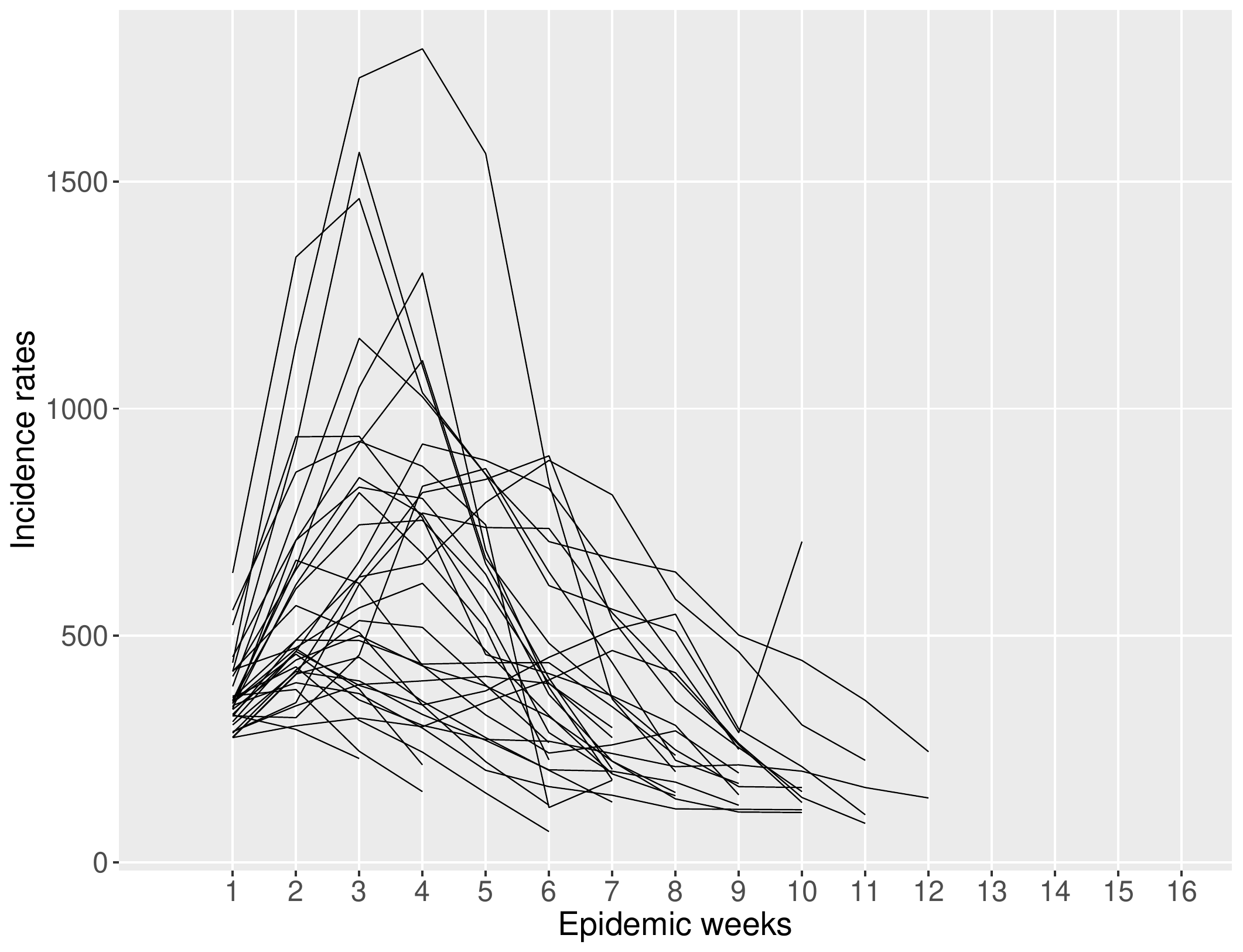}\\
			\label{fig:synch:epid:thres}\\
			a) & b)
		\end{tabular}
	\end{center}
	\caption{a) Weekly ILI incidence rates for epidemics identified with the Serfling method, and b) Weekly ILI incidence rates for epidemics obtained from the definition in this paper.}
	\label{fig:synch:epid}
\end{figure}

In this study, the 1985--2018 data were used for estimation and the 2019 data were kept as a test sample. Since we were interested in very high ILI incidence rates and relied on EVS methods, we  focused on the most active part of the epidemics. The Serfling method was thus adapted as follows. The start of the epidemic was set at the first of the first two consecutive weeks during which the ILI incidence rates exceeded { the 0.88-quantile (=272) of the 1985-2018 data}. The end was set at the end of the Serfling epidemic. {This quantile} was chosen to synchronise the peaks of the epidemics as much as possible, see Figure~\ref{fig:synch:epid}. This definition detected 34 epidemics between 1985 and 2018, and was thus in accordance with the Sentinelles network. The durations of the 34 epidemics ranged from 3 to 12 weeks in 2014 and 1985/{2018}, respectively.

The size of an epidemic was defined as the sum of the weekly ILI incidence rates from the start to the end of the epidemic. The smallest epidemic size was 847 in 2014 and the largest was 8,062 in 1989.

The size of an influenza epidemic depends on multiple factors including immunity and vaccination prevalence in the population. Some of these factors may likely be impacted by the sizes of past epidemics. Nevertheless, there is little indication that this translates into statistical dependence between epidemics--{ as} shown by the correlation and distance correlation plots in the Supplementary Material (Section~1)--{ and} we thus assumed that the ILI incidence rates of different epidemics were mutually independent.

\section{Results: prediction of  very high ILI loads on the { French} health care system}
\label{sec:results}

In this section, the methods described in Section~\ref{sec:methods} are applied to the Sentinelles ILI data. Recall that the 1985--2018 data were used for estimation and the 2019 data were kept as a test sample. 

For each epidemic, we shall refer to the first week of the epidemic as Week 1, the second week as Week 2 and so on until the end of the epidemic. For $j=1,2,3$ and $k=1985, \ldots, 2019$, we let $Y_j^k$ denote the ILI incidence rate of Week $j$ in year $k$, and $S^k$ denote the size of the epidemic of year $k$. We omit the index $k$ when we refer to a generic epidemic. 

To address Questions \ref{Q1}, \ref{Q2} and \ref{Q3}, we { choose to} focus on predictions for the Week 3 ILI incidence rate $Y_3$ and the epidemic size $S$ { given  observation of the Week 1 and 2 incidence rates. This choice was driven by both epidemiological and technical reasons. Policy makers will be predominantly interested in early predictions of the ongoing epidemic.} {   Thus, basing predictions on observation of the first two weeks was a reasonable compromise between accuracy of prediction and early prediction.  If a very high ILI incidence rate in Week 3 was predicted, this would suggested a rapid intensification of the ongoing epidemic, and prediction of very large epidemic size would mean that the ongoing epidemic will be severe.  Technically the number of parameters in models of dimension above 3 was too large for reliable computation, given the amount of available observations.}

\subsection{Question~\ref{Q1}: risk of very high ILI incidence rates and epidemic sizes over the following years}
\label{sec:riskfutureyears}


The univariate PoT method was applied to the Week 3 ILI incidence rates $(Y_3^{1985}, \ldots,Y_3^{2018})$ and to the epidemic sizes $(S^{1985},\ldots, S^{2018})$. 

As explained in Section~\ref{seq:predfutureyears}, the first step is to choose a suitably high threshold for each series of observations. The threshold must be high enough to ensure that the asymptotic model is valid, but low enough to yield a sufficient number of positive excesses. For the ILI rates, the threshold $u_{I}$ was chosen as the 0.9-quantile (=339) of the whole  1985-2018 series; for epidemic sizes, the threshold $u_S$ as the 0.6-quantile (=4,144) of the series of epidemic sizes from 1985 to 2018.  These thresholds yield 30 positive excesses for Week 3 ILI rates and 14 for epidemic sizes.

 
 One-dimensional GP distributions (Equation~\eqref{eq:GP}) were fitted to the ILI rate and size positive excesses. Likelihood ratio tests showed that the hypothesis $\gamma=0$ was not rejected ($p=0.64$, and $p=0.98$ respectively), so that cdf's were assumed  exponential. QQ-plots and estimates of scale parameters  are shown in the Supplementary Material (Section~2, Figure~2 and Table~1). 


Table~\ref{tab:estimates:risk} presents estimates of the levels $\widehat y_\alpha$ { such} that during the following year and { during} the following 10 years the Week 3 ILI incidence rate and the epidemic size will exceed { them} with given probability $1-\alpha$. These estimates were computed using Equation~\eqref{eq:quantile}. For Week 3 ILI incidence rates, $u_I=339$ yielded $\widehat p_{u_I} =0.88$ and for epidemic sizes $u_S=4,441$ yielded $\widehat p_{u_S} =0.41$.  
%


\begin{table}
\begin{center}
\caption{Estimated levels that the Week 3 ILI incidence rate and the epidemic size will exceed with either probability 10\% or 1\% during the following year and the following 10 years.}
  \label{tab:estimates:risk}
\vspace{4mm}
 \begin{tabular}{c|cccc}
&  one year &  one year &    10 years &   10 years\\
$1-\alpha$ & 10\%  & 1\% &  10\% & 1\%\\
\hline
Week 3 ILI incidence rates & 1,192 &  2,094 & 2,076 & 2,994 \\
Epidemic sizes & 6,165 & 9,452 & 9,385 & 12,733\\
 \end{tabular}
 \end{center}
\end{table}

{ For the 2019 epidemic, the Week 3 ILI incidence rate was 599 and the epidemic size was 1,505. The observations were well below the 90\% estimated quantile for the Week 3 ILI incidence rate and the epidemic size given in Table \ref{tab:estimates:risk}. For example, it was estimated that there is a 10\% probability that the epidemic size will exceed 9,385 at least once during the next 10 years. }

\subsection{Question~\ref{Q2}: real-time prediction of very high ILI incidence rates and epidemic sizes}
\label{subsec:realtimepredictionresults}

\sloppy  In this section, the multivariate PoT method is applied to $(\mathbf Y^{1985}, \ldots,\mathbf Y^{2018})$ and $(\mathbf S^{1985}, \ldots,\mathbf S^{2018})$ where $\mathbf Y^{k}= (Y^{k}_1,Y^{k}_2, Y^{k}_3)$ and $\mathbf S^{k}= (Y^{k}_1,Y^{k}_2, S^{k})$ for $k=1985,\ldots, 2018$.

The thresholds $u_I$ of $339$ for the ILI incidence rates $Y_1$, $Y_2$ and $Y_3$ and $u_S$ of $4,144$ for the epidemic size $S$ were as in the previous section. For both $(\mathbf Y^{1985}, \ldots,\mathbf Y^{2018})$ and $(\mathbf S^{1985}, \ldots,\mathbf S^{2018})$, there were 32 positive excess vectors as defined in Section~\ref{subsec:realtimepredictiont}.  In order to meet the assumption of standard exponential marginals, positive excess vectors were standardized by dividing each component by the corresponding scale parameter estimates (Table~1, Supplementary Material).


To define a multivariate GP model, the distribution of the generator $\mathbf U=(U_1,U_2,U_3)$ must be specified. Assuming that the three components $U_1$, $U_2$, $U_3$ of $\mathbf U$ were mutually independent, three families of marginal distributions were considered: Gumbel, reverse exponential, and reverse Gumbel distributions. The corresponding models were fitted to the 32 positive excess vectors. Formulas for the densities $h_{\boldsymbol{U}}$ for the three GP models are given in the Supplementary Material (Section~4).


Table~\ref{tab:aic} shows that both in terms of AIC and BIC, the best fit was that of the GP family with Gumbel generator, for both $\mathbf Y$ and $\mathbf S$. According to Equation~\eqref{eq:urepresentation}, the corresponding density is 
\begin{equation}
h_{\mathbf U}(\mathbf x) = \frac{\int_0^\infty \prod_{i=1}^3 \alpha_i (t \mathe^{x_i-\beta_i})^{-\alpha_i}\mathe^{-(t\mathe^{x_i-\beta_i})^{-\alpha_i}}\mathd t}{\int_0^\infty \left(1- \prod_{i=1}^3 \mathe^{-(t/\mathe^{\beta_j})^{-\alpha_j}} \right) \mathd t} \, , 
\end{equation}
with $\alpha_1, \alpha_2, \alpha_3 >1$ and $\beta_1, \beta_2, \beta_3 \in \mathbb R$ (for further details see Supplementary Material, Section~4 and \citep{kiriliouk2019peaks}). In the sequel, we refer to this GP model as the Gumbel model. The estimated parameters of the Gumbel model are given in the Supplementary Material (Table~2). More parsimonious sub-models were considered and consistently rejected on the basis of AIC, BIC and log-likelihood ratio tests (Supplementary Material, Table~3).

\begin{table}
\caption{AIC and BIC of GP models for Week 3 ILI incidence rates and for epidemic sizes}
 \label{tab:aic}
\begin{center}
 \begin{tabular}{c|ccc}
Generator &  Gumbel & Reverse exponential & Reverse Gumbel \\
\hline
AIC &  194 & 2189 & 208\\
BIC &  202 & 2196 & 215 \\
 \end{tabular}\\~\\
 a) Week 3 ILI incidence rates\\~\\
 \end{center}
 \begin{center}
 \begin{tabular}{c|ccc}
Generator &  Gumbel & Reverse exponential & Reverse Gumbel \\
\hline
AIC & 227 & 2240 & 268\\
BIC & 234 & 2248 & 275 \\
 \end{tabular}\\~\\
 b) Epidemic sizes
 \end{center}
\end{table}





The estimated model was used to provide estimates of the probability that $Y_3$ and $S$ will exceed a specified level given that $Y_1$ and $Y_2$ have been observed. Since Equation~\eqref{eq:conditionalprobability} is valid for positive excess vectors only, two situations must be considered
\begin{enumerate}
	\item[i)] If at least one of $Y_1$ or $Y_2$ exceeds its threshold, Equation~\eqref{eq:conditionalprobability} is valid whatever $Y_3$ {(or $S$)}. 
	\item[ii)] If neither $Y_1$ nor $Y_2$ exceeds its threshold, then whether the excess vector will be positive or not is unknown. In this case, the right-hand side of Equation~\eqref{eq:conditionalprobability} must be multiplied by the probability that $Y_3$ (or $S$) exceeds its threshold given that $Y_1$ and $Y_2$ do not exceed theirs. The corresponding empirical probability was 0.33 for both $Y_3$ and $S$. 
\end{enumerate}

The procedure is illustrated in Table~\ref{tab:pred:2019} which presents predictions for the 2019 epidemic. The largest observed Week 3 ILI incidence rate between 1985 and 2018 was 1,729. The table shows the estimated probabilities that the 2019 Week 3 ILI incidence rate exceeds a fraction $\kappa$ of the 1985-2018 maximum ILI incidence rate 1,729 for $\kappa= 0.5, 0.75, 0.95, 1$. Estimates for epidemic sizes are presented similarly with a largest observed epidemic size of { 8,062}. The prediction probabilities, even for the lowest level were quite small, and in fact this level was not exceeded in 2019. For the 2019 epidemic the Weeks 1, 2 and 3 ILI incidence rates were {366, 540, and 599}, respectively, and the epidemic size was {1,505}.

\begin{table}
 \begin{center}
  \caption{Prediction probabilities for the 2019 epidemic of exceedances of levels 1,729$\times \kappa$ for Week 3 ILI incidence rates and { 8,062}$\times \kappa$ for epidemic sizes.}
    \label{tab:pred:2019}
\vspace{4mm}

 \begin{tabular}{c|cccc|cccc}
 & \multicolumn{4}{c|}{Week 3 ILI incidence rates} & \multicolumn{4}{c}{Epidemic sizes}\\
\hline
$\kappa$ & 0.5 & 0.75 & 0.95 & 1 & 0.5 & 0.75 & 0.95 & 1 \\
 Level & 864 & 1,297 & 1,643 & 1,729 &  4,031 & 6,046 & 7,659 & 8,062\\
\hline
Probability & 0.185 & 0.012 & 0.001 & 0.0007 & 0.026 & 0.008 & 0.003 & 0.002\\
 \end{tabular}
 \end{center}
 \end{table}

\subsection{Question~\ref{Q3}: real-time prediction of anomalous epidemics}\label{subsec:results:anomaly}

The quantiles of the GP negative log-likelihood were obtained as follows: the estimated Gumbel model was used to generate 1,500 datasets, each consisting of 33 three-dimensional positive excess vectors. For each simulated dataset, a Gumbel model was fitted to the 32 first vectors and the estimated negative log-likelihood was computed at the 33rd vector. The significance levels obtained as the empirical quantiles of these negative log-likelihoods are shown in Table~\ref{tab:significancelevels}. 

\begin{table}
\begin{center}
  \caption{Quantiles of the estimated negative log-likelihood.}
   \label{tab:significancelevels}
\vspace{4mm}
 \begin{tabular}{c|cccc}
Significance level & 10\% & 5\% & 1\% & 0.1\%  \\
\hline
Quantile  & 4.72 & 5.60 & 7.79 &  14.50\\
 \end{tabular}
\end{center}
\end{table}

\begin{figure}
\begin{center}
 \includegraphics[width = 0.75\textwidth]{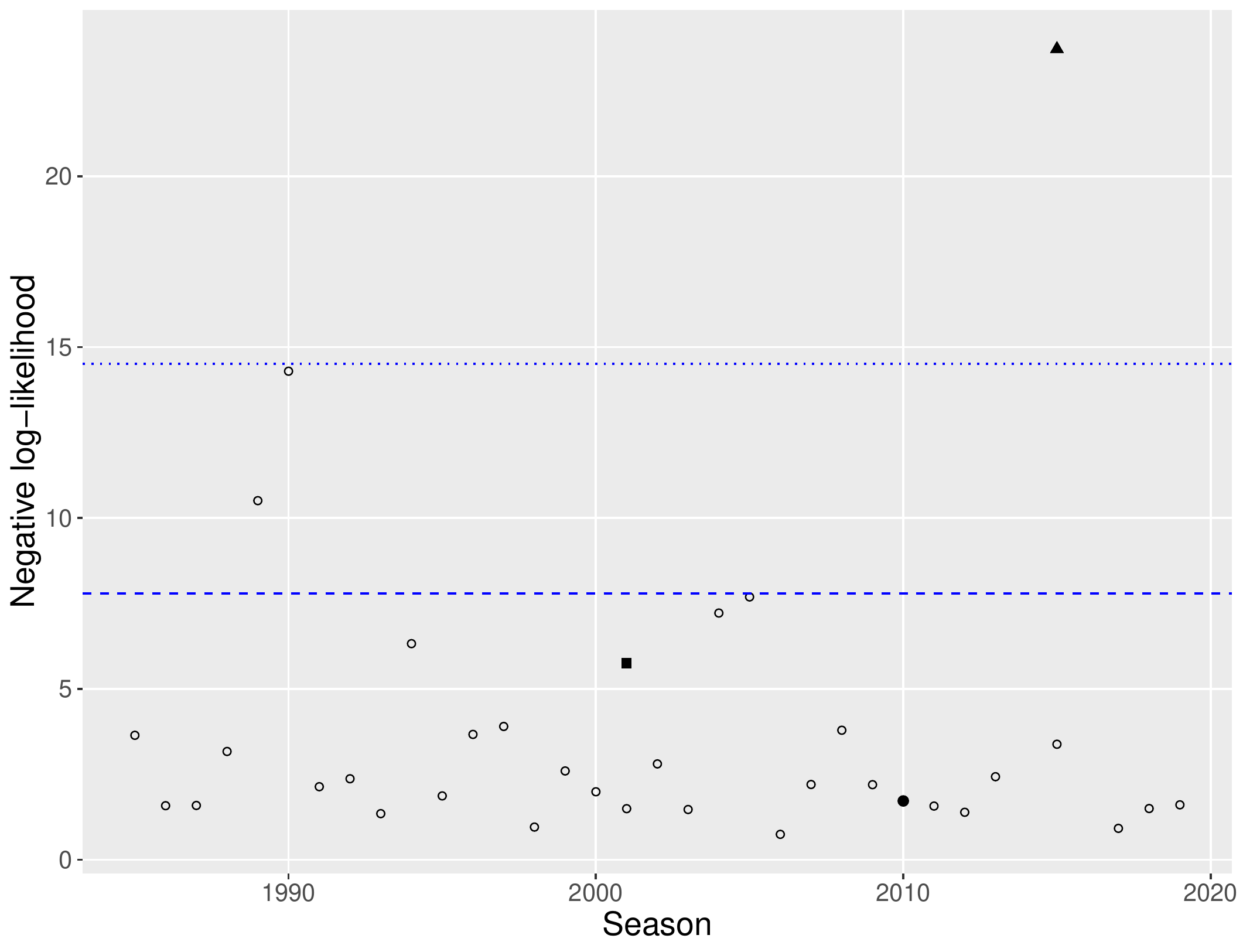}
\end{center}
 \caption{Leave-one-out negative Gumbel log-likelihoods for the ILI incidence rates of Weeks 1-3 of the 33 epidemics with positive excesss vectors observed between 1985 and 2019 (open circles). Swine flu pandemic (closed circle), simulated point with very large third component (closed square), and simulated anomalous point (closed triangle). Dashed and dotted blue lines are quantiles for the 1\% and 0.1\% significance levels.}
\label{fig:loo:llk}
\end{figure}

To illustrate the meaning of ``anomalous'', we simulated two extra positive excess vectors: (i) a positive excess vector with a very high third component equal to the 0.99-quantile of the third component of simulated positive excess vectors and (ii) an anomalous positive excess vector obtained from (i) by multiplying the first and the third components by 1.5 and the second by 0.5. Figure~\ref{fig:loo:llk} shows that only the anomalous point (ii) exceeds the 0.999-quantile. Interestingly, the 2009-10 swine flu pandemic was not unusual compared to the other epidemics. 







\subsection{Quality of real-time predictions}
\label{sec:accuracy}

Quality of real-time predictions was assessed on both the Sentinelles 1985-2019 series and on simulated data. For purpose of comparison, prediction probabilities were also estimated using a standard logistic regression model, with $Y_1$ and $Y_2$ as covariates and response variable coded as 1 in the presence of an exceedance and 0 otherwise.  Levels used in this section are those defined in Table~\ref{tab:pred:2019} (Section~\ref{subsec:realtimepredictionresults}). 


\paragraph{Assessment on the 1985-2019 ILI data}

A leave-one-out procedure was performed on the 1985--2019 epidemics yielding 35 estimates of prediction probabilities for the two lower levels ($\kappa =0.5,0.75$). Prediction probabilities for the two higher levels could not be estimated since there was only one exceedance for $\kappa = 0.95$ and none for $\kappa =1$. 

\begin{figure}
 \begin{center}
 \begin{tabular}{cc}
 \includegraphics[width = 0.38\textwidth]{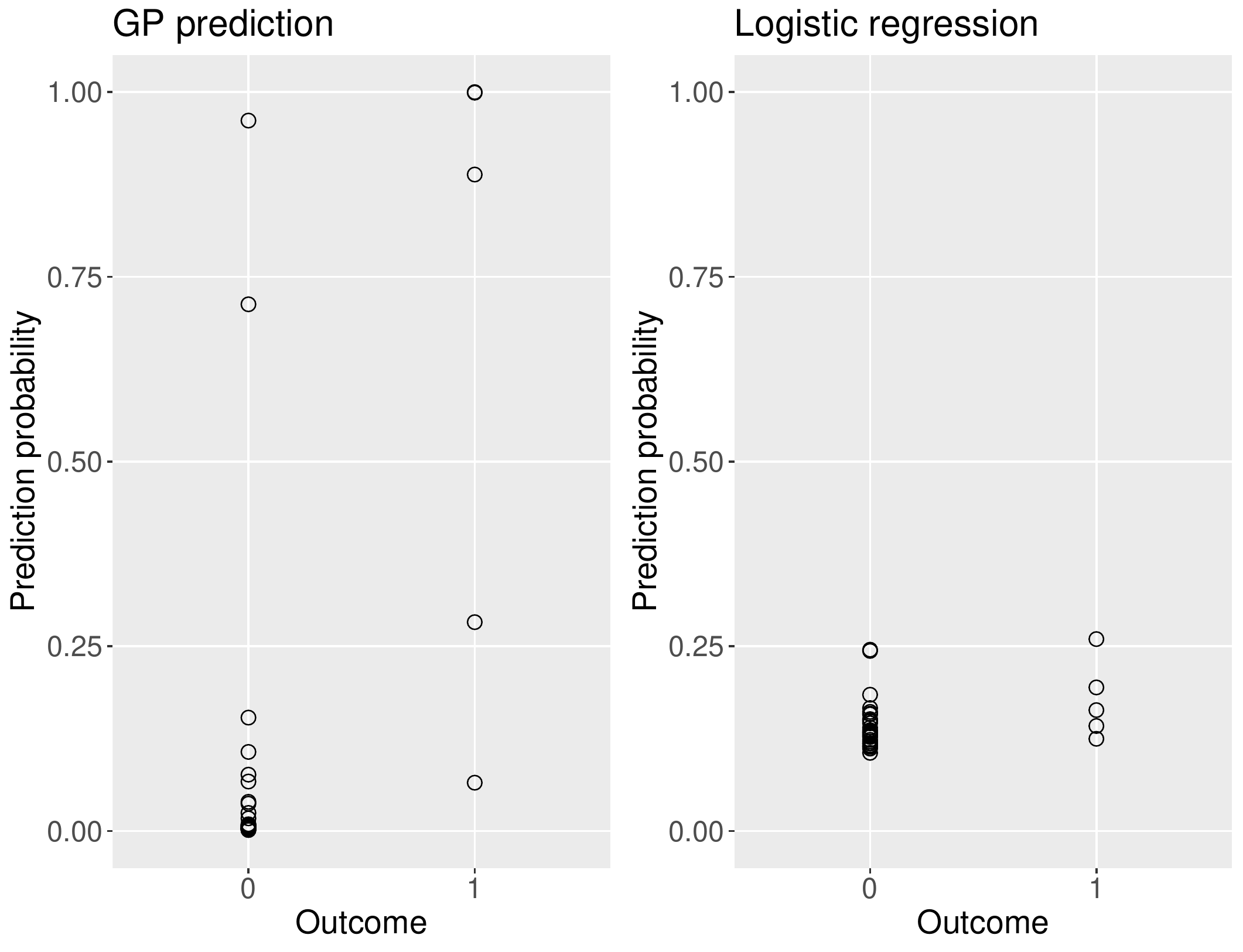}&
 \includegraphics[width = 0.38\textwidth]{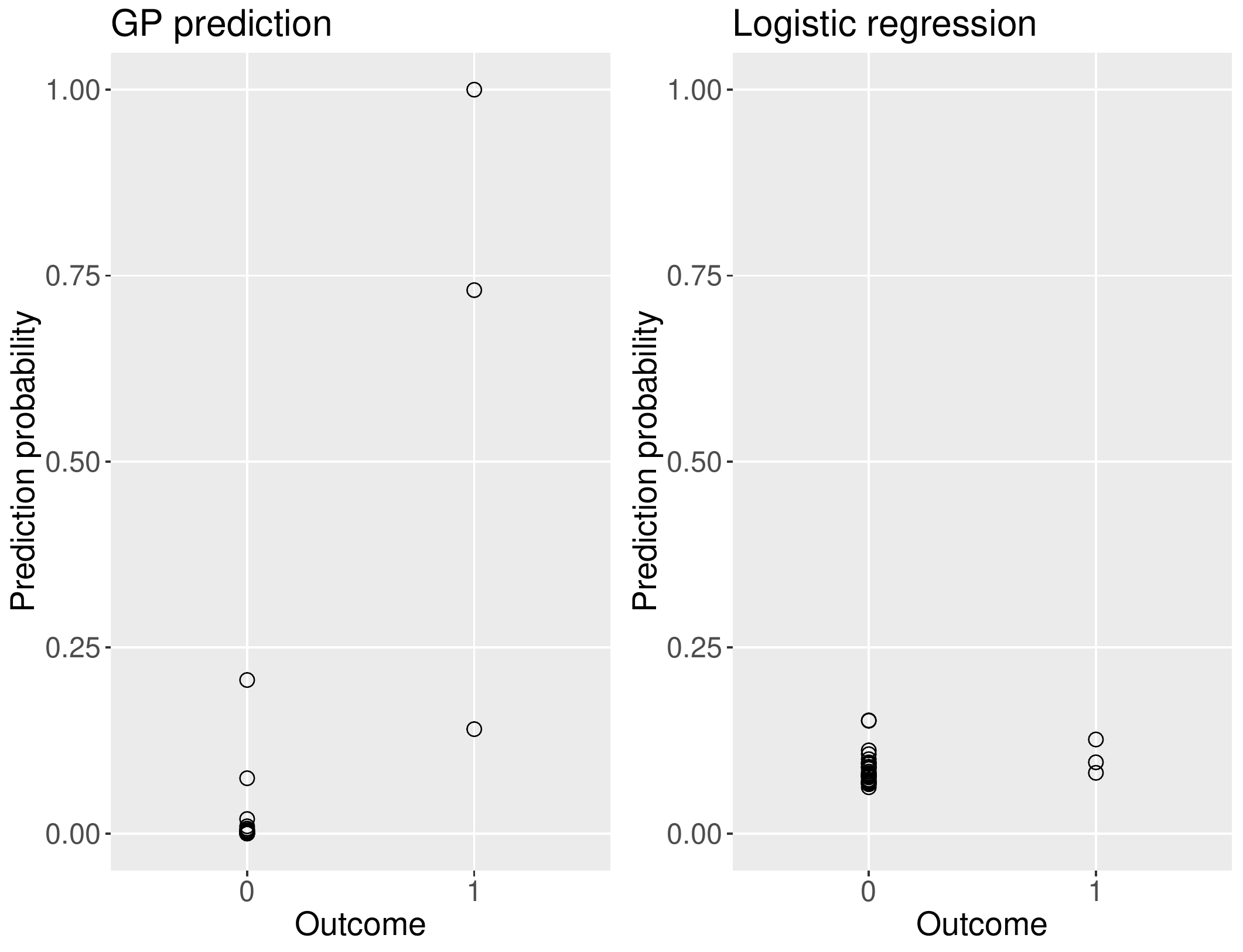}
  \\
   a)  & b)   \\
\includegraphics[width = 0.38\textwidth]{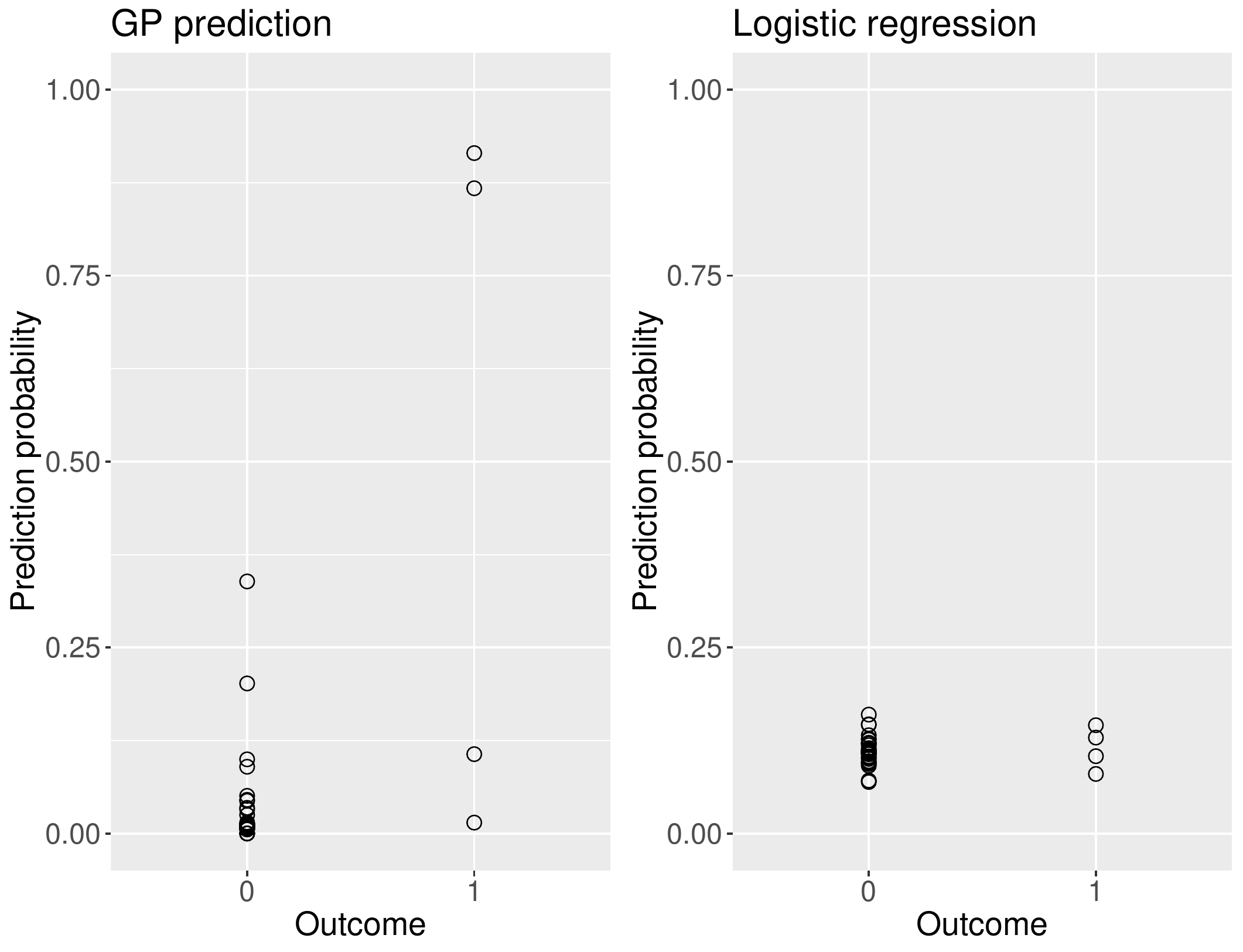}&
\includegraphics[width = 0.38\textwidth]{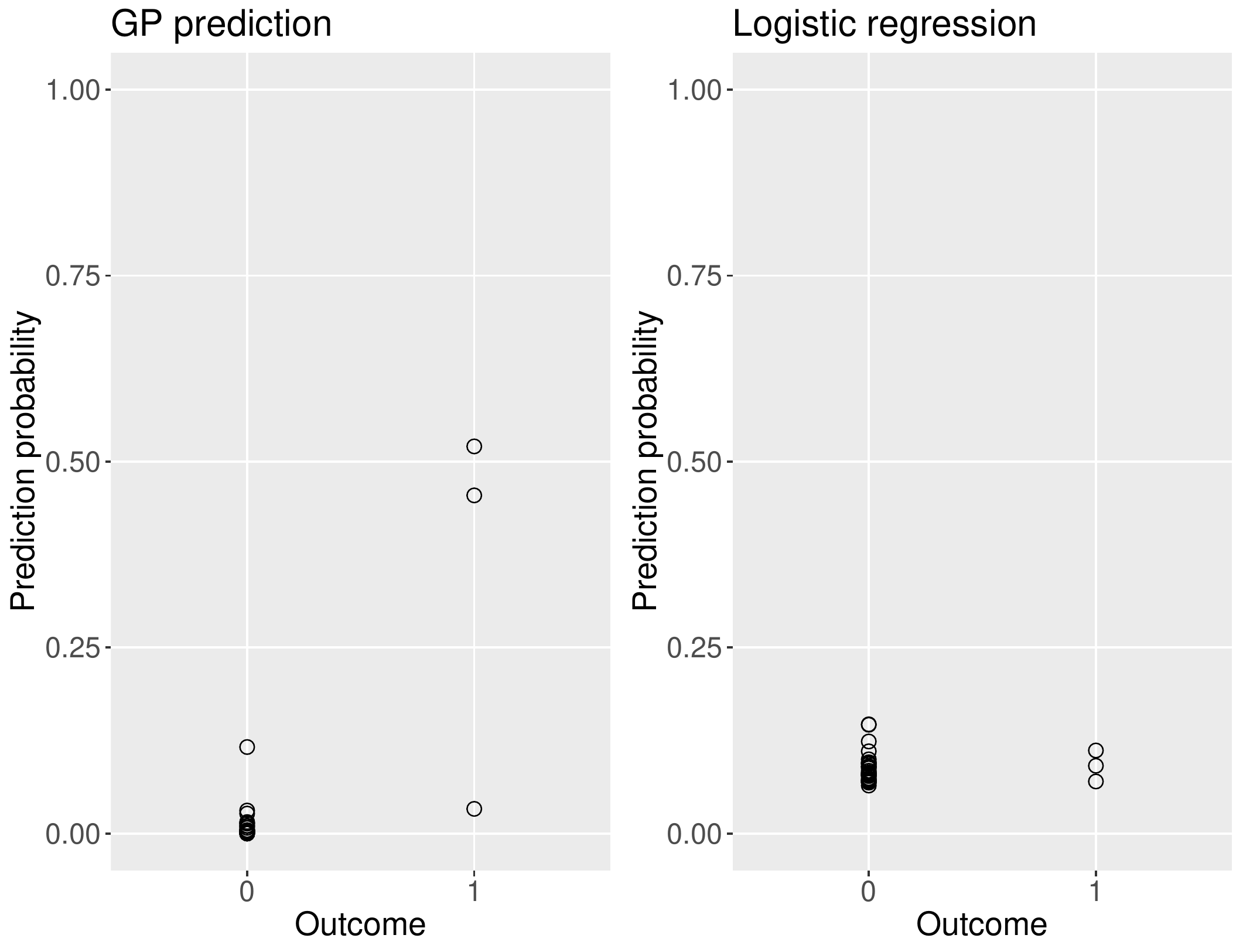}\\
 c) &  d)
 \end{tabular}
\end{center}
 \caption{Prediction probabilities of level exceedances for the 1985--2019 ILI incidence rates obtained from the Gumbel model and from logistic regression using a leave-one-out procedure. Outcome is  0 if there was no exceedance and 1 otherwise. a) Week 3 ILI incidence rates, Level = 816 ($\kappa = 0.5$) b) Week 3 ILI incidence rates, Level = 1,224 ($\kappa = 0.75$) c) Epidemic sizes, Level = 4,031 ($\kappa = 0.5$) d) Epidemic sizes, Level = 6,046 ($\kappa = 0.75$) }
\label{fig:brier:crossvalidation}
\end{figure}

Figure~\ref{fig:brier:crossvalidation} shows the prediction probabilities of level exceedances stratified according to whether an exceedance was observed or not. Contrary to GP prediction, logistic regression was never able to discriminate between the two outcomes. Table~\ref{tab:brier:crossvalidation} presents the corresponding standardized Brier scores and confirms that GP prediction performs much better than the logistic regression. 

Precision-Recall curves and Average Precision scores are not shown since they were uninformative due to insufficient data.  

  

\begin{table}[!h]
  \caption{Standardized Brier scores derived from a leave-one-out procedure on observed data for the GP prediction and the logistic regression for predictions of exceedances of 1,729$\times \kappa$ for Week 3 ILI incidence rates and { 8,062}$\times \kappa$ for epidemic sizes.}
  \label{tab:brier:crossvalidation}
\vspace{4mm}
{\small
\begin{center}
\begin{tabular}{c|cc|cc}
 & \multicolumn{2}{c|}{Week 3 ILI incidence rates} & \multicolumn{2}{c}{Epidemic sizes}\\
\hline
$\kappa$ & 0.5 & 0.75  & 0.5 & 0.75 \\
 Level & 816 & 1,224  & 4,031 & 6,046 \\
\hline
GP prediction & 0.33 &  0.69  & 0.44 &  0.46\\
 Logistic & 0.06 & 0.02 & 0.005 &  0.002 \\
 \end{tabular}


 \end{center}
}
\end{table}

\paragraph{Assessment on simulated data}


 1,500 datasets consisting of 33 three-dimensional vectors were simulated from the estimated Gumbel models for Week 3 ILI incidence rates and epidemic sizes, respectively. The simulations were carried out following Section~7 of \citep{rootzen2018multivariate}. A Gumbel model was fitted to the first 32 vectors of each dataset and the estimated model was then used to predict the third component of the 33rd vector, conditionally on the first two components. 

\begin{figure}
 \begin{center}
  \vspace{4mm}

 \begin{tabular}{cc}
 \includegraphics[width = 0.38\textwidth]{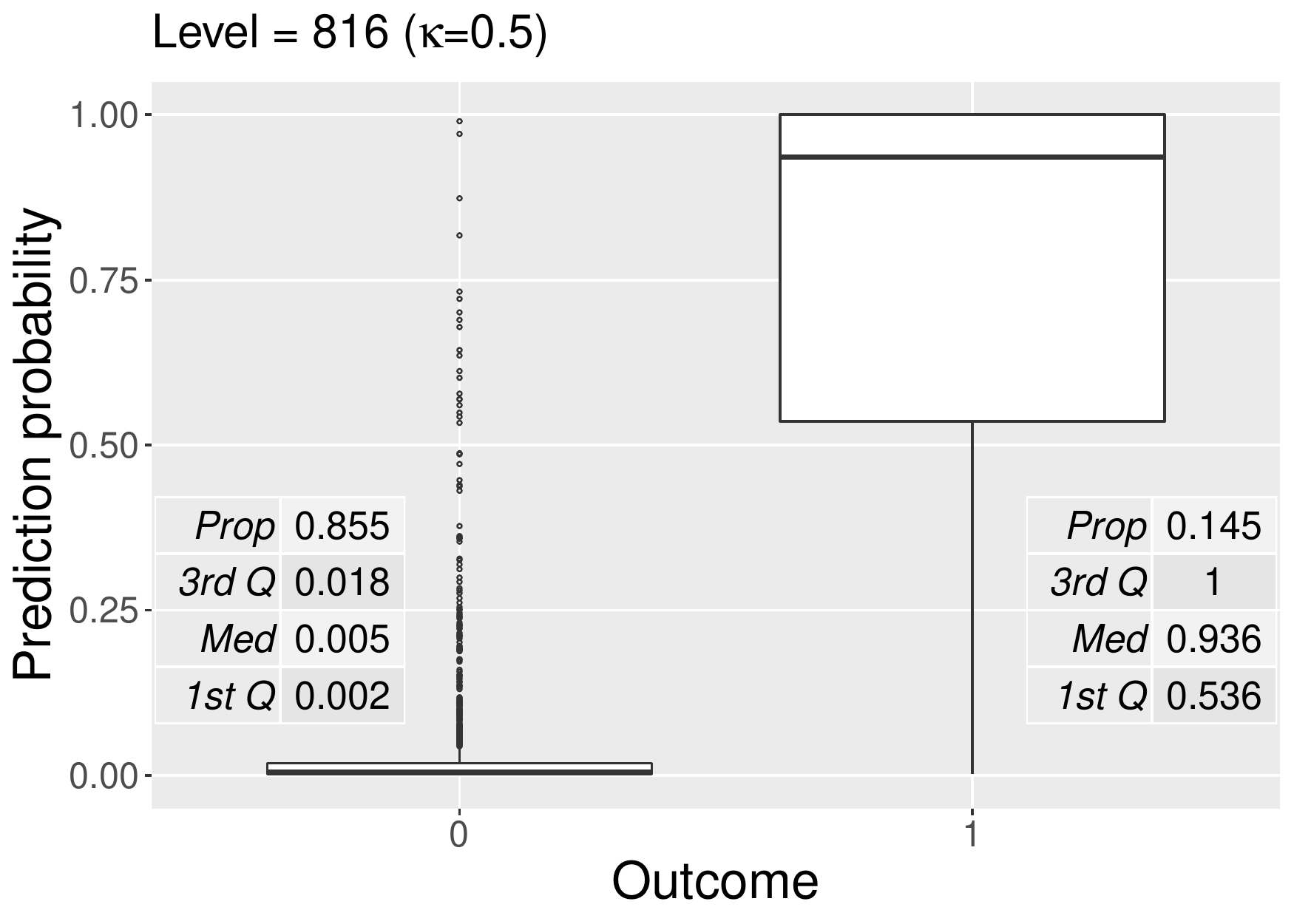}&
 \includegraphics[width = 0.38\textwidth]{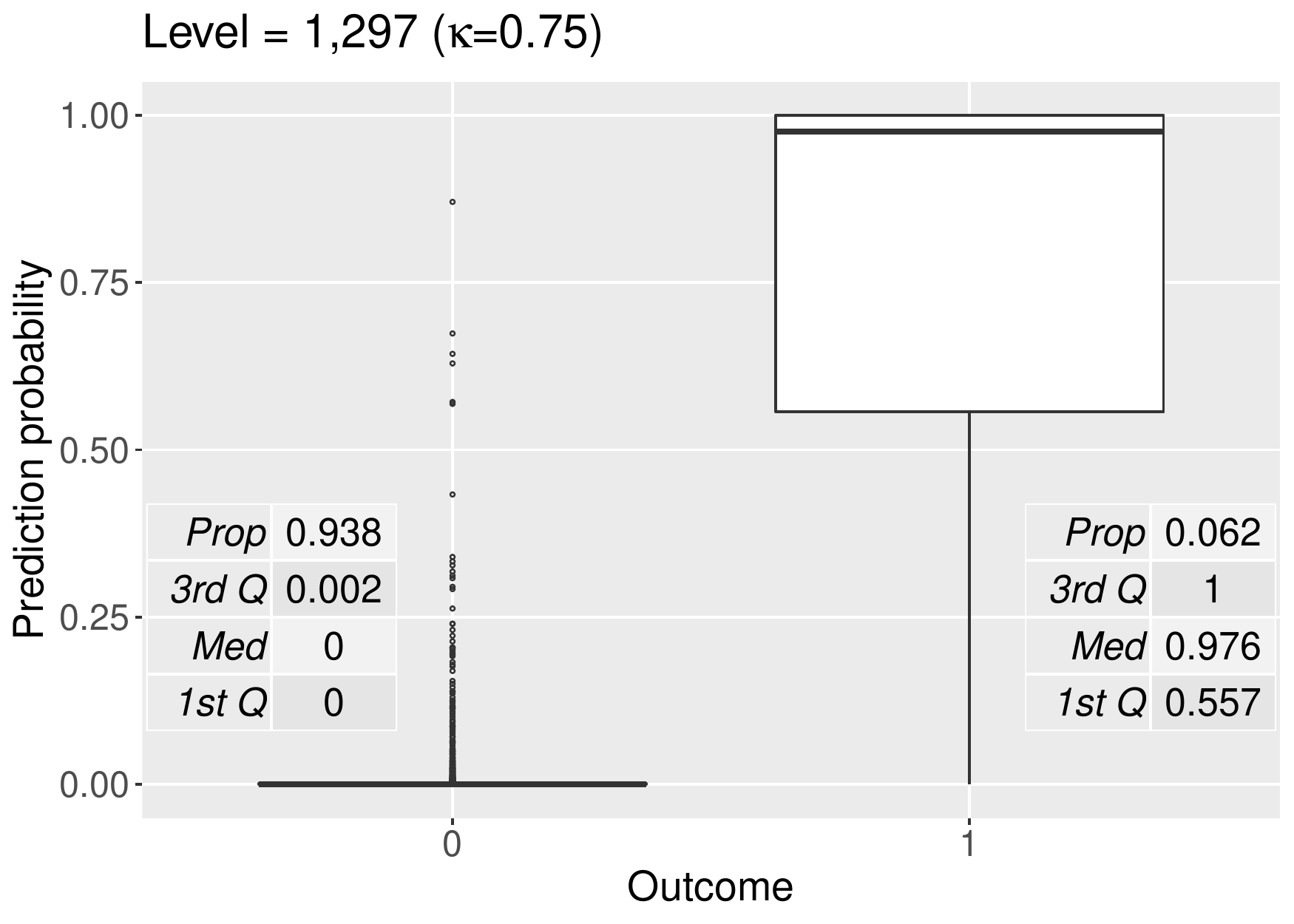}\\
  \includegraphics[width = 0.38\textwidth]{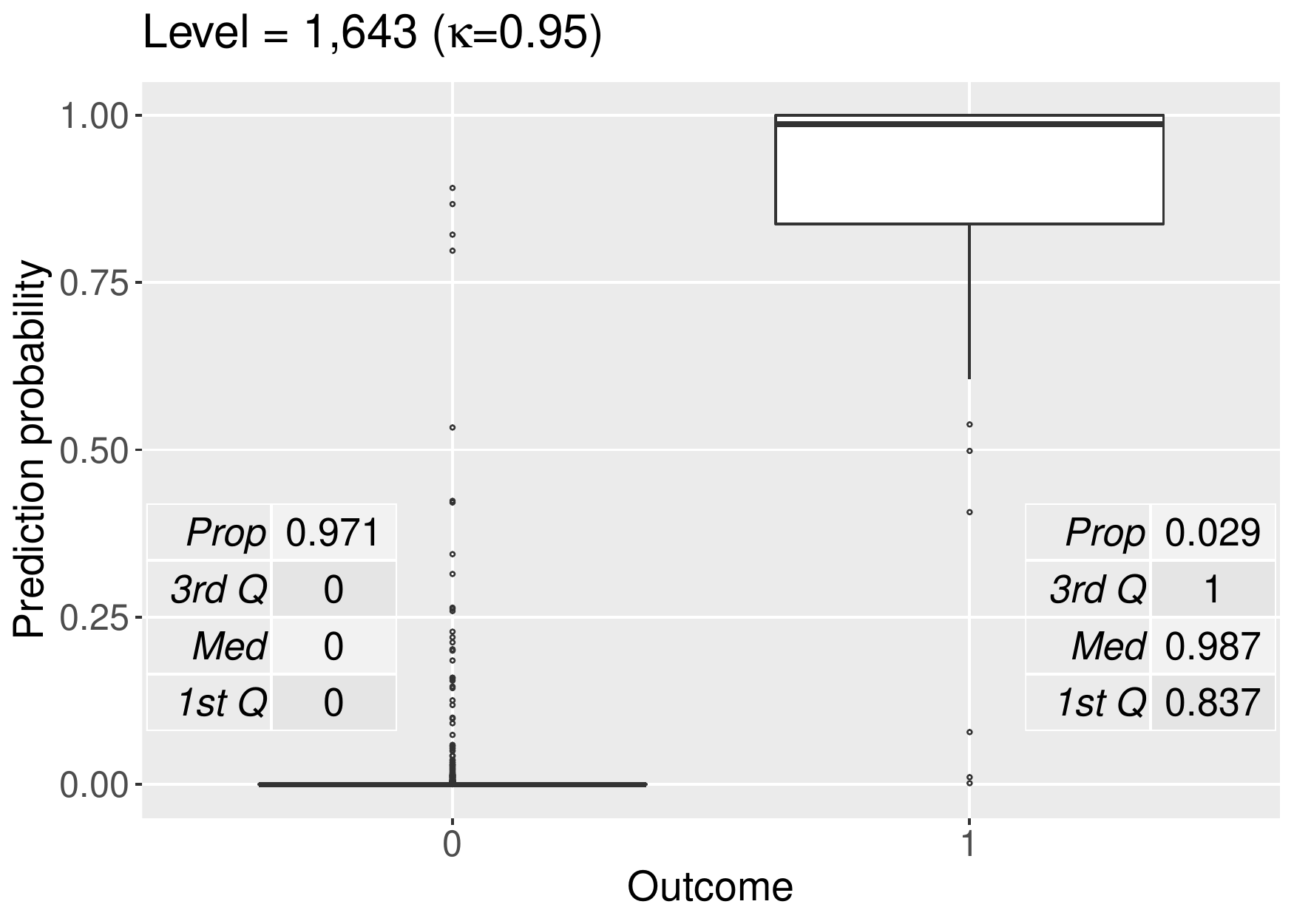}&
\includegraphics[width = 0.38\textwidth]{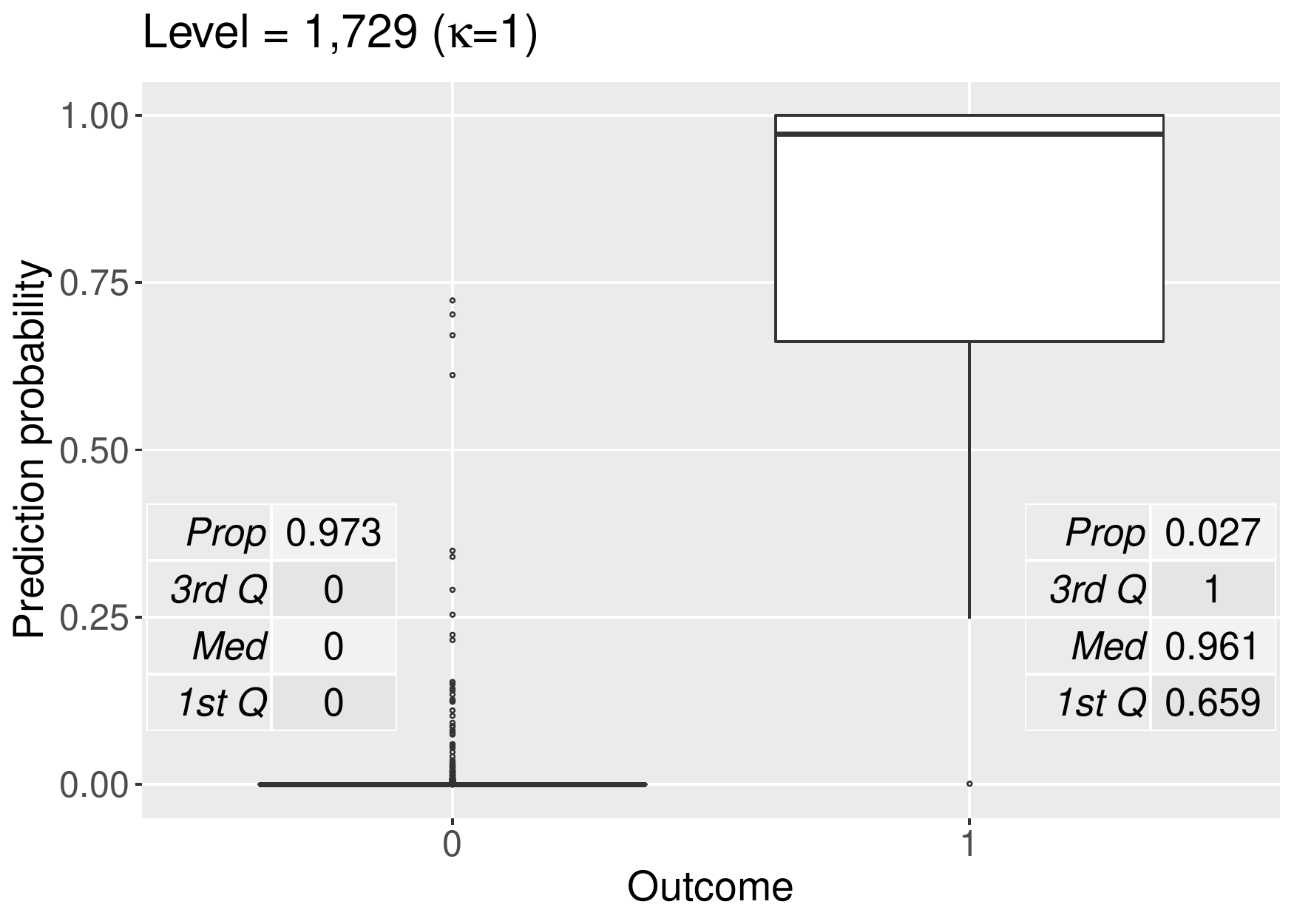}\\
\multicolumn{2}{c}{a)}\\
 \includegraphics[width = 0.38\textwidth]{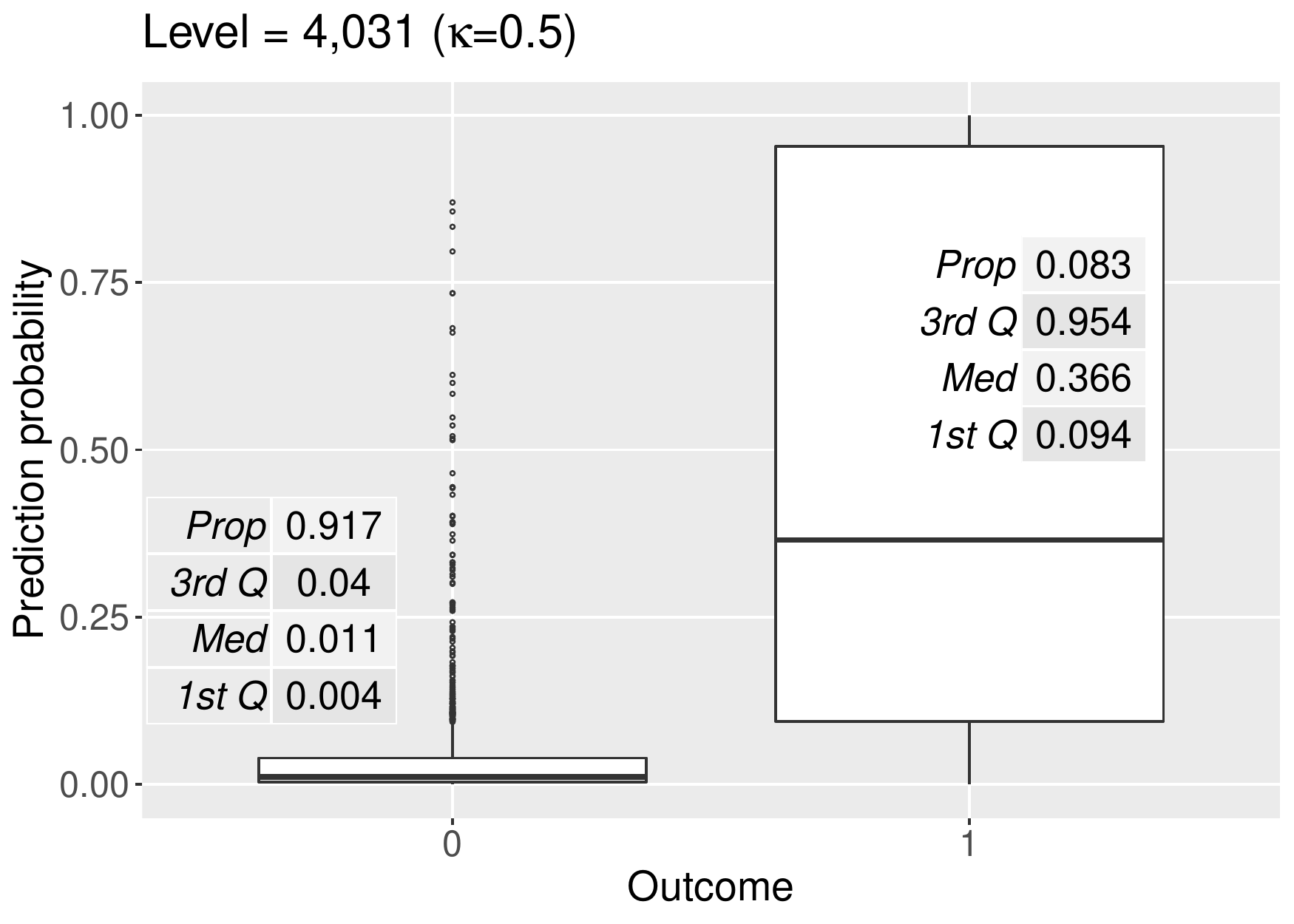}&
 \includegraphics[width = 0.38\textwidth]{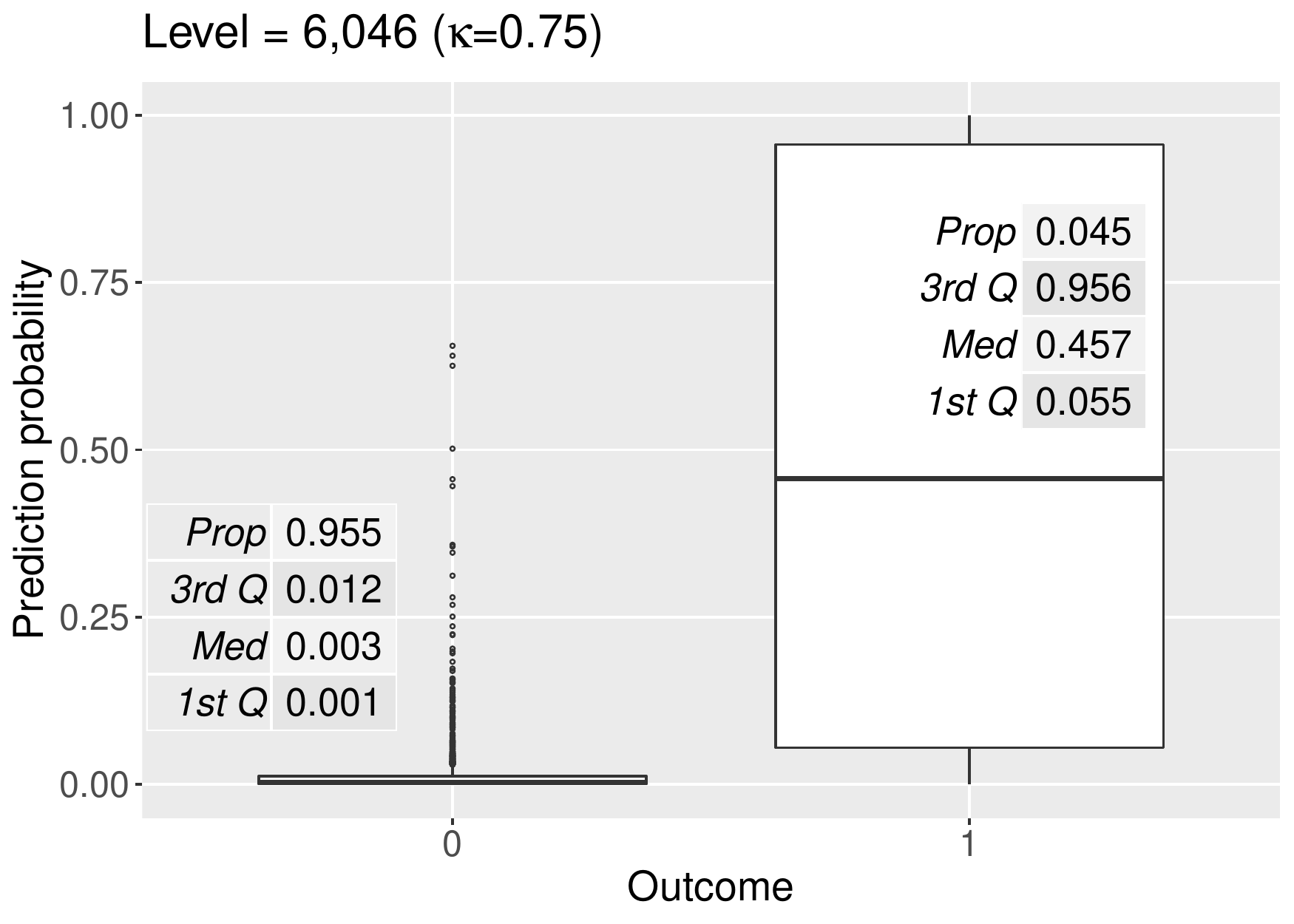}\\
  \includegraphics[width = 0.38\textwidth]{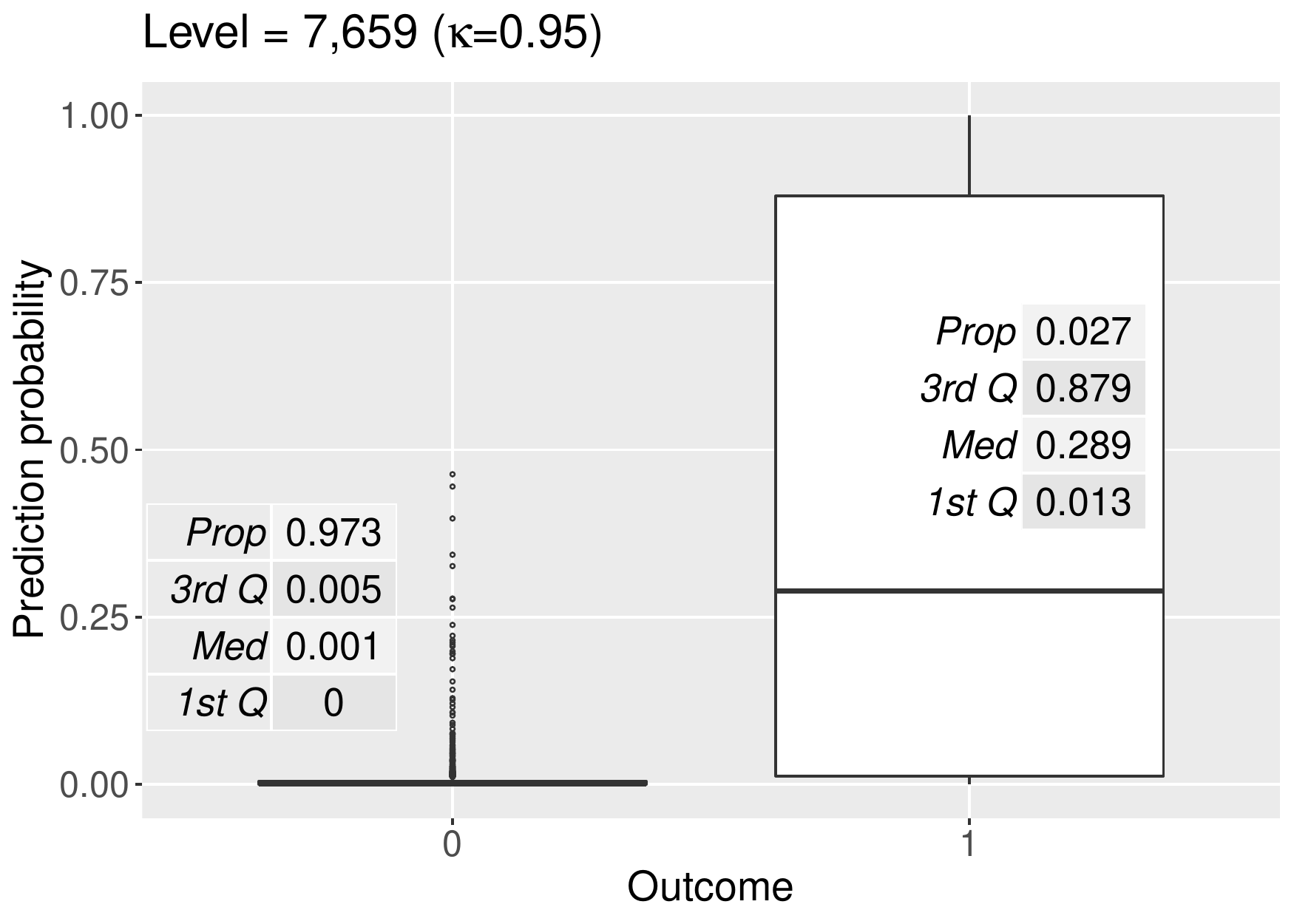}&
\includegraphics[width = 0.38\textwidth]{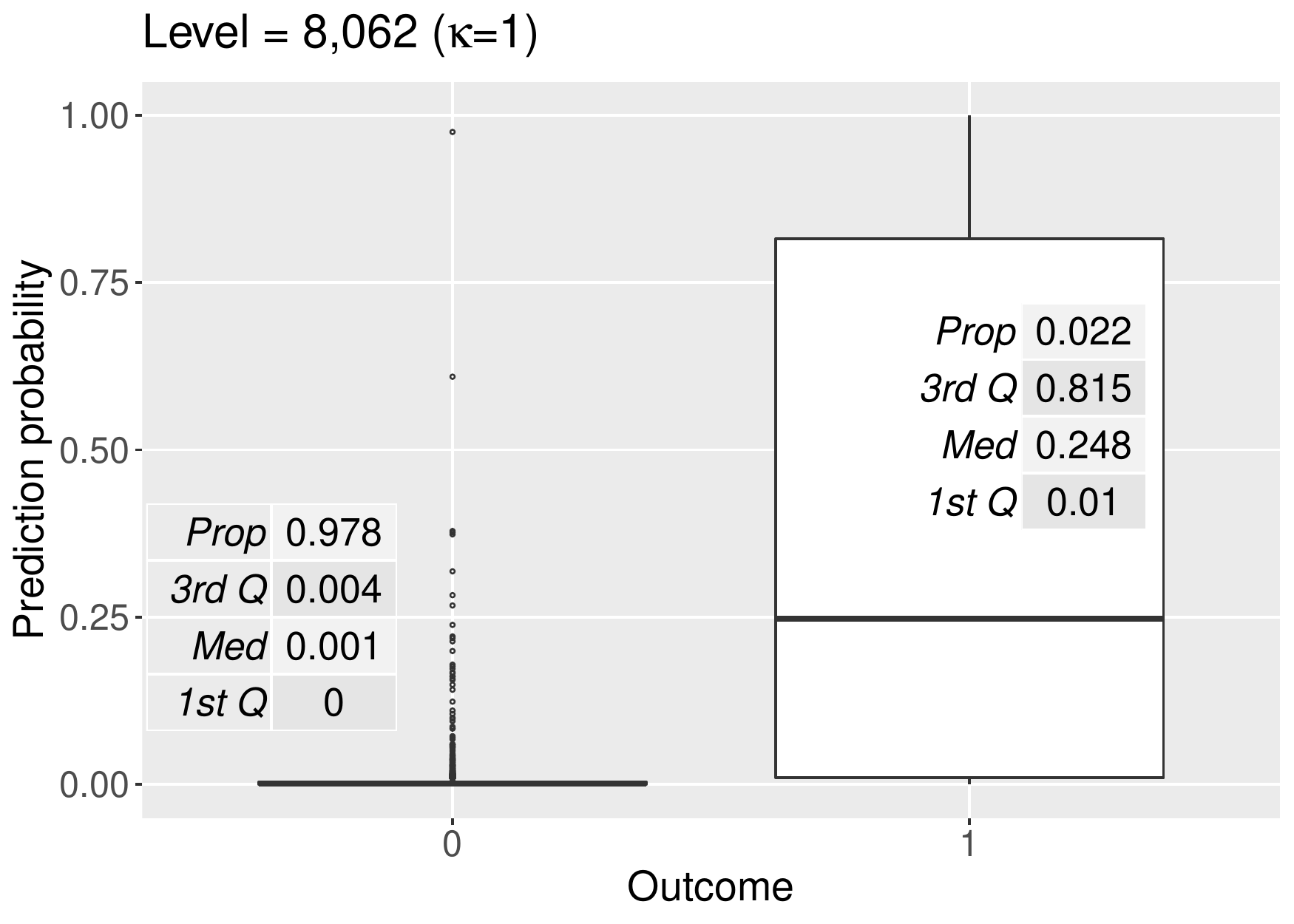}\\
\multicolumn{2}{c}{b)}
   \end{tabular}
\end{center}
 \caption{Boxplots of prediction probabilities of level exceedances for simulated data obtained from the fitted Gumbel model for a) Week 3 ILI incidence rates and b) epidemic sizes. Outcome is equal to 0 if there was no exceedance and 1 otherwise. { The tables give the proportions of 0-outcomes and of 1-outcomes, and the quartiles of the prediction probabilities}.}
\label{fig:brier:simulated:gumbel}
\end{figure}

{ Figure~\ref{fig:brier:simulated:gumbel} shows boxplots of the prediction probabilities stratified according to whether an exceedance was observed or not, for the GP prediction for both Week 3 ILI incidence rates and epidemic sizes for the four levels ($\kappa=0.5, 0.75, 0.95, 1$).   A table is associated with each boxplot. These tables first give the proportion of 0-outcomes and of 1-outcomes, which shows that the classes are indeed very imbalanced, and then the quartiles of the prediction probabilities. One can see that the boxes for the 0-outcomes are very concentrated and close to 0 while the boxes for the 1-outcomes are larger. There is no overlap between the two boxes. Moreover,} the widths of the boxes indicate that the performance of the GP prediction is better for incidence rates than for epidemic sizes.  

\begin{figure}
 \begin{center}
  \vspace{4mm}
 \begin{tabular}{cc}
 \includegraphics[width = 0.38\textwidth]{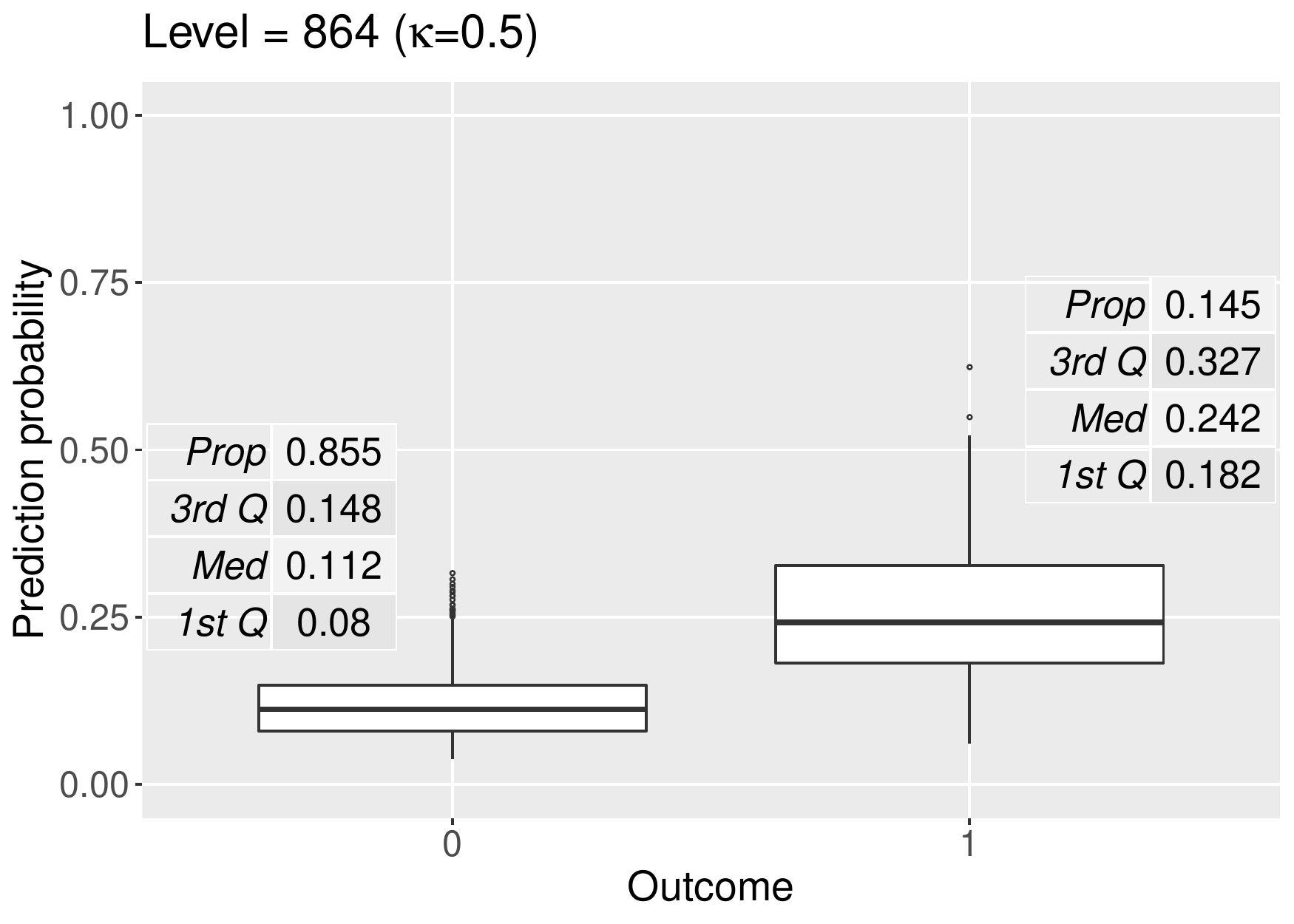}&
 \includegraphics[width = 0.38\textwidth]{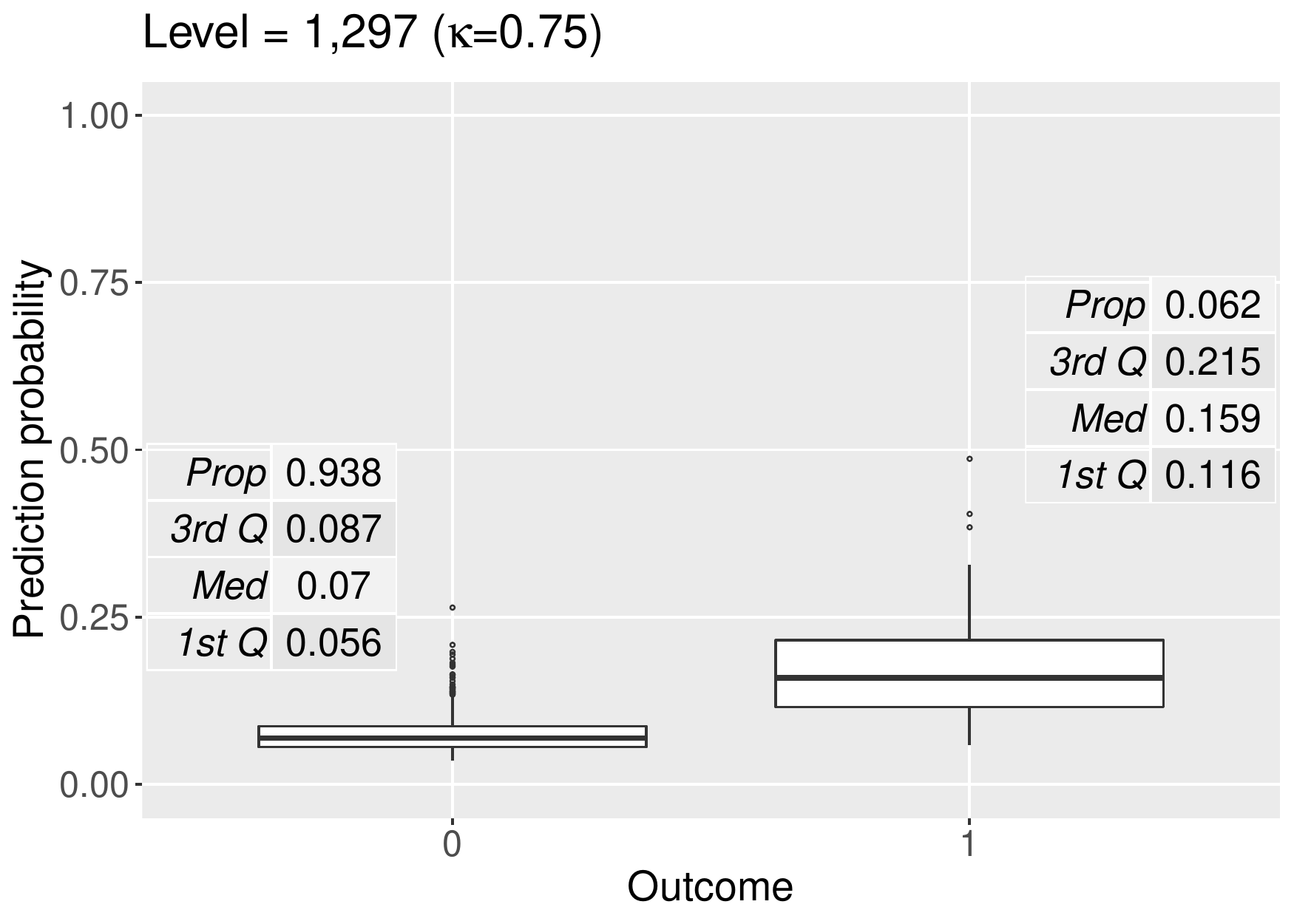}\\
 \multicolumn{2}{c}{a)} \\
 \includegraphics[width = 0.38\textwidth]{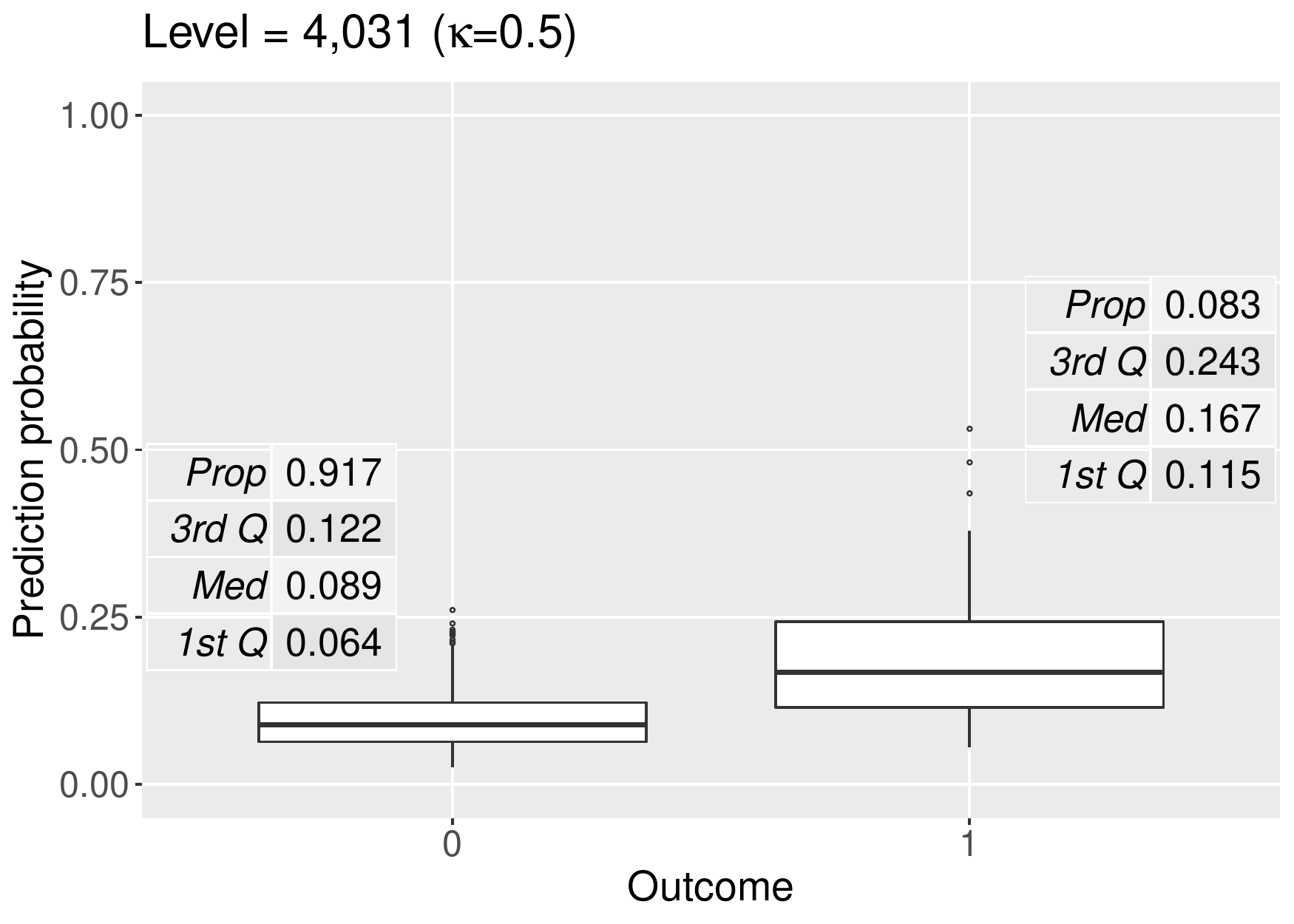}&
 \includegraphics[width = 0.38\textwidth]{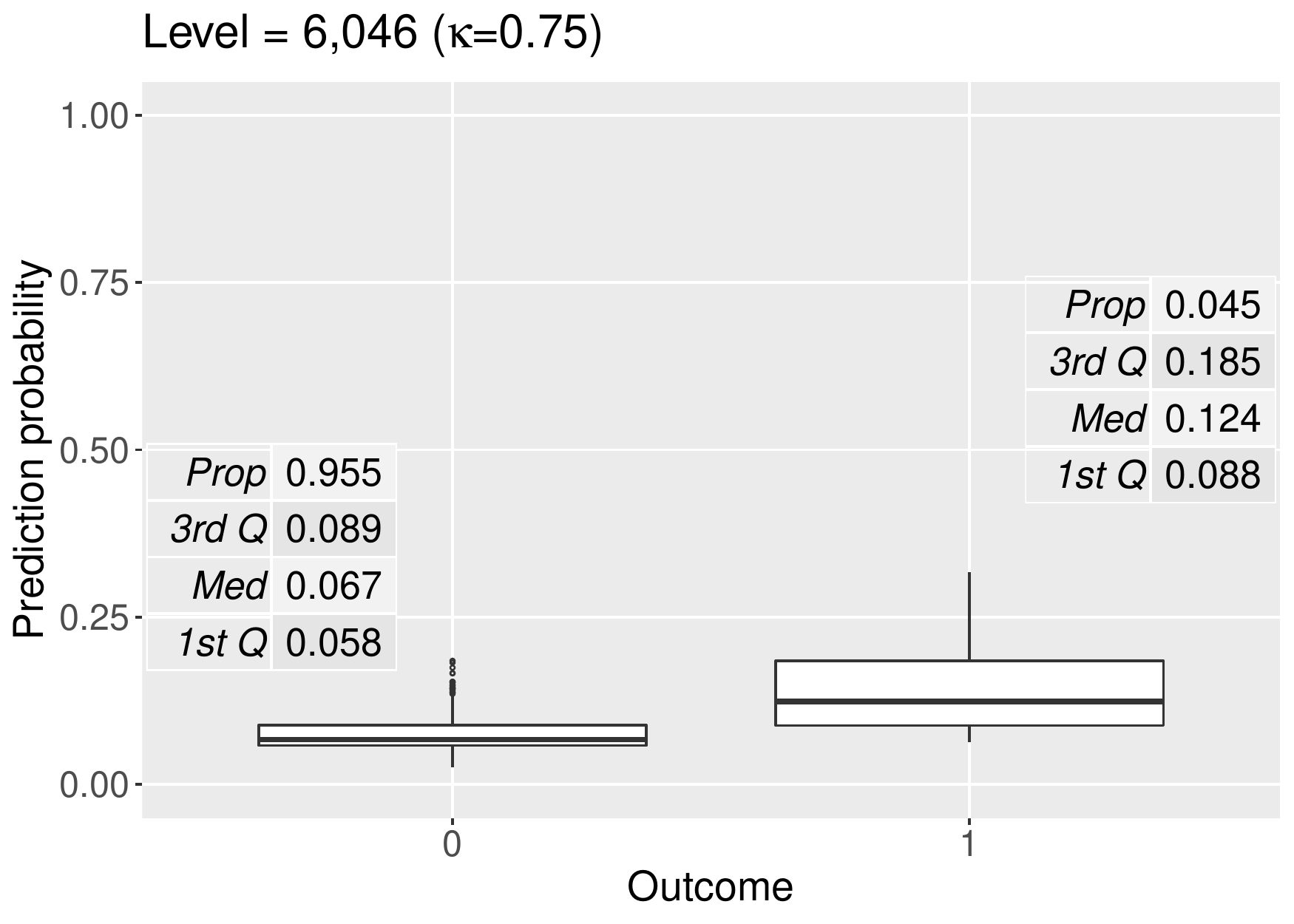}\\
 \multicolumn{2}{c}{b)}
   \end{tabular}
\end{center}
 \caption{Boxplots of prediction probabilities of level exceedances for simulated data obtained from the logistic regression  for a) Week 3 ILI incidence rates and b) epidemic sizes. Outcome is equal to 0 if there was no exceedance and 1 otherwise. { The tables give the proportions of 0-outcomes and of 1-outcomes, and the quartiles of the prediction probabilities}.}
\label{fig:brier:simulated:logistic}
\end{figure}

The boxplots for predictions obtained from the logistic regression are shown in Figure~\ref{fig:brier:simulated:logistic} for the two lower levels. Predictions could not be made for the higher levels since, for these levels, there were too few exceedances in the simulated data to allow estimation of the parameters. The quality of the GP prediction is much better than that of the logistic regression.

\begin{figure}
	\begin{center}
		\begin{tabular}{cc}
			\includegraphics[width = 0.39\textwidth]{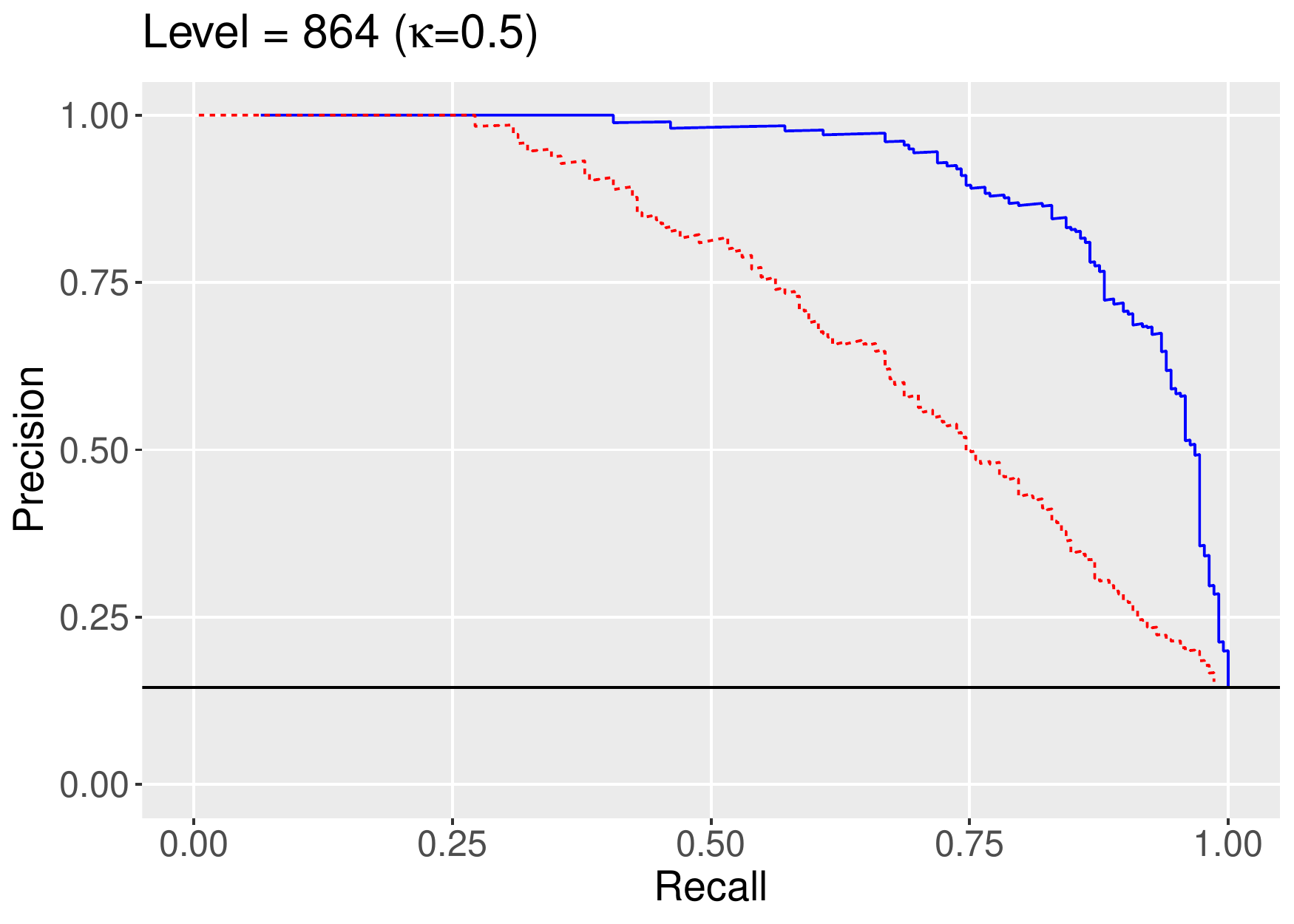}&
			\includegraphics[width = 0.39\textwidth]{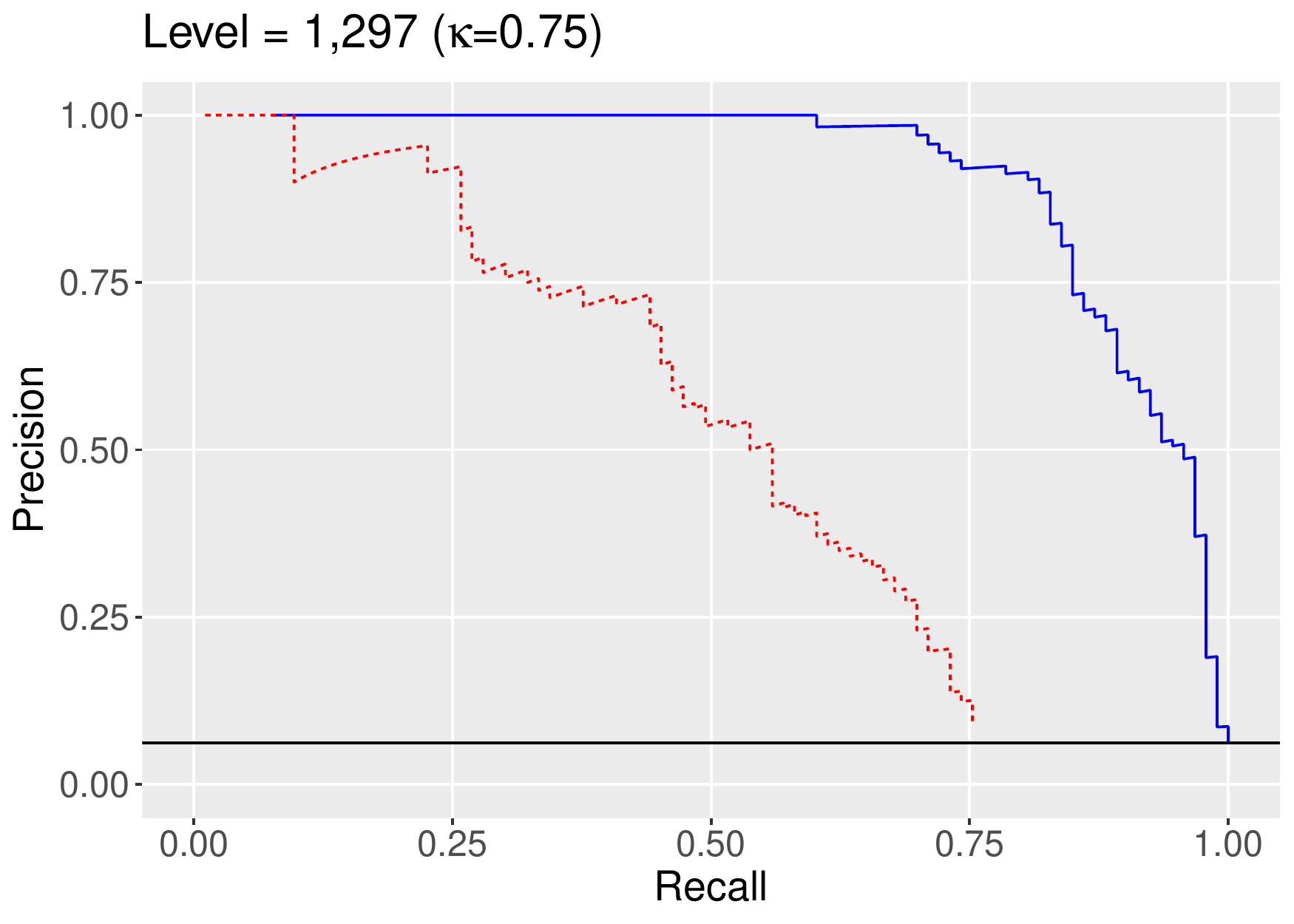}
			\\
			\includegraphics[width = 0.39\textwidth]{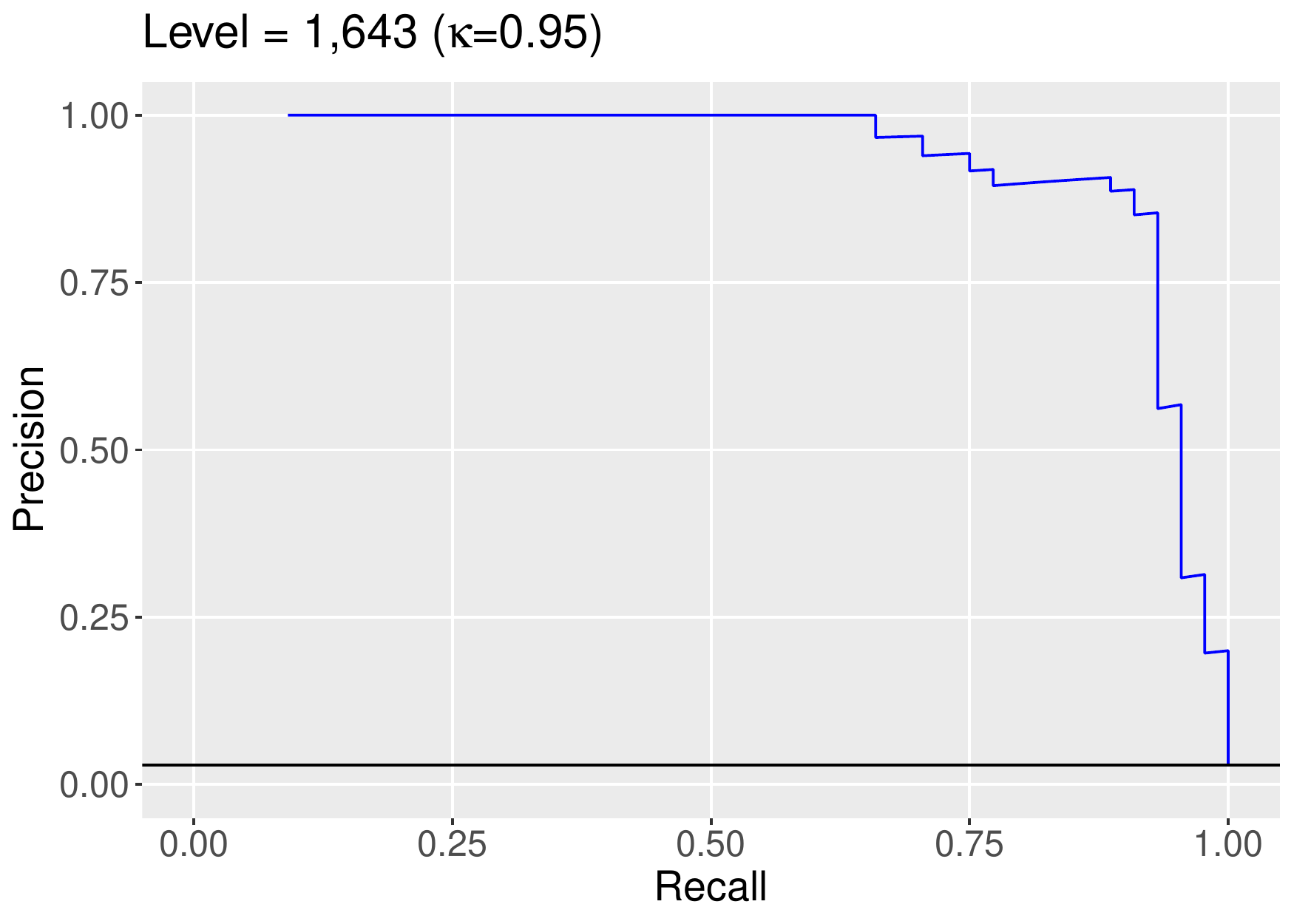}&
			\includegraphics[width = 0.39\textwidth]{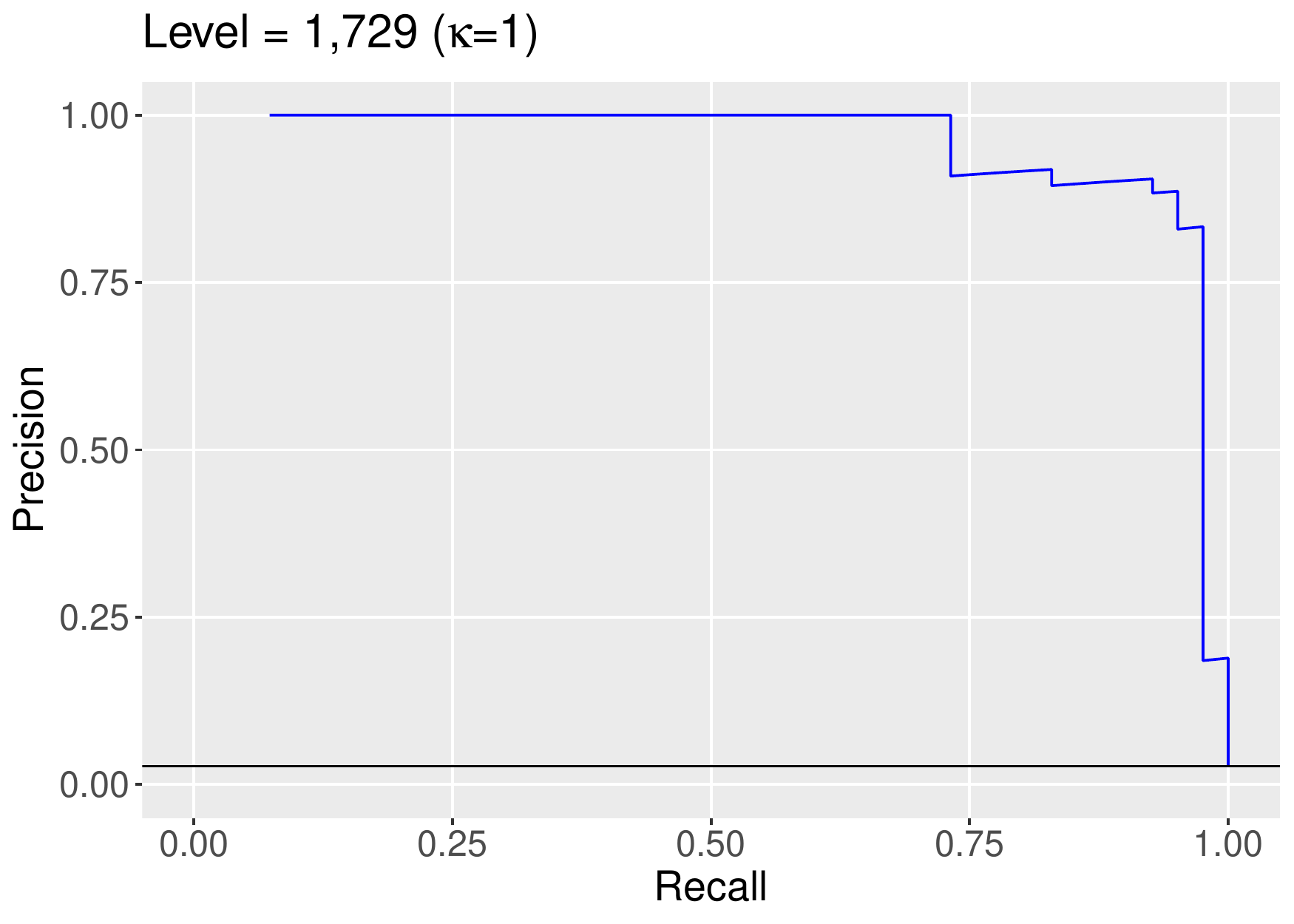}\\
			\multicolumn{2}{c}{a)}\\~\\
						\includegraphics[width = 0.39\textwidth]{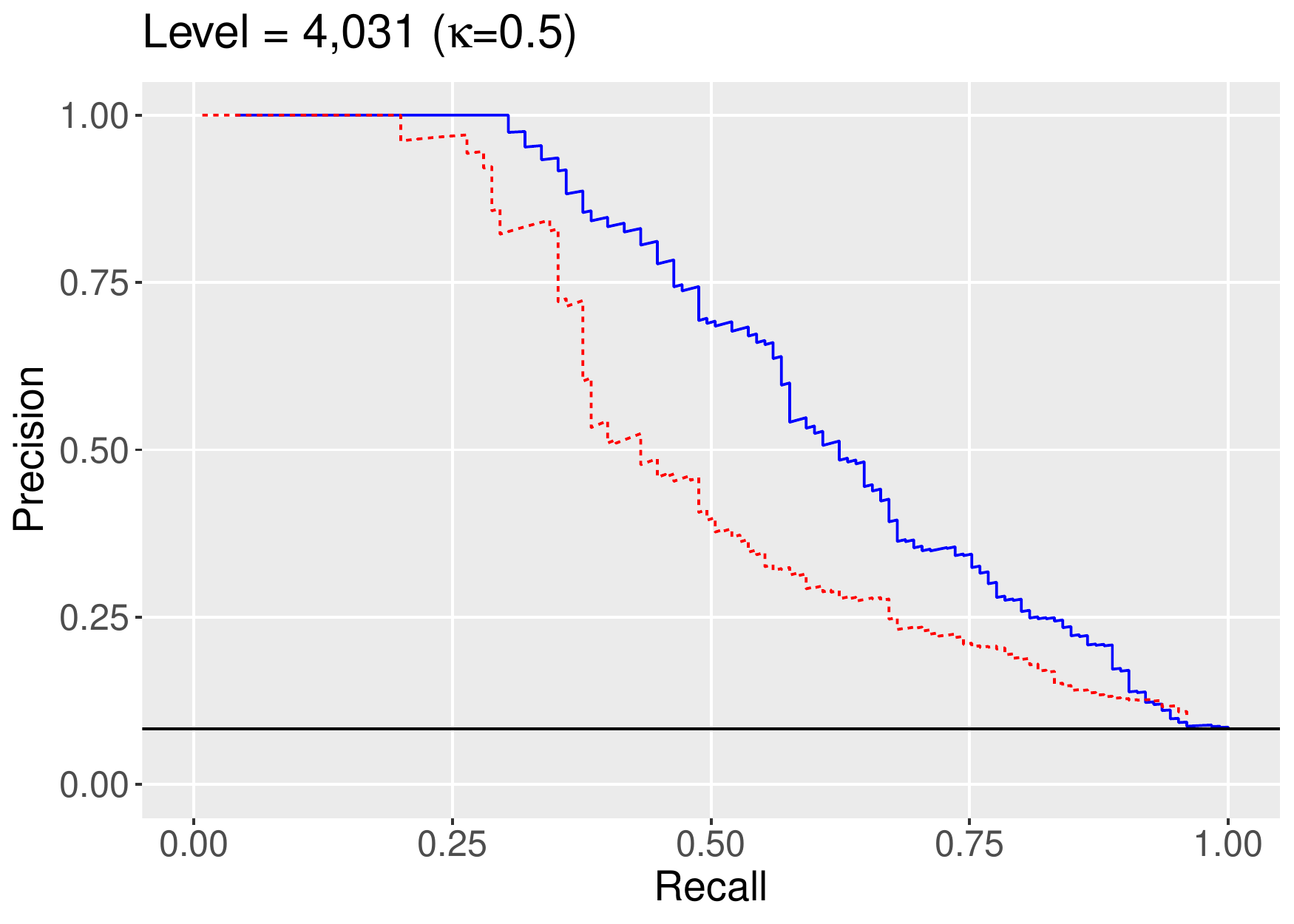}&
			\includegraphics[width = 0.39\textwidth]{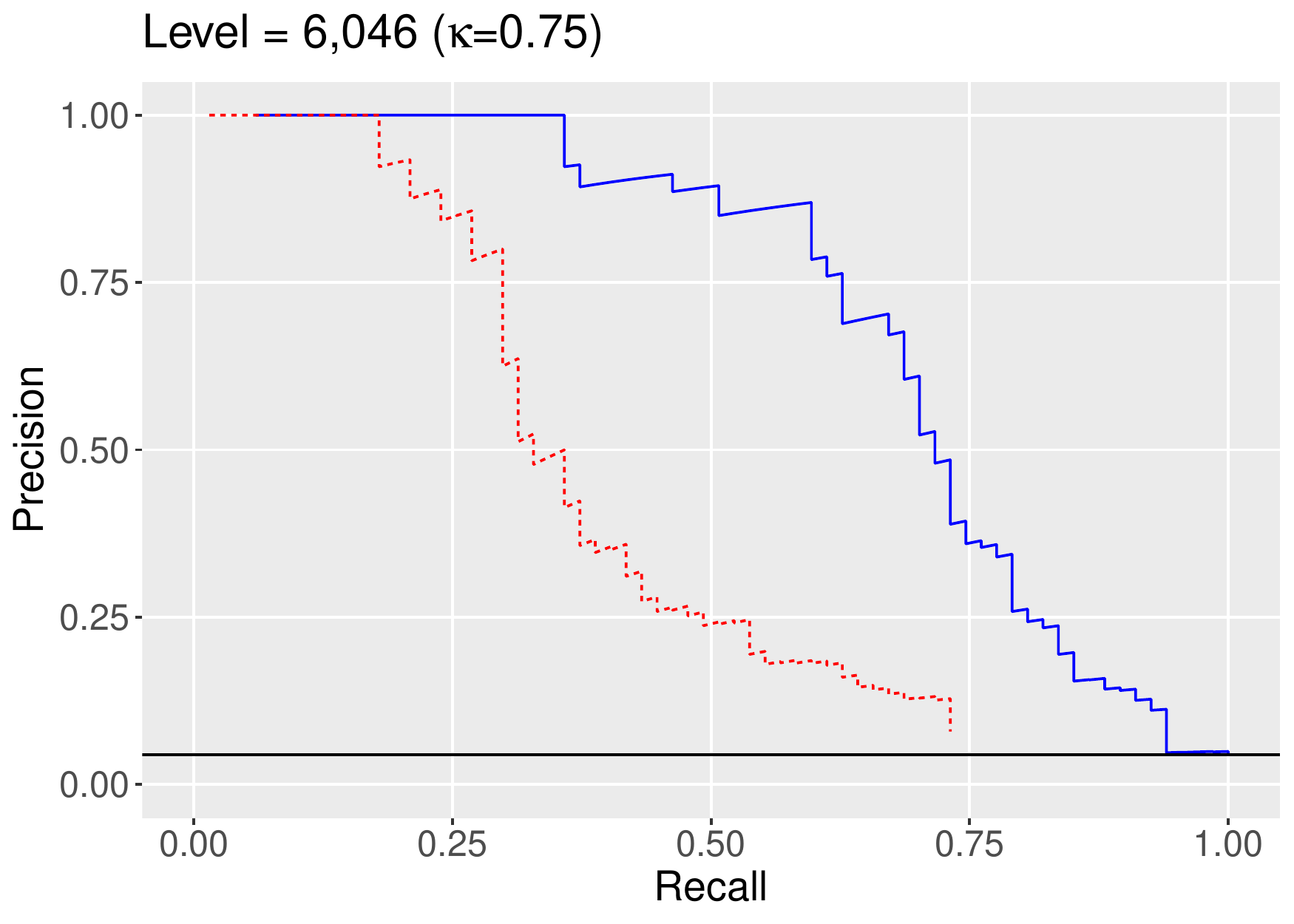}
			\\
			\includegraphics[width = 0.39\textwidth]{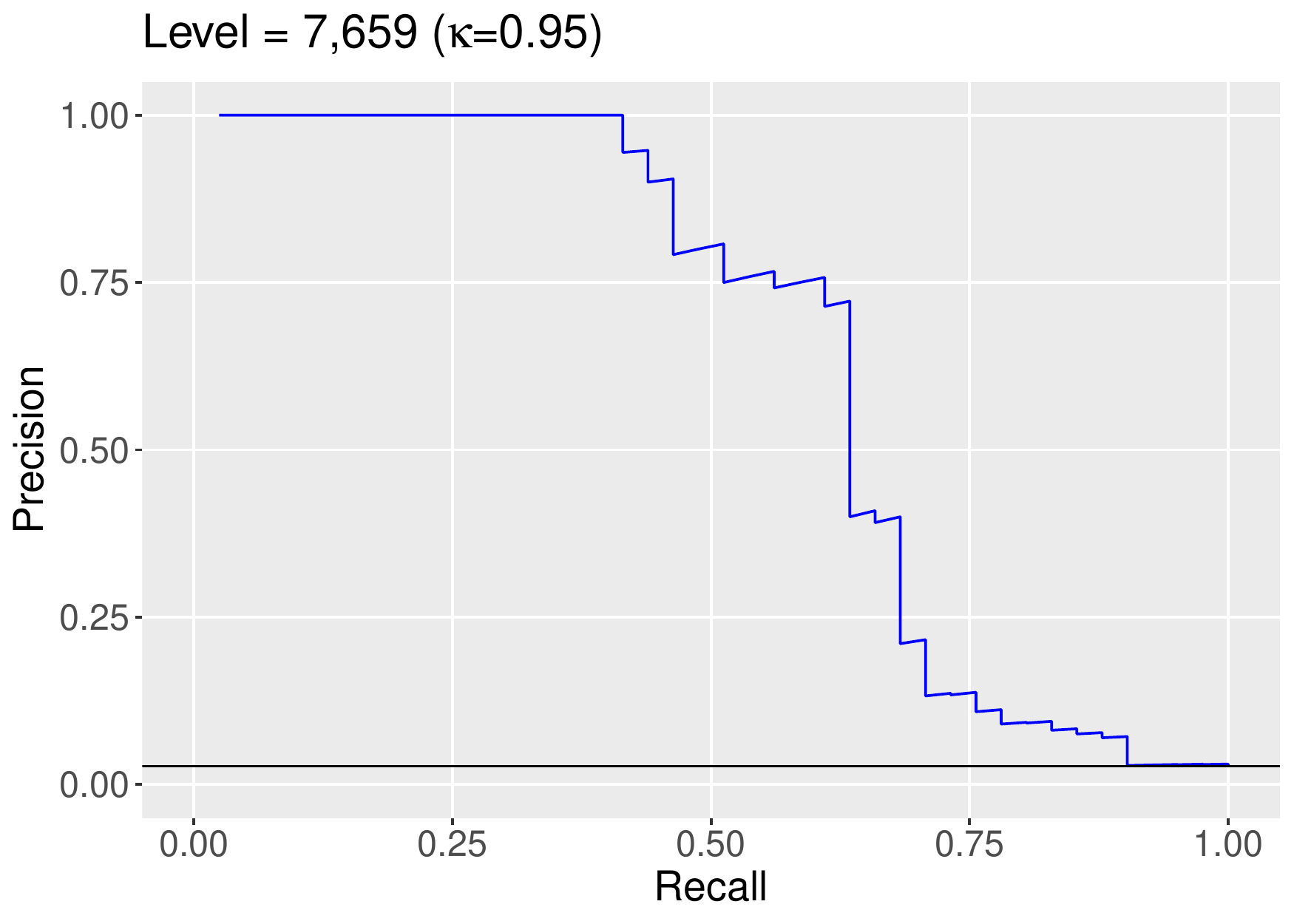}&
			\includegraphics[width = 0.39\textwidth]{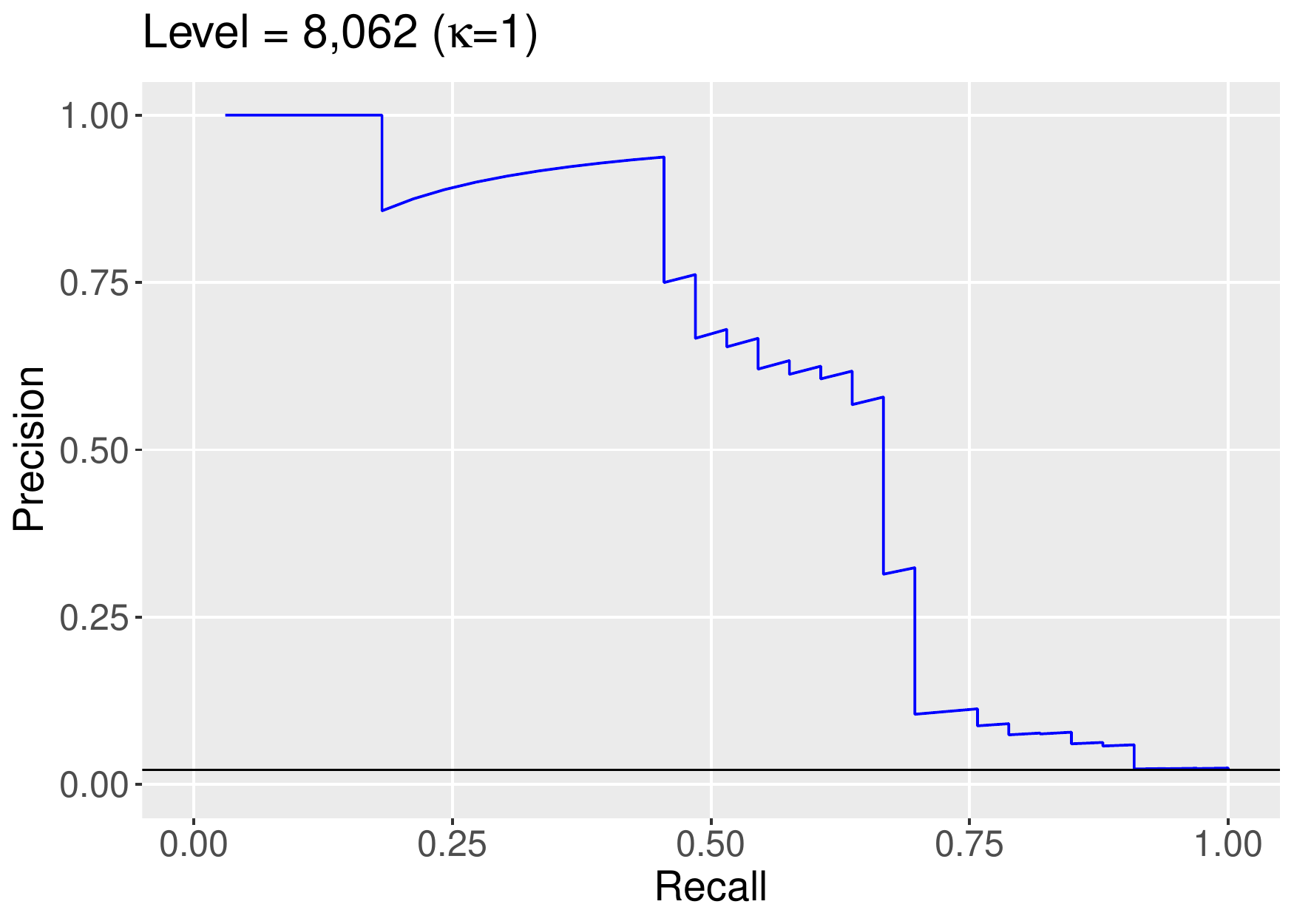}\\
			\multicolumn{2}{c}{b)}
			
		\end{tabular}
	\end{center}
	\caption{Precision-Recall curves for prediction of a) the Week 3 ILI incidence rates and b) epidemic sizes for both the Gumbel model (plain blue line) and the logistic regression (dashed red line).}
	\label{fig:brier:simulated:precision-recall}
\end{figure}

Standardized Brier scores and Average Precision scores for predictions with GP and logistic predictions are presented in Table~\ref{tab:brier:simulation}, and Precision-Recall curves are shown in Figure~\ref{fig:brier:simulated:precision-recall}.

\begin{table}[!h]
\begin{center}
   \caption{Standardized Brier scores and Average Precision scores for the GP prediction, the logistic regression and the true model for predictions of exceedances of 1,729$\times \kappa$ for Week 3 ILI incidence rates and { 8,062}$\times \kappa$ for epidemic sizes for the simulated data.}
   \label{tab:brier:simulation}
  \vspace{4mm}

\begin{tabular}{c|c|cccc}

& $\kappa$ & 0.5 & 0.75 &  0.95  & 1 \\  
& Level & 816 & 1,224 & 1,551 & 1,632 \\
\hline
 \multirow{2}{*}{Brier scores}&GP prediction & 0.72 & 0.75 & 0.80 & 0.84  \\
 &Logistic &0.19 & -0.18 & - & -  \\  
 \hline
 \multirow{2}{*}{Average Precision scores}&GP prediction & 0.92 & 0.91 & 0.93 & 0.96 \\
 &Logistic &0.72 & 0.51 & - & -  \\  
 \end{tabular}\\~\\
a) Week 3 ILI incidence rates\\~\\
 \end{center}

  \begin{center}

  \begin{tabular}{c|c|cccc}

& $\kappa$ & 0.5 & 0.75 &  0.95  & 1 \\  
& Level & { 4,031} & { 6,046} & {7,659} & {8,062}\\
\hline
 \multirow{2}{*}{Brier scores}&GP prediction & 0.40 & 0.51 & 0.47 & 0.44  \\
 &Logistic &-0.03 & -0.50 & - & - \\  
 \hline
 \multirow{2}{*}{Average Precision scores}&GP prediction & 0.64 & 0.71 & 0.64 & 0.60   \\
 &Logistic &0.52 & 0.40 & - & - \\  
\end{tabular}\\~\\


  b) Epidemic sizes
  \end{center}
\end{table}

For the simulated data, the true parameters for the Gumbel model are known. The prediction probabilities obtained from the true model{, perhaps surprisingly, give results which are not noticeably better than the results from fitted Gumbel model (Figure~3, Supplementary Material).}

\newpage
\section{Conclusion}
\label{sec:conclusions}


  
 {   Extreme events that can put high stress on the health care system are a major concern for resource planning in public health.  Both real-time prediction of the development of an ongoing epidemic and prediction of sizes of future epidemics are important for policy makers. 
 	
 In Section~\ref{sec:riskfutureyears} we used existing univariate EVS to estimate risks of very high ILI incidence rates and of extreme epidemic sizes in France  during the following year, and during the following 10 year, as input for long term resource planning. 

A main result of this paper is the development of a methodology for real-time prediction of  risks as early as at the start of an epidemic. The prediction  method is based on  estimates of conditional probabilities of exceedance of a very high level,  using recent results  on multivariate GP distributions \citep{rootzen2018multivariate,kiriliouk2019peaks}. Sections \ref{subsec:realtimepredictionresults} and \ref{sec:accuracy} indicate that our predictions work well for the  French Sentinelles data.
 We chose to focus on prediction of very high levels for the third week of the epidemic and of epidemic size, given observation of the first two weeks of the epidemic. Policy makers may be most interested predicting the overall severity of epidemic as measured by epidemic size. However, according to Section \ref{sec:accuracy}, our method gives better predictions for the Week 3 incidence rates than for epidemic sizes. This is as expected since the epidemic size is the sum of the incidence rates from the beginning to the end of the epidemic and predicting it hence requires prediction further into the future than just to the following week . 
 
 The choice of 3-dimensional GP models was motivated both by epidemiological and by technical considerations.   The methods can be used to predict any other characteristic of the epidemic, e.g. the fourth week given the first two weeks, and in other dimension if the number of available data allows it, and provided dependence is strong enough to make prediction is possible. We also developed a method to use observation of the first three weeks of an epidemic to warn decision makers if something anomalous and potentially dangerous was happening.}

{ Real-time prediction provides an important support for health care resource planning. However, our methodology is only useful for diseases, like influenza, for which historical data are available but does not apply to emerging diseases such as Covid-19. }

{   To a large extent, assessment of predictions of extreme events remains an open issue since  by definition there are always few observations of  very extreme events, which does not fit well into standard theory. Here we used an assessment strategy which uses standardized Brier scores, Precision-Recall curves and Average Precision scores} \citep{steyerberg2010assessing,brownlee2020imbalanced,saito2015precision}.

\paragraph{Acknowledgment:}
 We thank Tom Britton, Anna Kiriliouk, Andreas Pettersson, Thordis Torainsdottir and Jenny Wadsworth for help and comments, {and two referees and an associate editor for comments which have lead to important improvements of the paper.}

\paragraph{R codes:} The data and the \texttt{R} codes are publicly available at \url{github.com/maudmhthomas/predict_extremeinfluenza}. Numerical optimizations used the R-function \texttt{optim} with the 1-dimensional integrals in the likelihoods calculated by the \texttt{R}-package \texttt{pracma}.

\bibliographystyle{abbrvnat}
\bibliography{biblioMaud}
\end{document}


\title{Web-based supporting materials for ``Real-time prediction of severe influenza epidemics using Extreme Value Statistics'' }
\author{ by Maud Thomas and Holger Rootz\'en}

\date{}

\maketitle

\section{Mutual independence of ILI incidence rates}

\begin{figure}[H]
	\begin{center}
		\vspace{4mm}
		\begin{tabular}{cc}
			\includegraphics[width = 0.45\textwidth]{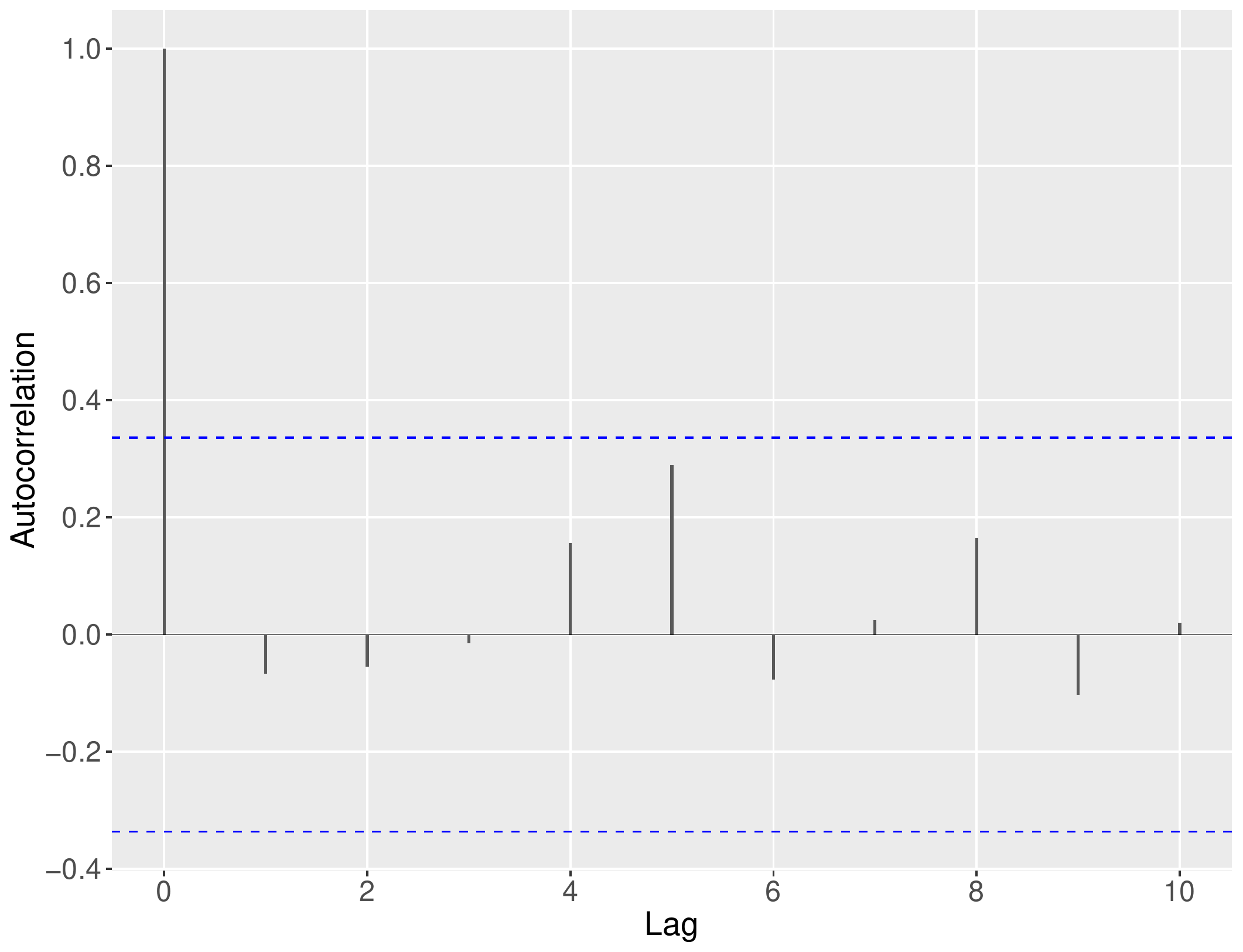}&
			\includegraphics[width = 0.45\textwidth]{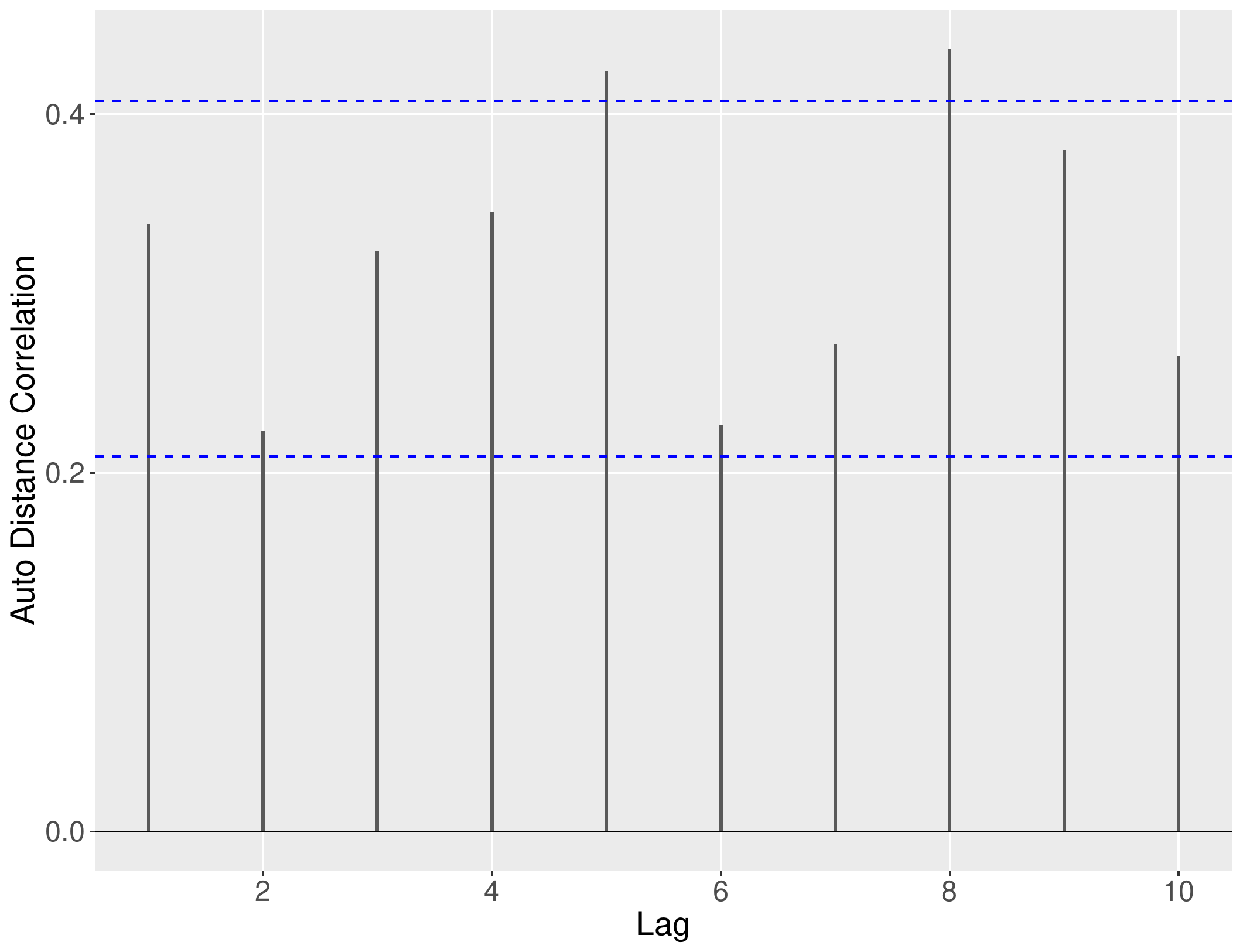}\\
			a) ACF for Week 3 ILI incidence rates & b) ADCF for Week 3 ILI incidence rates \\
		\end{tabular}\\~\\
	\end{center}
	\begin{center}
		\vspace{4mm}
		\begin{tabular}{cc}
			\includegraphics[width = 0.45\textwidth]{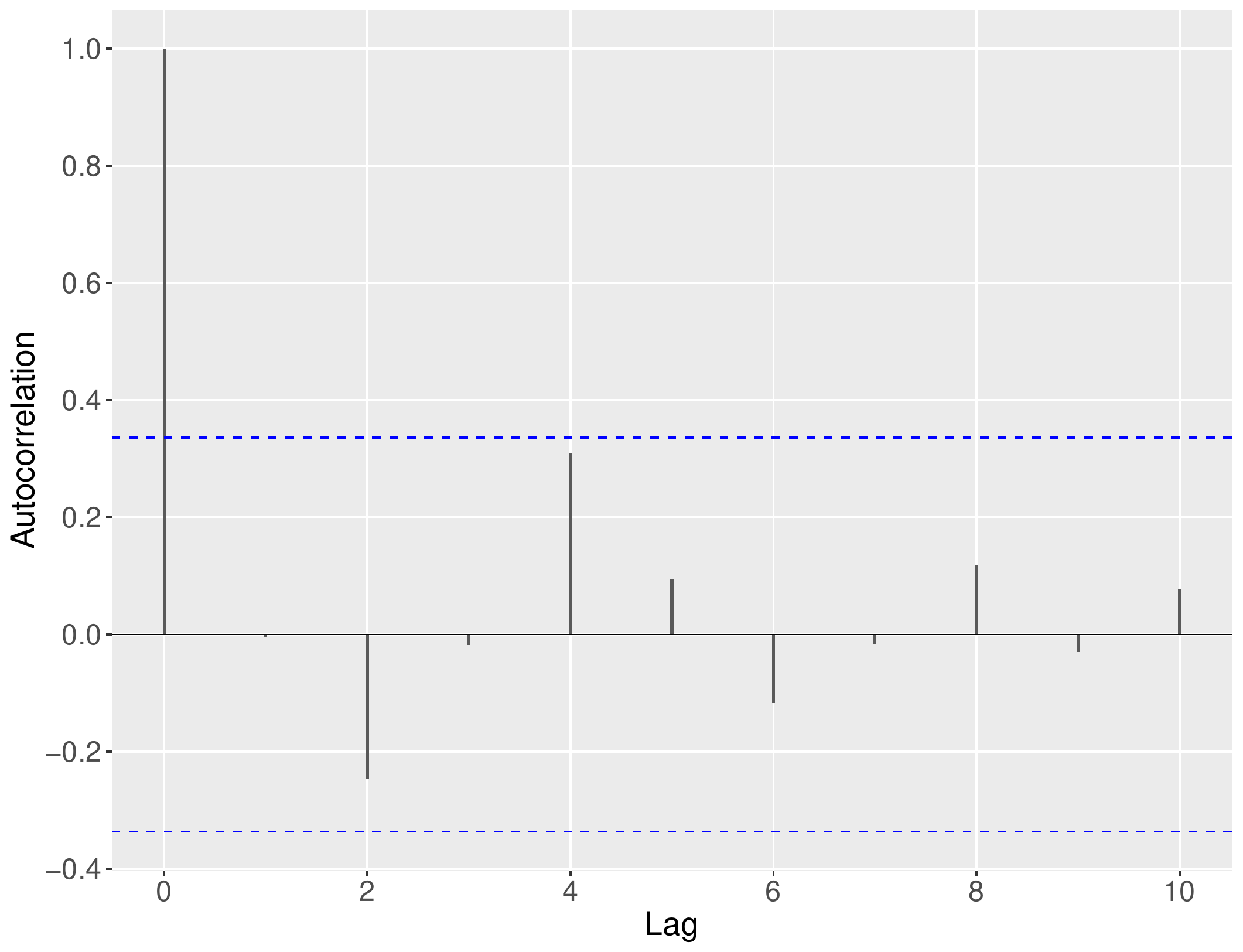}&
			\includegraphics[width = 0.45\textwidth]{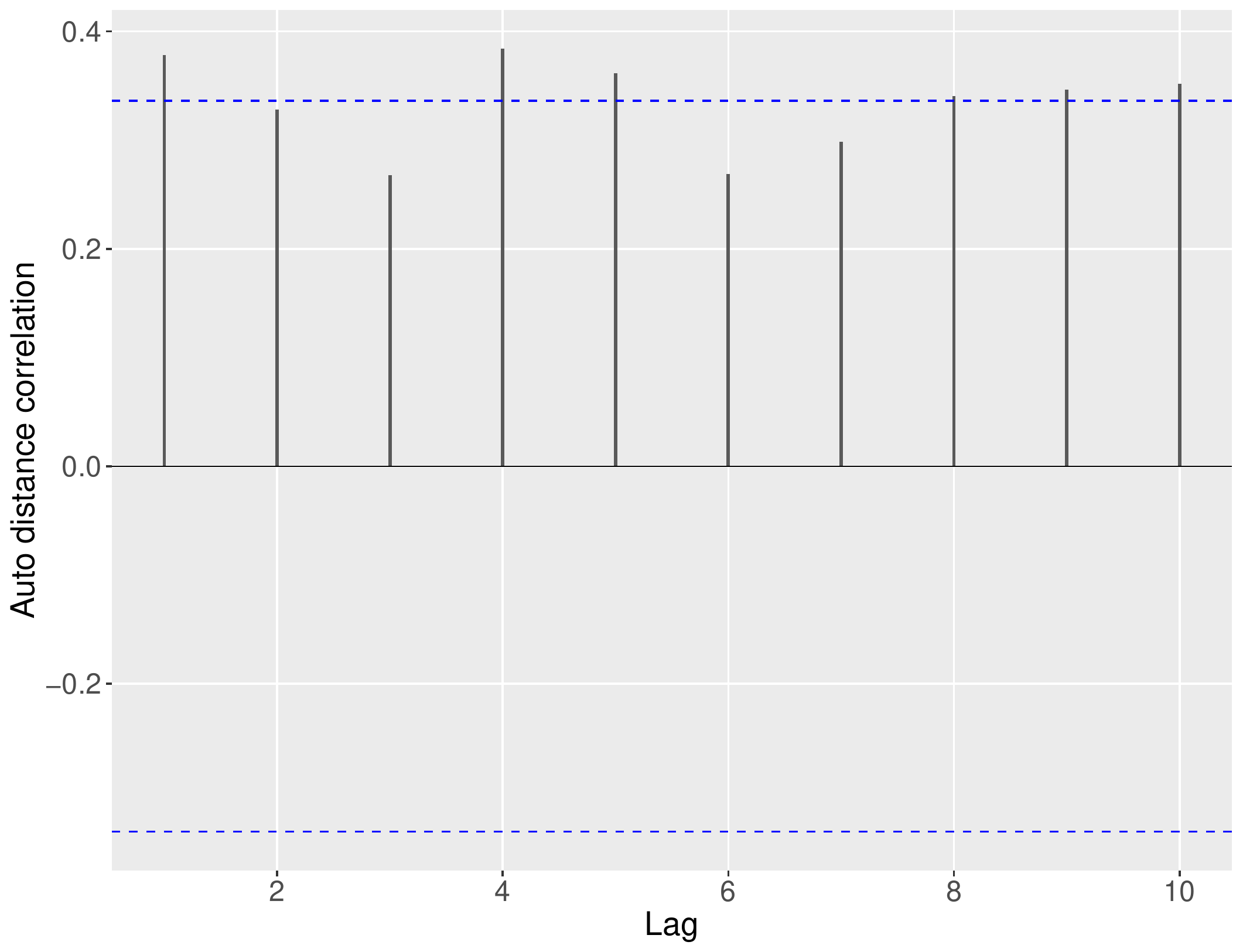}\\
			c) ACF for epidemic sizes & d) ADCF for epidemic sizes \\
		\end{tabular}
	\end{center}
	\caption{Auto Correlation Function (ACF) and Auto Distance Correlation Function (ADCF) for Week 3 ILI incidence rates and epidemic sizes, with blue lines showing significance level 0.95.}
	\label{fig:ac:acdc}
\end{figure}

We made autocorrelation, cross-correlation, auto-distance-correlation, and cross-distance-correlation plots of all pairs of the ILI incidence rates of Weeks 1, 2 and 3, and epidemic sizes. None of the plots suggested dependence. For  example, see Figure~\ref{fig:ac:acdc}.

\section{Marginal exponential fits}

\begin{figure}[!h]
 \begin{center}
 \begin{tabular}{cc}
 \includegraphics[width = 0.45 \textwidth]{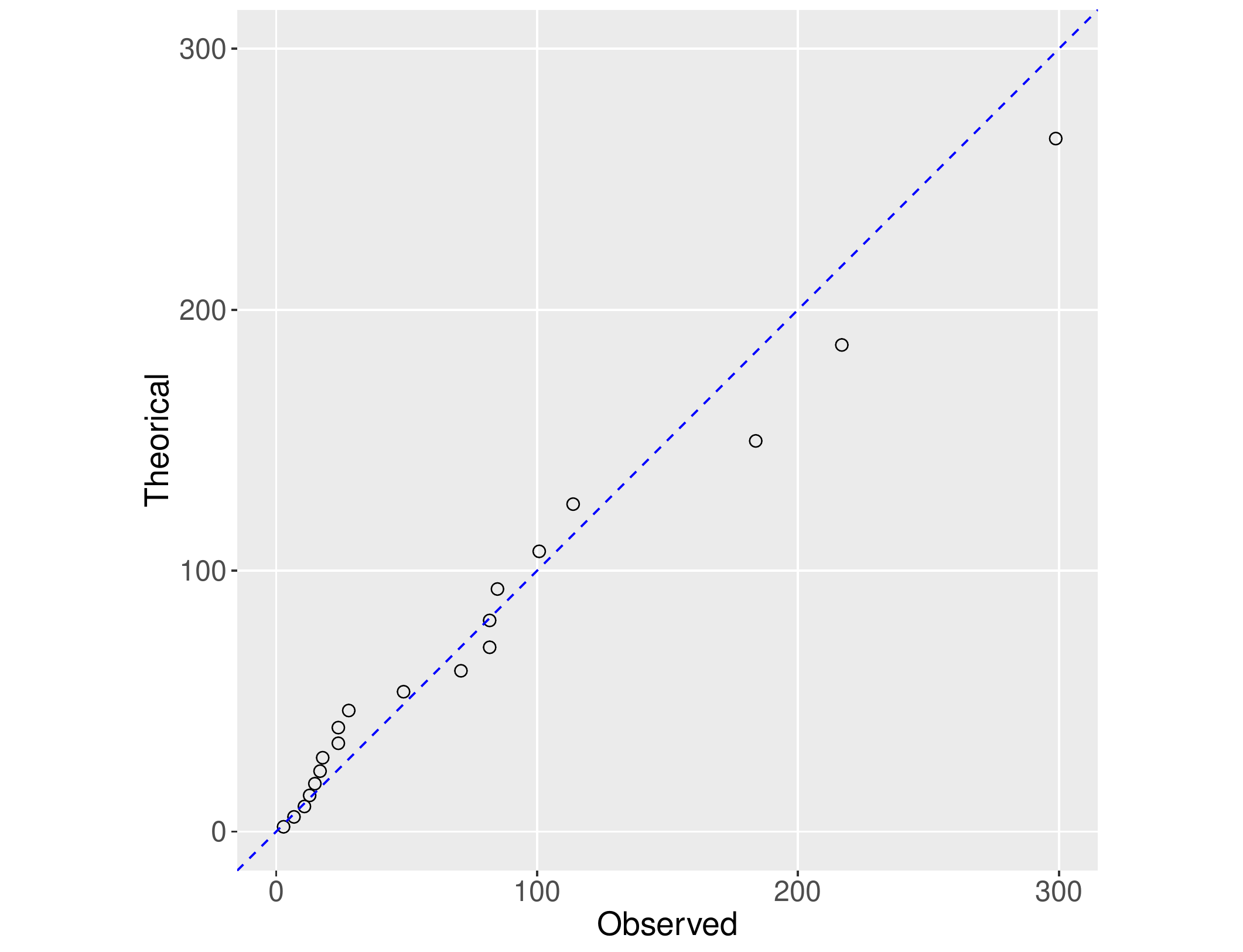}&
 \includegraphics[width = 0.45 \textwidth]{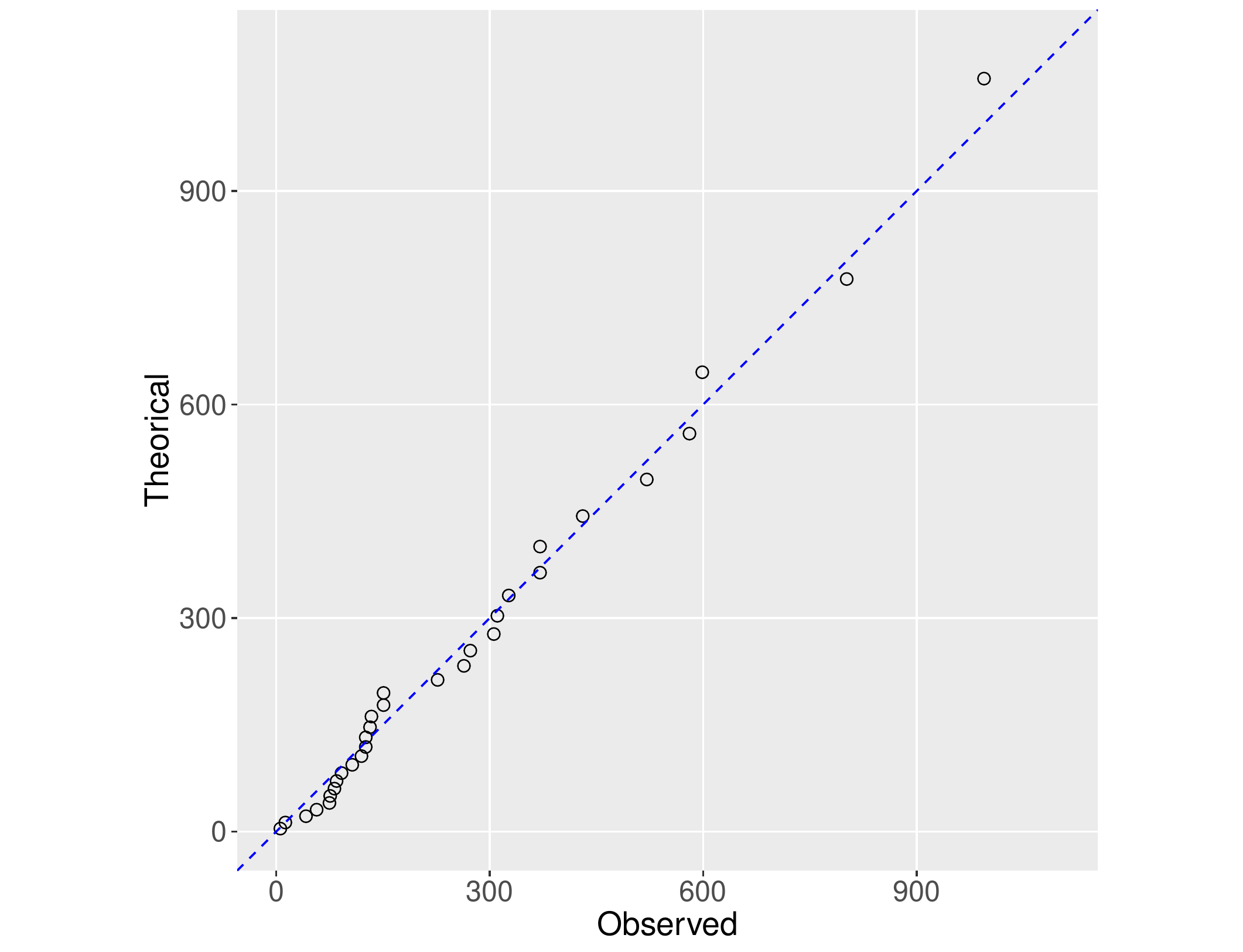}
  \\
   a) &  b)  \\~\\
\includegraphics[width = 0.45 \textwidth]{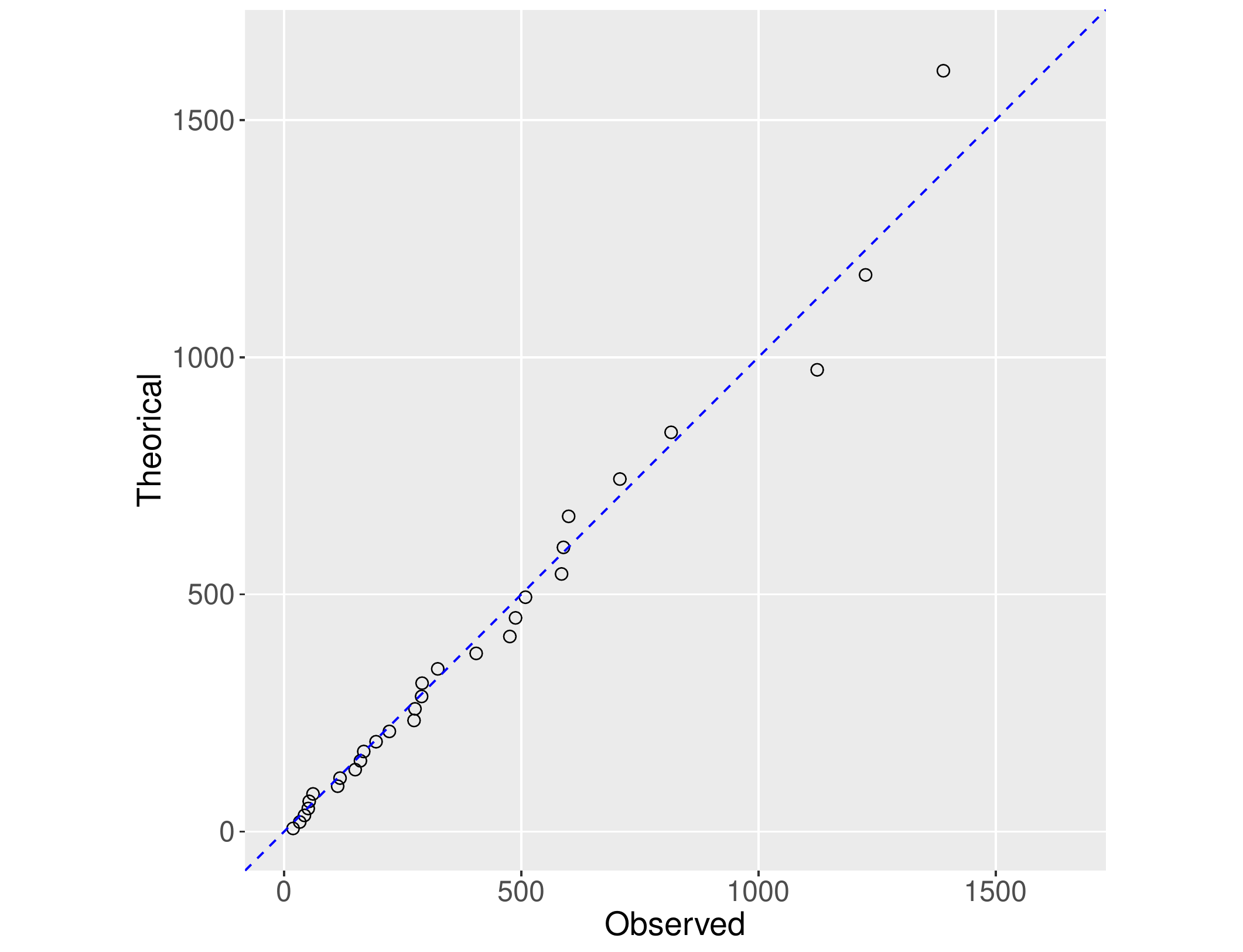}&
\includegraphics[width = 0.45 \textwidth]{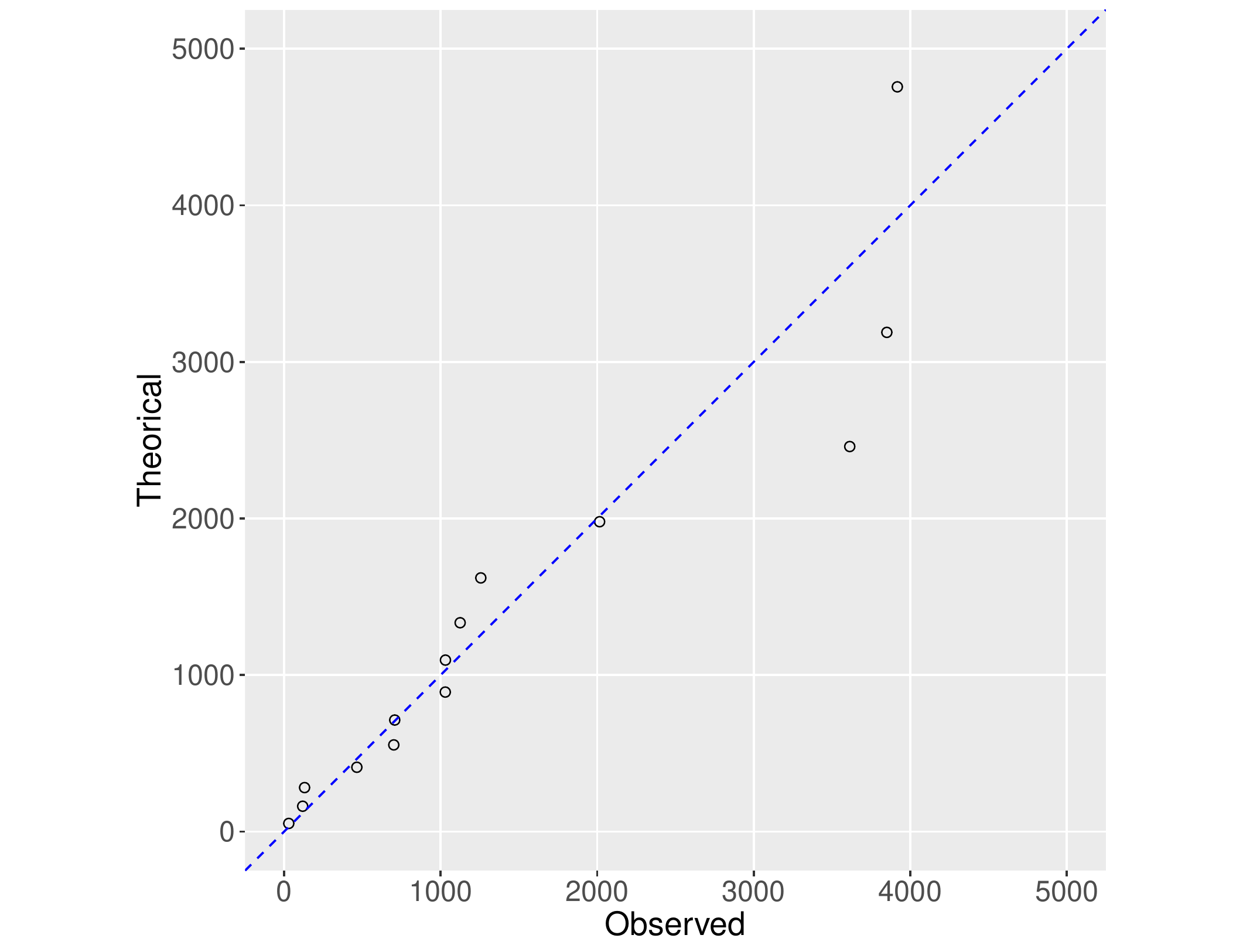}\\
 c)  & d)  \\
 \end{tabular}
\end{center}
 \caption{Quantile-quantile plots of the exponential fit of the  excesses of the 0.9-quantile of the ILI incidence rates for a) ILI incidence rates for Week 1, b) ILI incidence rates for Week 2, c) ILI incidence rates for Week 3 and of the  excesses of the 0.6-quantile of the sizes for d) the epidemic sizes.}
\label{fig:exp:qqplot}
\end{figure}

Table~\ref{tab:estimates:expo} gives the estimates of the scale parameters for the exponential fit of the Week 1, 2 and 3 ILI incidence rates and epidemic sizes and the corresponding QQ-plots are shown in Figure~\ref{fig:exp:qqplot}. The $p$-values for the likelihood ratio tests for the hypothesis $\gamma=0$ were $p=0.66$, $p=0.57$ and $p=0.64$, $p=0.98$ for the Week 1, 2 and 3 ILI incidence rates and epidemic sizes, respectively. 

\begin{table}[!h]
\begin{center}
\caption{Estimates of scale parameters of the positive  ILI rate excesses. CI is the 95\% confidence interval.}
\label{tab:estimates:expo}
\vspace{4mm}  
 \begin{tabular}{c|cccc}
& Week 1 ILI rates & Week 2 ILI  rates &  Week 3 ILI  rates & Epidemic sizes\\
\hline
Estimate & 72 & 256 & 392 & 1,428\\
 95\% CI & [40 ; 104] & [166; 347] & [252 ; 532]&[680 ; 2,175]\\
 \end{tabular}
 \end{center}
\end{table}

\section{Conditional prediction of extreme events}

\vspace*{2mm}
\noindent
Generalizing the notation in Section~3.2, write $\boldsymbol x_{i:j}=(x_i,\ldots,x_j)$ and $\boldsymbol x=(x_1,\ldots,x_d)$. Then, using the same conventions as in  Section~3.2, the U-representation of the density of d-dimensional GP distributions is 
\[
h_{\boldsymbol{U}}(\boldsymbol{x})= h_{\boldsymbol{U}}(\boldsymbol{x}; \boldsymbol{\sigma}, \boldsymbol{\gamma}  )= \frac{1_{\{\boldsymbol{x} \nleq \boldsymbol{0}\} }}{\esp[\mathe^{\max\boldsymbol{U}}]} \left(\prod_{j=1}^d \frac{1}{\sigma_j+\gamma_j x_j}\right)\int_{0}^\infty f_{\boldsymbol{U}}\left(\frac{1}{\boldsymbol{\gamma}}\log\left(1+\boldsymbol{\gamma}\frac{\boldsymbol{x}}{\boldsymbol{\sigma}} \right)+\log t\right) \mathd t
\]
where $\boldsymbol \sigma = (\sigma_1,\ldots, \sigma_d)$ and $\boldsymbol \gamma = (\gamma_1,\ldots, \gamma_d)$. Note that in the general case $h_{\boldsymbol{U}}$ depends both on the generator $\boldsymbol{U}$ and on the marginal parameters $\boldsymbol{\sigma}, \boldsymbol{\gamma}$.
For $\boldsymbol{\sigma} = \boldsymbol{1}, \boldsymbol{\gamma} = \boldsymbol{0}$ this expression reduces to Equation~(3) in Section~3.2. The next proposition gives the conditional prediction formulas for prediction of the exceedances of the $d$th component given the first $c$ components of a $d$-dimensional vector.

\begin{prp} \label{prp:cond:pred}
Let $\mathbf{X} = (x_1,\ldots,x_d)$ a $d$-dimensional GP vector with density $f_{\boldsymbol U}$ and let $1 \leq c < d$. The probability that $x_d$ will exceed a threshold $v_d$ given $\boldsymbol x_{1:c}=\boldsymbol x_{1:c}$ is given by
\begin{enumerate}
  \item[i)]  for $\boldsymbol x_{1:c} \nleq 0$ and $v_d \in \mathbb R$
\end{enumerate}
\begin{eqnarray*}
\mathbb{P}[X_d > v_d \mid \boldsymbol{x}_{1:c} =\boldsymbol x_{1:c}]
&=&
\begin{cases}
\frac{\int_{x_d>v_d} \int_{\boldsymbol{x}_{c+1:d-1}}  \phi(\boldsymbol{x}) \mathd \boldsymbol{x}_{c+1:d-1} \mathd x_d}{\int_{\boldsymbol{x}_{c+1:d-1}}  \phi(\boldsymbol{x}) \mathd \boldsymbol{x}_{c+1:d} }& \ \text{if $c<d-1$} \\~\\
\frac{\int_{x_d>v_d} \phi(\boldsymbol{x}) \mathd x_d}{\int_{\boldsymbol{x}_d}  \phi(\boldsymbol{x}) \mathd \boldsymbol{x}_{c+1:d}} & \ \text{if $c=d-1$;}
\end{cases}
\end{eqnarray*}
\begin{enumerate}
  \item[ii)]  for $\boldsymbol x_{1:c} \leq 0$ and $v_d >0$
\end{enumerate}
\begin{eqnarray*}
\mathbb{P}[X_d > v_d \mid \boldsymbol{x}_{1:c} =\boldsymbol x_{1:c}]
 &=&
\begin{cases}
\frac{\int_{x_d>v_d}\int_{\boldsymbol{x}_{c+1:d-1}} \phi(\boldsymbol{x}) \mathd \boldsymbol{x}_{c+1:d-1} \mathd x_d}{\int_{\boldsymbol{x}_{c+1:d-1} \nleq \boldsymbol 0} \phi(\boldsymbol{x}) \mathd \boldsymbol{x}_{c+1:d-1}} &\ \text{if $c<d-1$} \\~\\

\frac{\int_{x_d>v_d}  \phi(\boldsymbol x)\mathd x_d}{\int_{x_d>0}\phi(\boldsymbol x) \mathd x_d} & \ \text{if $c=d-1$;}
\end{cases}
\end{eqnarray*}
\begin{enumerate}
  \item[iii)]  for $\boldsymbol x_{1:c} \leq 0$ and $v_d\leq 0$
\end{enumerate}
\begin{eqnarray*}
\mathbb{P}[X_d > v_d \mid \boldsymbol{x}_{1:c} =\boldsymbol x_{1:c}]
&=&
\begin{cases}
\begin{aligned}
&\frac{\int_{x_d>0}  \int_{\boldsymbol{x}_{c+1:d-1}\leq \boldsymbol 0}   \phi(\boldsymbol{x}) \mathd \boldsymbol{x}_{c+1:d-1} \mathd x_d}{\int_{\boldsymbol{x}_{c+1:d-1} \nleq \boldsymbol 0}  \phi(\boldsymbol{x}) \mathd \boldsymbol{x}_{c+1:d-1}}\\
&+\frac{\int_{x_d>v_d} \int_{\boldsymbol{x}_{c+1:d-1}\nleq \boldsymbol 0}   \phi(\boldsymbol{x}) \mathd \boldsymbol{x}_{c+1:d-1}\mathd x_d }{\int_{\boldsymbol{x}_{c+1:d-1} \nleq \boldsymbol 0}   \phi(\boldsymbol{x}) \mathd \boldsymbol{x}_{c+1:d-1} }
\end{aligned}& \ \text{if $c<d-1$} \\~\\
1 & \ \text{if $c=d-1$.}
\end{cases}
\end{eqnarray*}
\end{prp}

\begin{exe}  For $d=3$ and $c=2$ we get
\begin{eqnarray*}
\mathbb{P}\left[X_3 > v_3 \mid \mathbf{X}_{1:2}= \mathbf{x}_{1:2} \right] =
\begin{cases}
\frac{\int_{x_3>v_3}   \phi(\boldsymbol x)\mathd x_3}{\int_{x_3}\phi(\boldsymbol x) \mathd x_3} & \ \text{if $\mathbf{x}_{1:2} \not \leq \mathbf{0}$ and $v_3 \in \mathbb{R}$;} \\~\\
\frac{\int_{x_3>v_3}   \phi(\boldsymbol x)\mathd x_3}{\int_{x_3>0}\phi(\boldsymbol x) \mathd x_3} & \ \text{if $\mathbf{x}_{1:2}  \leq \mathbf{0}$ and $v_3>0$;} \\~\\
1 & \ \text{if $\mathbf{x}_{1:2}  \leq \mathbf{0}$ and $v_3\leq0$.}
\end{cases}
\end{eqnarray*}
\end{exe}

\section{Probability density functions for the 3-dimensional GP distributions}\label{an:densities:formulas}

In this section, we recall the density functions for the 3-dimensional GP distributions that we use in Section~4.2 (for further details see Section~4.1. in [Kiriliouk et al. 2019]).

Let $\mathbf U=(U_1,U_2,U_3)$ be a 3-dimensional random vector such that $\esp[\mathe^{\max U}]< \infty$, let $f_{\mathbf U}$ be the probability density function of $\mathbf U$ and $f_i$ the density function of its $i$-th component $U_i$, for $i=1,2,3$. 
\paragraph{Generators with independent Gumbel components}
\vspace*{2mm}\quad \\

If $\mathbf U=(U_1,U_2,U_3)$ with $U_i,i=1,2,3$ independent and distributed according to a Gumbel distribution with parameters $\alpha_i$ and $\beta_i$. The marginal density $f_i$ is given by 
\begin{equation}\label{eq:gumbel} 
f_i(x_i)= \alpha_i\exp \left(-\alpha_i(x_i-\beta_i)\right)\exp \left(-\exp \left(-\alpha_i(x_i-\beta_i)\right)\right),  \qquad \alpha_i>0,\,\beta_i\in\mathbb{R}
\end{equation}
where the requirement $\esp[\mathe^{\max \boldsymbol{U}}]< \infty$ imposes the restriction $\alpha_i> \boldsymbol{1}$ for $i=1,2,3$. The density of a 3-dimensional GP distribution with independent Gumbel generators is then given by 
\begin{equation}
h_{\mathbf U}(\mathbf x) = \frac{\int_0^\infty \prod_{i=1}^3 \alpha_i (t \mathe^{x_i-\beta_i})^{-\alpha_i}\mathe^{-(t\mathe^{x_i-\beta_i})^{-\alpha_i}}\mathd t}{\int_0^\infty \left(1- \prod_{i=1}^3 \mathe^{-(t/\mathe^{\beta_j})^{-\alpha_j}} \right) \mathd t} \,. 
\end{equation}

\paragraph{Generators with independent reverse exponential components}
\vspace*{2mm}
\noindent

If $\mathbf U=(U_1,U_2,U_3)$ with $U_i,i=1,2,3$ independent and distributed according to a reverse exponential distribution with parameters $\alpha_i$ and $\beta_i$. The marginal density $f_i$ is given by 
\begin{equation}\label{eq:reverse:expo} 
f_i(x_i)=\exp\left\{(x_i + \beta_i)/\alpha_i \right\}/\alpha_i  \qquad x_i \in (-\infty, -\beta_i), \, \alpha_i>0,\,\beta_i\in\mathbb{R} \, . 
\end{equation}
 The density of a 3-dimensional GP distribution with independent reverse exponential generators is then given by 
\begin{equation}
h_{\mathbf U}(\mathbf x) = \frac{\left(\mathe^{-\max(\mathbf x + \boldsymbol \beta)}\right)^{\sum_{i=1}^3 1/\alpha_i+1} }{\int_0^{\infty} \left(1 - \prod_{i=1}^3 \min\left(\mathe^{\beta_i}t, 1 \right)\right)^{1/\alpha_i} \mathd t}\frac{1}{1+\sum_{i=1}^3 1/\alpha_i} \prod_{i=1}^3 \frac{1}{\alpha_i} \left(\mathe^{x_i + \beta_i} \right)^{1/\alpha_i}\, ,
\end{equation}
where $\boldsymbol \beta = \left( \beta_1, \beta_2, \beta_3 \right)$.

\paragraph{Generators with independent reverse Gumbel components}
\vspace*{2mm}
\noindent

If $\mathbf U=(U_1,U_2,U_3)$ with $U_i,i=1,2,3$ independent and distributed according to a reverse Gumbel distribution with parameters $\alpha_i$ and $\beta_i$. The marginal density $f_i$ is given by 
\begin{equation*}
f_j(x_j) = \alpha_j\exp\{\alpha_j(x_j -\beta_j)\}\exp[-\exp\{\alpha_j(x_j -\beta_j)\}], \qquad \alpha_j>0,\,\beta_j\in\mathbb{R}.
\end{equation*}
Calculations very similar to the derivation of Equation~(7.2) in [Kiriliouk et al., 2019] show that the resulting density of a 3-dimensional GP distribution with independent reverse exponential generators is then given by 
\begin{equation}\label{density:revgumbel}
h_{\boldsymbol U}(\boldsymbol x, \boldsymbol 1, \boldsymbol 0) = \frac{\int_{0}^\infty \prod_{i=1}^d \alpha_j \left(t \mathe^{x_j- \beta_j} \right)^{\alpha_j}\exp\left(-\left(t \mathe^{x_j- \beta_j}\right)^{\alpha_j}\right)\mathd t}{\int_{0}^\infty \left(1 -  \prod_{i=1}^n \mathe^{-(t\mathe^{-\beta_j})^{\alpha_j}}\right)\mathd t}.
\end{equation}


\section{Model simplification and estimates of parameters of the Gumbel model}

To ensure identifiability, the first component of the location parameter $\beta_1$ was fixed, equal to 0. Table~\ref{tab:standard:model} gives the estimated parameters for the selected Gumbel model. Table~\ref{tab:simplify:model} gives AIC, BIC and log-likelihood tests for simplification of the Gumbel model and indicates that the full model (M1) should not be  simplified further.

\begin{table}[!h]
\begin{center}
\caption{Estimates of parameters of the Gumbel model.}
 \label{tab:standard:model}
\vspace{4mm}
 \begin{tabular}{c|ccccc}
Parameter & $\alpha_1$ & $\alpha_2$ & $\alpha_3$ & $\beta_2$ & $\beta_3$  \\
\hline
Week 3 ILI incidence rates & 2.22 & 10.37 & 3.21 & 0.84 & 0.59 \\
Epidemic sizes & 2.22 & 38.77 & 1.76 & 0.89 & -0.70
 \end{tabular}
 \end{center}
\end{table}

\begin{table}[!h]
\begin{center}
 \caption{AIC, BIC and LR tests for simplification of the Gumbel model. }
  \label{tab:simplify:model}
 \vspace{4mm}
 \begin{tabular}{cccccc}
Model & Parameters  & AIC & BIC & LR $p$-values  \\
\hline
M1 & $\alpha_1,\alpha_2,\alpha_3,\beta_2,\beta_3$ &  194 & 202&  - \\
M2 & $\alpha_1,\alpha_2,\alpha_3$ & 257& 265 & $<10^{-3}$ \\
M3 & $\alpha,\beta_2,\beta_3$ &  220& 227 &   $<10^{-3}$ \\
M4 & $\alpha$ & 250 & 252 &  $<10^{-3}$ \\
 \end{tabular}\\~\\
 a)   Week 3 ILI incidence rates \\~\\
 \end{center}
 \begin{center}
 \begin{tabular}{cccccc}

Model & Parameters  & AIC & BIC & LR p-values  \\
\hline
M1 & $\alpha_1,\alpha_2,\alpha_3,\beta_2,\beta_3$ & 227 & 234 &  - \\
M2 & $\alpha_1,\alpha_2,\alpha_3$ & 334 & 341 &  $<10^{-3}$ \\
M3 & $\alpha,\beta_2,\beta_3$ &  258 & 266 &  $<10^{-3}$ \\
M4 & $\alpha$ &  311 & 312 & $<10^{-3}$ \\
 \end{tabular}\\~\\
 b) Epidemic sizes
 \end{center}
\end{table}

\section{Estimation of prediction probabilities from the true model on simulated data}

\begin{figure}[!h]
 \begin{center}
 \begin{tabular}{cc}
 \includegraphics[width = 0.36\textwidth]{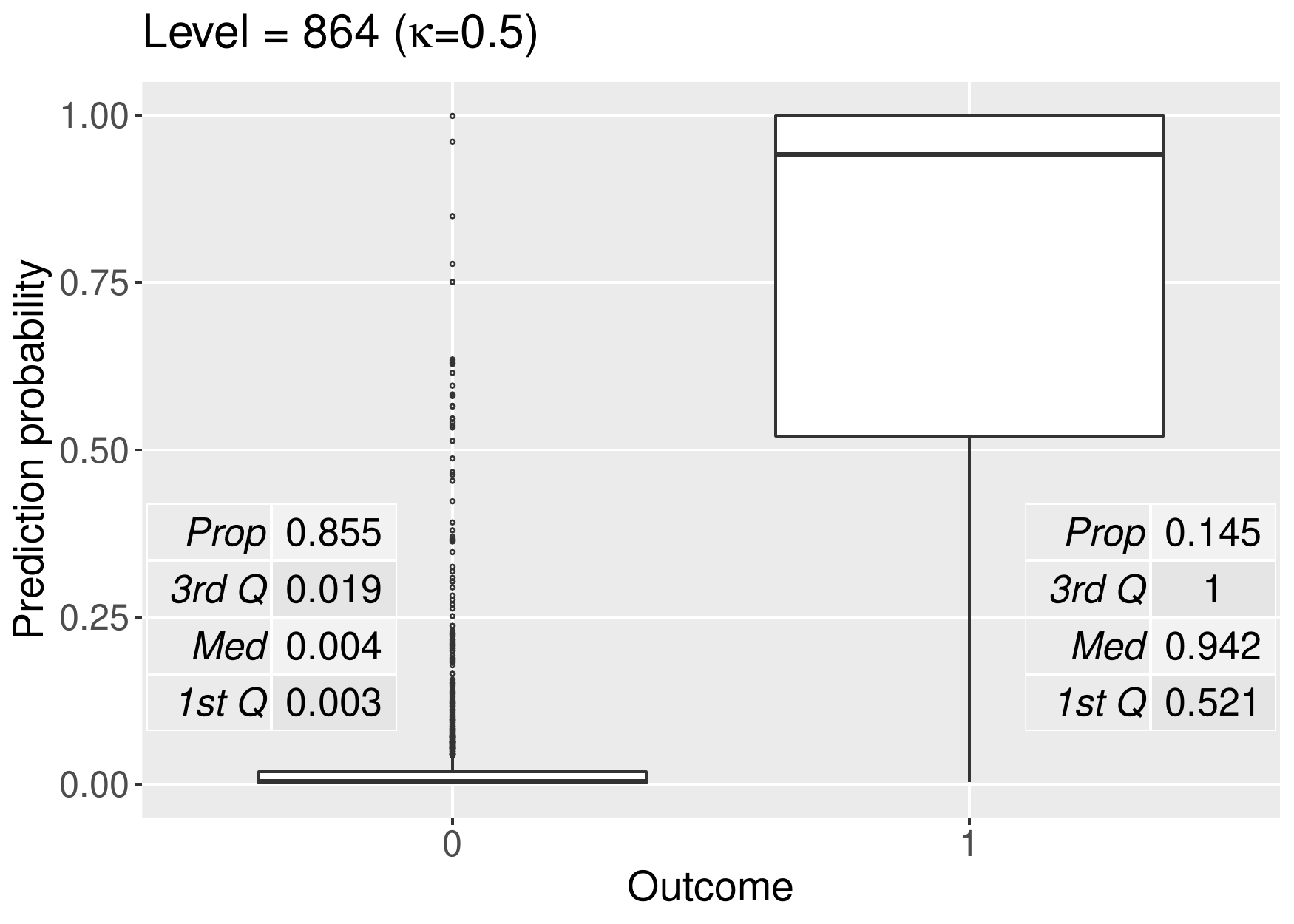}&
 \includegraphics[width = 0.32\textwidth]{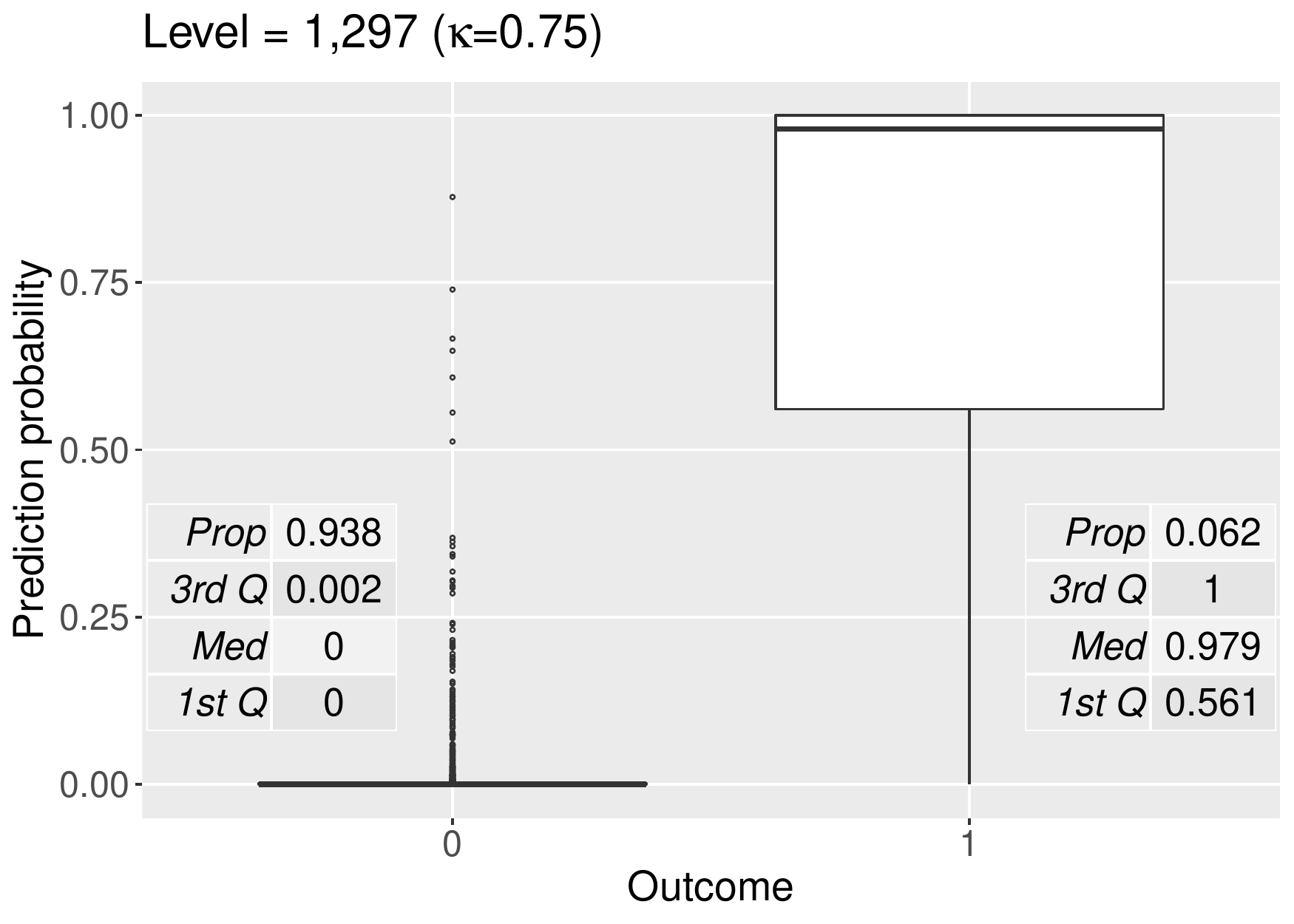}\\
  \includegraphics[width = 0.32\textwidth]{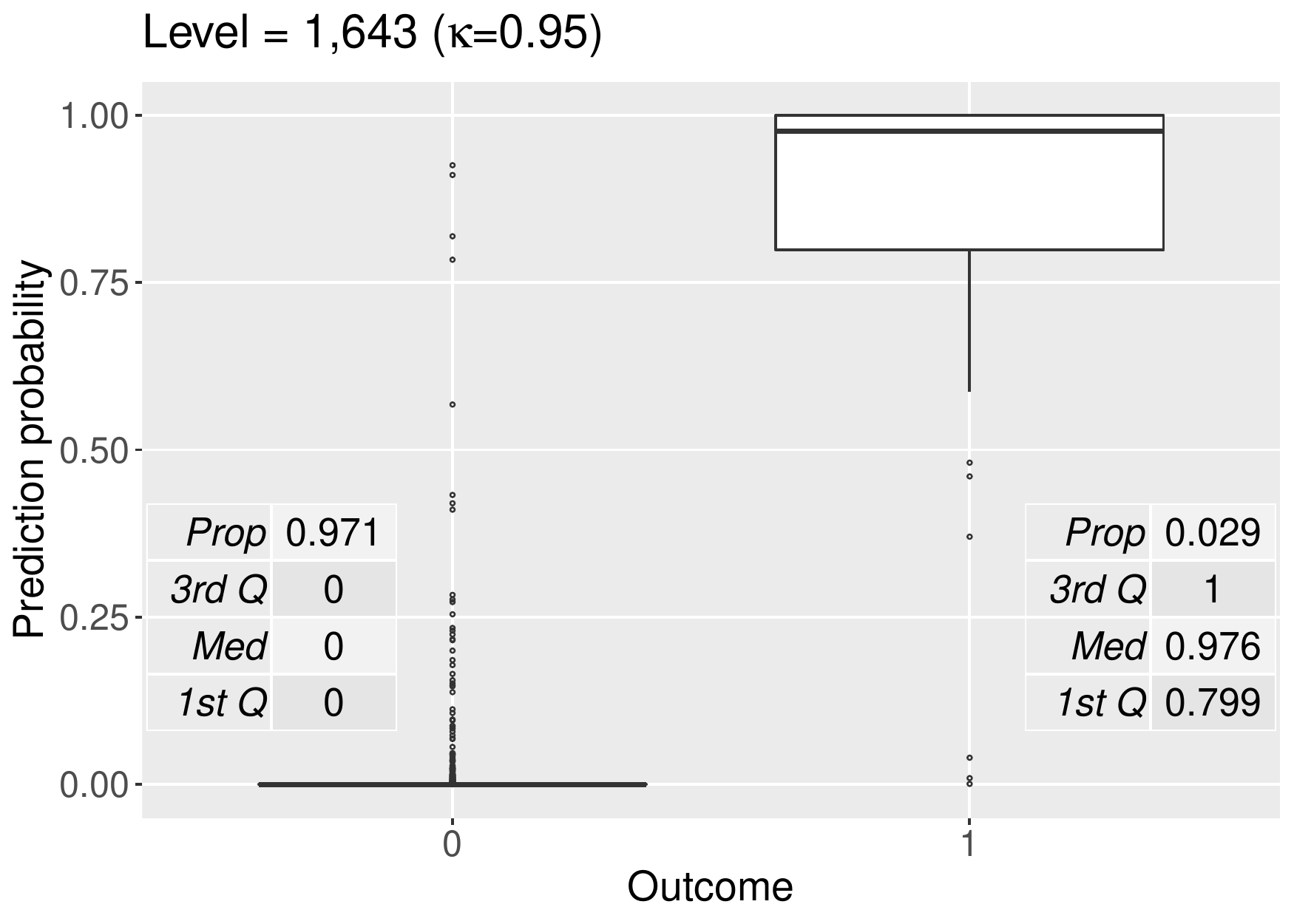}&
\includegraphics[width = 0.32\textwidth]{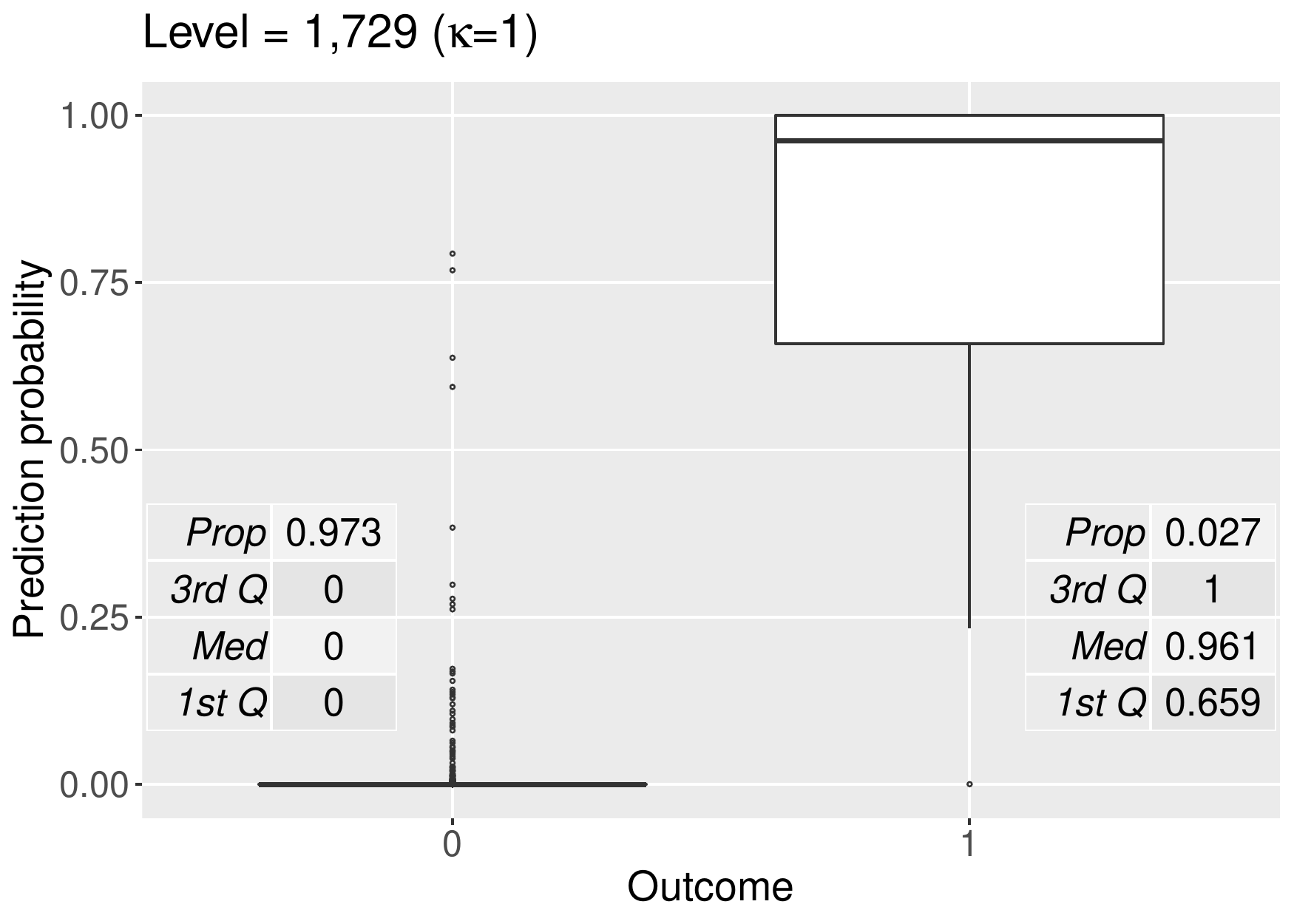}\\
\multicolumn{2}{c}{a)}\\
 \includegraphics[width = 0.32\textwidth]{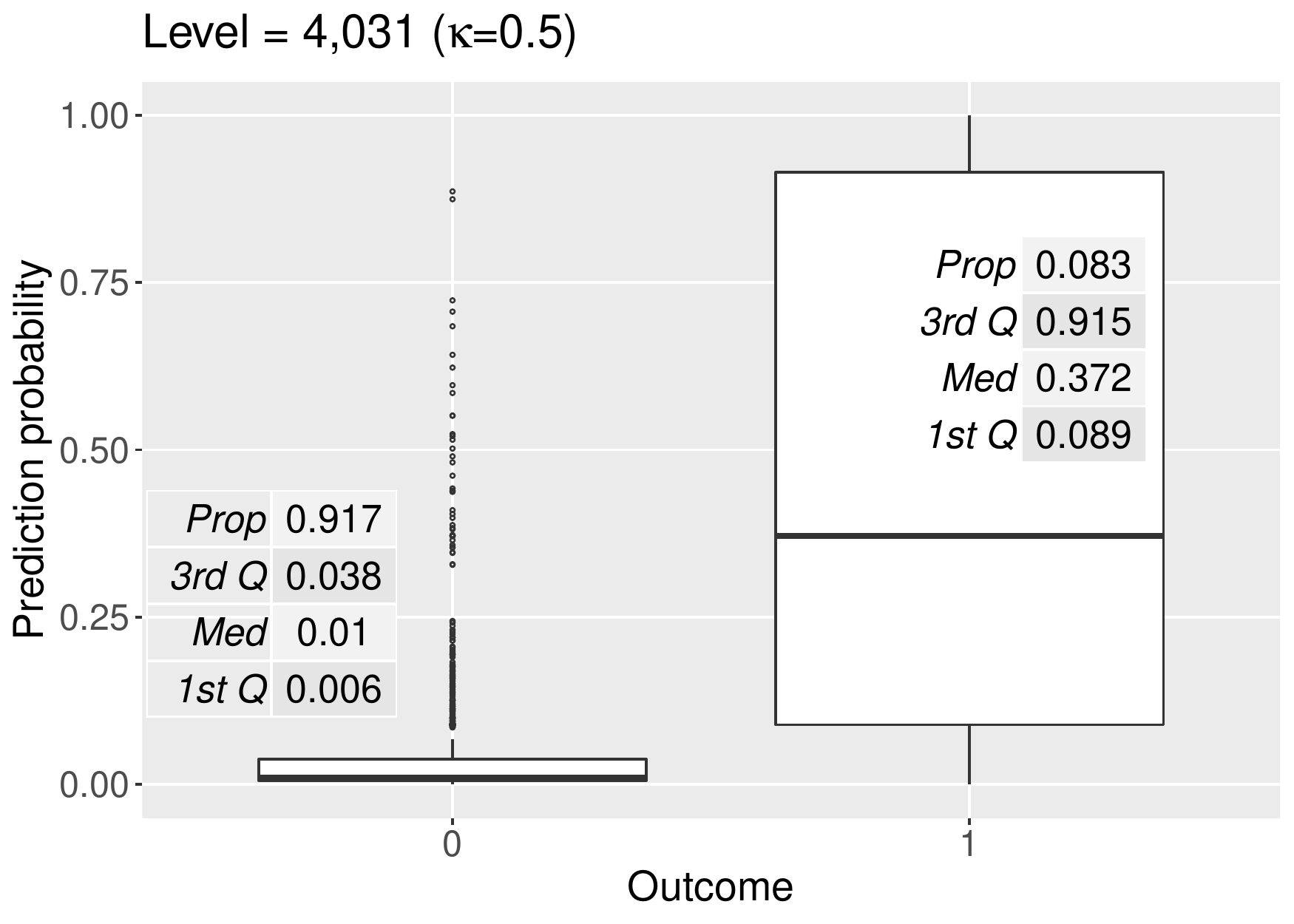}&
 \includegraphics[width = 0.32\textwidth]{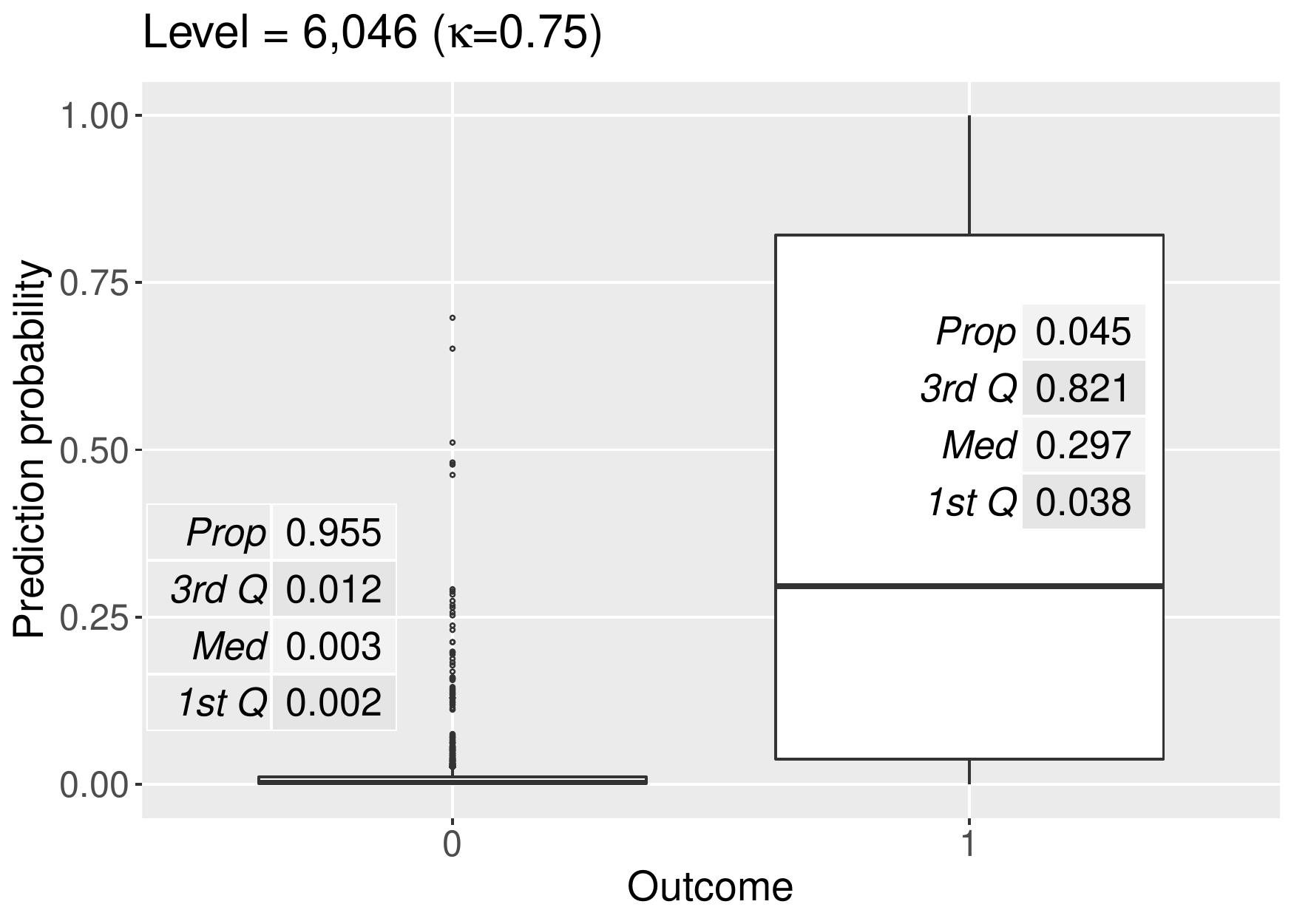}\\
  \includegraphics[width = 0.32\textwidth]{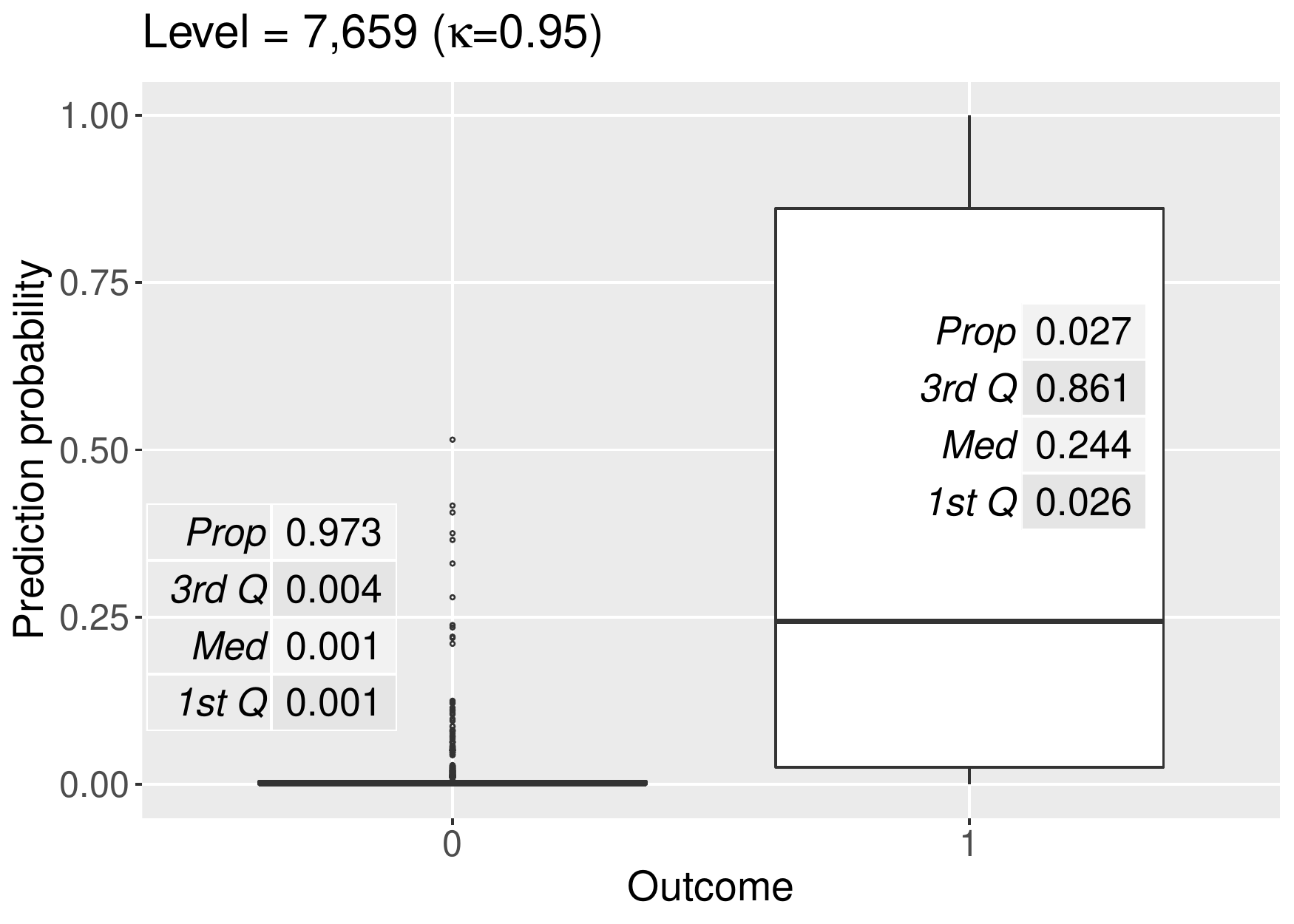}&
\includegraphics[width = 0.32\textwidth]{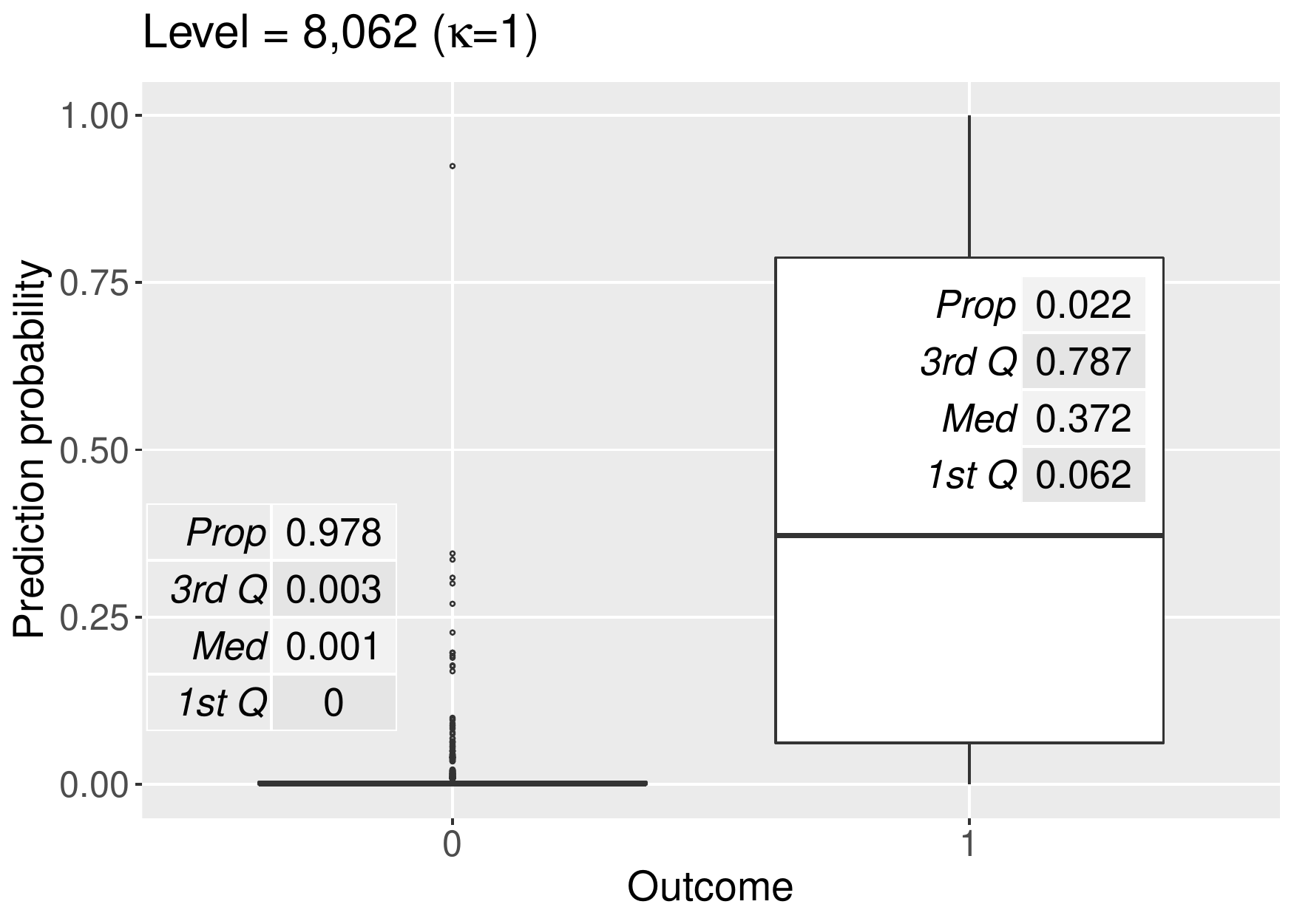}\\
\multicolumn{2}{c}{b)}
   \end{tabular}
\end{center}
 \caption{Boxplots of prediction probabilities of level exceedances for simulated data obtained from the Gumbel model with the true parameters for a) Week 3 ILI incidence rates and epidemic sizes. Outcome is equal to 0 if there was no exceedance and 1 otherwise. The tables give the proportions of 0-outcomes and of 1-outcomes, and the quartiles of the prediction probabilities.}
\label{fig:brier:simulated:gumbel}
\end{figure}

For the simulated data presented in Section~5.2, the true parameters for the Gumbel model are known. Figure~\ref{fig:brier:simulated:gumbel} shows the boxplots of the prediction probabilites obtained from the model with the true parameters. 

